\documentclass[12pt,a4paper]{article}
\usepackage{amssymb,enumitem,booktabs,subfig,xcolor,microtype,setspace,bm,longtable,paralist,dsfont}
\usepackage[top=1in, bottom=1in, left=1in, right=1in]{geometry}
\usepackage{natbib,url,palatino,eulervm}
\usepackage[pdftex,colorlinks=true,hypertexnames=false]{hyperref}
\definecolor{darkblue}{rgb}{0,0,.6}
\hypersetup{citecolor=darkblue,linkcolor=darkblue,urlcolor=darkblue}
\usepackage{amsmath,orcidlink,rotating,tikz,amsfonts,epsfig}
\usepackage{graphics,graphicx,lscape}
\usepackage[normalem]{ulem}
\usetikzlibrary[decorations.shapes]
\usetikzlibrary[petri]
\usetikzlibrary{positioning}
\usepackage[linewidth=1pt]{mdframed}
\allowdisplaybreaks[4]

\usepackage[font=small]{caption}
\newsavebox\CBox
\def\textBF#1{\sbox\CBox{#1}\resizebox{\wd\CBox}{\ht\CBox}{\textbf{#1}}}

\providecommand{\U}[1]{\protect\rule{.1in}{.1in}}
\renewcommand{\baselinestretch}{1.2}
\graphicspath{{plots/}}

\newcommand{\blind}{0}

\usepackage{amsthm,thmtools}
\usepackage{mathrsfs}

\makeatletter
\def\th@newremark{\th@remark\thm@headfont{\bfseries}}
\makeatletter
\theoremstyle{newremark}

\declaretheoremstyle[spaceabove=6pt, spacebelow=6pt, headfont=\bfseries, notefont=\mdseries, notebraces={(}{)}, bodyfont=\normalfont,
postheadspace=0.5em]{mystyle}

\makeatletter
\newcommand*{\addFileDependency}[1]{\typeout{(#1)}\@addtofilelist{#1}\IfFileExists{#1}{}{\typeout{No file #1.}}}
\makeatother
\newcommand{\Rlogo}{\protect\includegraphics[height=1.8ex,keepaspectratio]{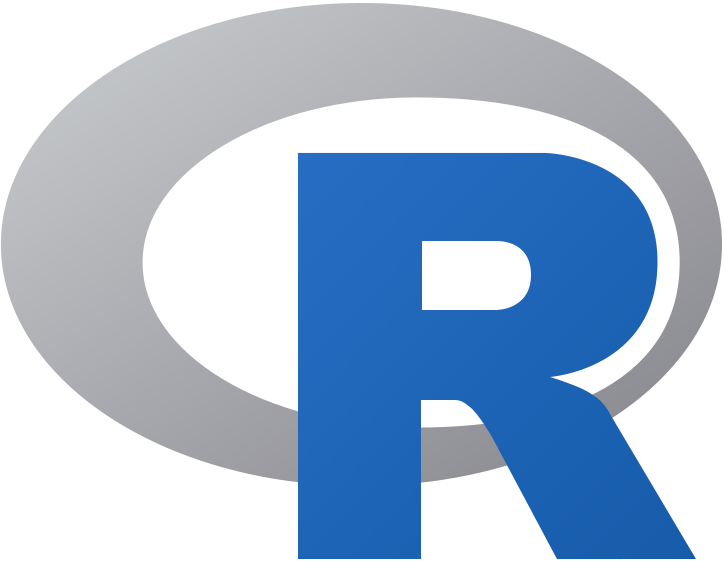}}
\begin{document}

\def\spacingset#1{\renewcommand{\baselinestretch}
{#1}\small\normalsize} \spacingset{1}

\if0\blind
{
\title{\bf Forecasting age distribution of deaths across countries: Life expectancy and annuity valuation}
\author{
\normalsize  Han Lin Shang \orcidlink{0000-0003-1769-6430} \\
\normalsize  Department of Actuarial Studies and Business Analytics \\
\normalsize  Macquarie University \\
\\
\normalsize  Steven Haberman \orcidlink{0000-0003-2269-9759} \\
\normalsize  Bayes Business School \\ 
\normalsize  City St George's, University of London
}
\date{}
\maketitle
} \fi

\if1\blind
{
\title{\bf Forecasting age distribution of deaths across countries: Life expectancy and annuity valuation}
\author{}
\date{}
\maketitle
} \fi

\begin{abstract}
In this paper, we provide a comprehensive cross-country validation study of compositional mortality modeling and forecasting methods. Thus, we consider two one-to-one transformations: the cumulative distribution function and the centered log-ratio transformation in compositional data analysis. Between the two transformations, the cumulative distribution function provides a scale-free way to visualize the gender gap and cross-country heterogeneity in the probability of dying by sex and country. Drawing on age-specific period life-table death counts from 24 countries in the \cite{HMD25}, we assess and compare the point and interval forecast accuracy of the two transformations, using the same forecasting method. Enhancing the forecast accuracy of period life-table death counts is of significant value to demographers, who rely on such forecasts to estimate survival probabilities and life expectancy, and to actuaries, who use them to price annuities across various entry ages and maturities.

\vspace{.1in}

\end{abstract}

\newpage
\spacingset{1.47}

\section{Introduction}\label{sec:intro}

Actuaries and demographers have long been engaged in mortality modeling and forecasting, dating back to the early 20\textsuperscript{th} century, as a response to the financial challenges posed by rapid mortality improvements. These improvements affected the accuracy of annuity calculations and the sustainability of government pension systems \citep{Pollard87}. Forecasting life tables for annuitants was a key topic at the 5\textsuperscript{th} International Congress of Actuaries in Berlin in 1906 \citep{CW35}. In the field of statistical demography, numerous researchers have proposed innovative approaches to modeling and forecasting mortality by age, sex, and cause, across time and space. Special emphasis has been placed on statistical time-series extrapolation models \citep[see, e.g.,][for comprehensive reviews]{Booth06, BT08, CBD08, SBH11, BCB23}. Within mortality modeling, at least three instruments are commonly used: age-specific mortality rates, survival probabilities, and age distribution of death counts. \cite{BKO+19} provides a detailed analysis of these mortality instruments, highlighting their respective advantages and limitations. All three instruments can be derived from one another using standard life-table relationships \citep[see][Chapter~3]{PHG01}. Computationally, the conversion among these instruments can be achieved using the \textit{LifeTable} function in the MortalityLaws package \citep{Pascariu25}.

Among the three mortality instruments, we focus on life-table death counts ($d_x$) and model their evolution over time and entire ages from $x=0$ to 110+. The $d_x$ is designed for a synthetic cohort (often normalized), not registered deaths, and that once one works on the $d_x$ scale, the mapping to actuarial quantities is straightforward. In particular,
\begin{equation*}
l_{x+1} = l_x-d_x, \qquad q_x = \frac{d_x}{l_x}, \qquad p_x = 1-q_x,
\end{equation*}
so survival probabilities and annuity factor follow directly from the number of people alive $l_x$ sequence (if needed, one can obtain age-specific mortality $m_x$ from the probability of dying $q_x$ using standard approximation in the \textit{LifeTable} function).

Since it does not depend on exposure to risk, these synthetic death counts reveal a clear shifting pattern, with deaths progressively moving from younger to older ages. Between cohort and period life-table death counts, we choose to work with the latter, as period life tables are not only complete, but also reflect the prevailing mortality conditions within a given time frame \citep[see also][]{Oeppen08, BCO+17}. A time series of period life-table death counts enables the analysis of temporal changes in key longevity indicators, such as the modal age at death \citep[see, e.g.,][]{CanudasRomo10}, as well as measures of lifespan variability, including standard deviation, interquartile range, and Gini coefficient \citep[see, e.g.,][]{VZV11, VC13}.

In demography, \cite{Oeppen08} introduced the use of principal component analysis to model and forecast the age distribution of deaths within a compositional data analysis framework, treating age-specific life-table death counts as compositional data. As with all compositional data, these counts are subject to inherent constraints, typically bounded between zero and a fixed upper limit. Consequently, the sample space of compositional data is a simplex, defined as
\begin{equation*}
\mathbb{S}^{D-1}=\left\{(d_1,\dots,d_D)^{\top},\; d_x\geq 0, \; \sum^D_{x=1}d_x = c\right\},
\end{equation*}
where $\mathbb{S}$ denotes a simplex, which is a $(D-1)$-dimensional subset of real-valued space $R^{D-1}$, $^{\top}$ denotes vector transpose and $c$ is a fixed constant, set typically to one (portions) \citep{SDG+15}, $10^6$ parts per million \citep{SDG+15}, or $10^5$ in the life-table death counts.

Due to the non-negativity and summability constraints inherent in compositional data, conventional linear techniques may be inappropriate, as they do not account for the unique variance-covariance structure of such data. A common strategy is first to transform the data to remove the summability constraint, thereby enabling the application of linear methods. One widely used approach is the family of log-ratio transformations, particularly the centered log-ratio (CLR) transformation \citep[see, e.g.,][]{Aitchison86}. An alternative is the cumulative distribution function (CDF) transformation, which leverages monotonicity to effectively handle the presence of zero values \citep[see][]{SH25}.

This paper makes three key contributions. Based on the early work of \cite{SH25b, SH25}, we evaluate and compare the point forecast accuracy of the CLR and CDF transformations, using data from 24 countries in the \cite{HMD25}, with time series starting in or before 1950. Although the existing literature has largely focused on point forecasts, we implement a computationally efficient approach for constructing pointwise prediction intervals and compare the interval forecast accuracy between the CLR and CDF transformations. From a forecasting perspective, this study presents a comprehensive comparison between countries and further explores the heterogeneity of mortality, as well as its implications for forecasting life expectancy and pricing annuities. In addition, the CDF transformation provides a scale-free measure to visually compare the probability of dying between genders and across countries.

The remainder of this paper is structured as follows. Section~\ref{sec:2} describes the multi-country data sets used in the analysis. Section~\ref{sec:3} revisits the CLR and CDF transformations and introduces a forecasting method based on principal component analysis. In Section~\ref{sec:3_4}, we introduce new visualization plots, based on the CDF transformation, to compare the age distribution of deaths between sexes in the same country and across countries. In Section~\ref{sec:4}, we assess point forecast accuracy using the Kullback-Leibler and Jensen-Shannon divergences. For evaluating interval forecast accuracy, we consider the empirical coverage probability and its absolute deviation from the nominal coverage level, as well as the interval score proposed by \cite{GR07}. Section~\ref{sec:5} applies the CDF transformation to estimate the single-premium temporary immediate annuity prices for various entry ages and maturities for female and male populations in 24 countries. Section~\ref{sec:6} concludes with a summary and suggestions for future extensions of the proposed methodology.

\section{Age distribution of death counts}\label{sec:2}

The data sets used in this study are sourced from the \cite{HMD25}. We consider single-year ages ranging from 0 to 110. Of the 41 predominantly developed countries, we select 24 for which life-table death counts are available from 1950 or earlier. This results in 48 sex-specific populations used for analysis and comparison. The selected countries and their corresponding sample periods are listed in Table~\ref{tab:1}.
\begin{table}[!htb]
\centering
\tabcolsep 0.11in
\caption{Countries and their respective sample periods. For the annuity price calculations, we use the official interest rate as of May 18, 2025.}\label{tab:1}
\begin{small}
\begin{tabular}{@{}llccllcc@{}}
\toprule
Country & Code & Period & Interest rate & Country & Code & Period & Interest rate\\
\midrule
Australia 	& AUS 	& 1921--2021 	& 4.10\% & Italy 			& ITA 	& 1872--2022 & 2.25\% \\
Austria 	& AUT 	& 1947--2023 	& 2.25\% & Japan 			& JPN 	& 1947--2023 & 0.50\% \\
Belgium 	& BEL 	& 1919--2023 	& 2.25\% & Netherlands 		& NLD 	& 1850--2022 & 2.25\% \\
Bulgaria 	& BGR 	& 1947--2021 	& 2.24\% & Norway 			& NOR 	& 1846--2023 & 4.50\% \\
Canada 	& CAN 	& 1921--2022 	& 4.50\% & New Zealand 		& NZ 	& 1948--2021 & 5.50\% \\
Czech 	& CZE 	& 1950--2021 	& 3.50\% & Portugal 			& PRT 	& 1940--2023 & 2.25\% \\
Denmark 	& DEN	&1835--2024 	& 2.25\% & Slovakia 			& SVK 	& 1950--2019 & 2.25\% \\
Finland 	& FIN 	& 1878--2023 	& 2.25\% & Spain 			& SPA 	& 1908--2023 & 2.25\% \\
France 	& FRA 	& 1816--2022 	& 2.25\% & Sweden 			& SWE 	& 1751--2023 & 4.00\% \\
Hungary 	& HUN 	& 1950--2020 	& 6.50\% & Switzerland 		& CHE 	& 1876--2023 & 1.75\% \\
Iceland 	& ICE 	& 1838--2023 	& 7.75\% & United Kingdom 	& UK 	& 1922--2022 & 5.25\% \\
Ireland 	& IRE 	& 1950--2022 	& 2.25\% & United States 		& USA 	& 1933--2023 & 5.25\% \\
\bottomrule
\end{tabular}
\end{small}
\end{table}

\newpage

The life-table death counts are non-negative and sum to a radix (i.e., a population experiencing 100,000 births annually) for each year. For the life-table death counts, there are 111 ages: 0, 1, $\dots$, 109, 110+. Due to rounding, there may be zero counts for higher ages in some years. To rectify this problem, we work with the probability of dying (i.e., $q_x$) and the radix of the life-table to recalculate our estimated death counts (up to six decimal places). In doing so, we obtain more detailed death counts than the ones reported. Note that $q_x$ is the mortality instrument studied by \cite{CBD06}.

\begin{figure}[!htb]
\centering
\includegraphics[width=8.74cm]{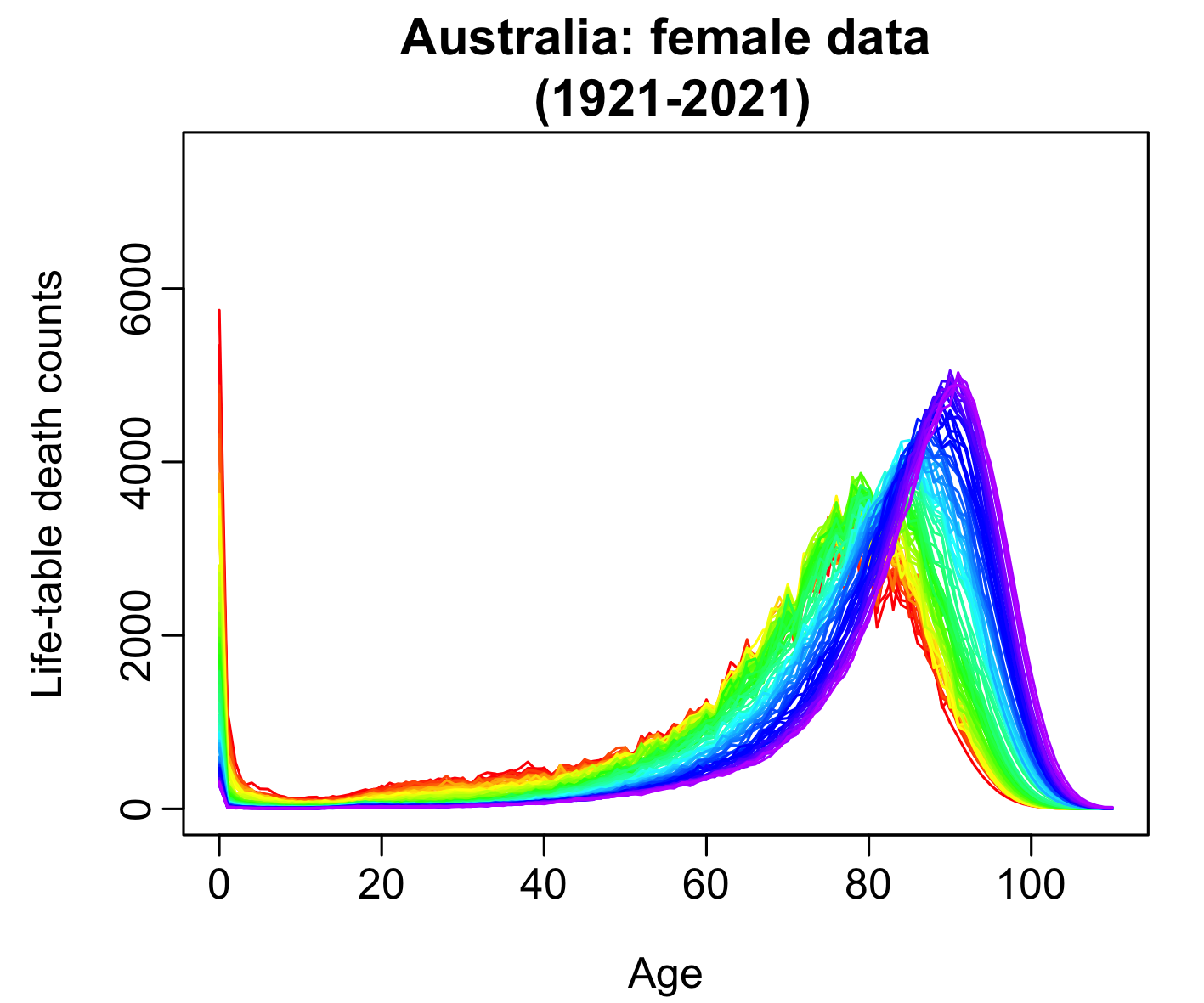}
\quad
\includegraphics[width=8.74cm]{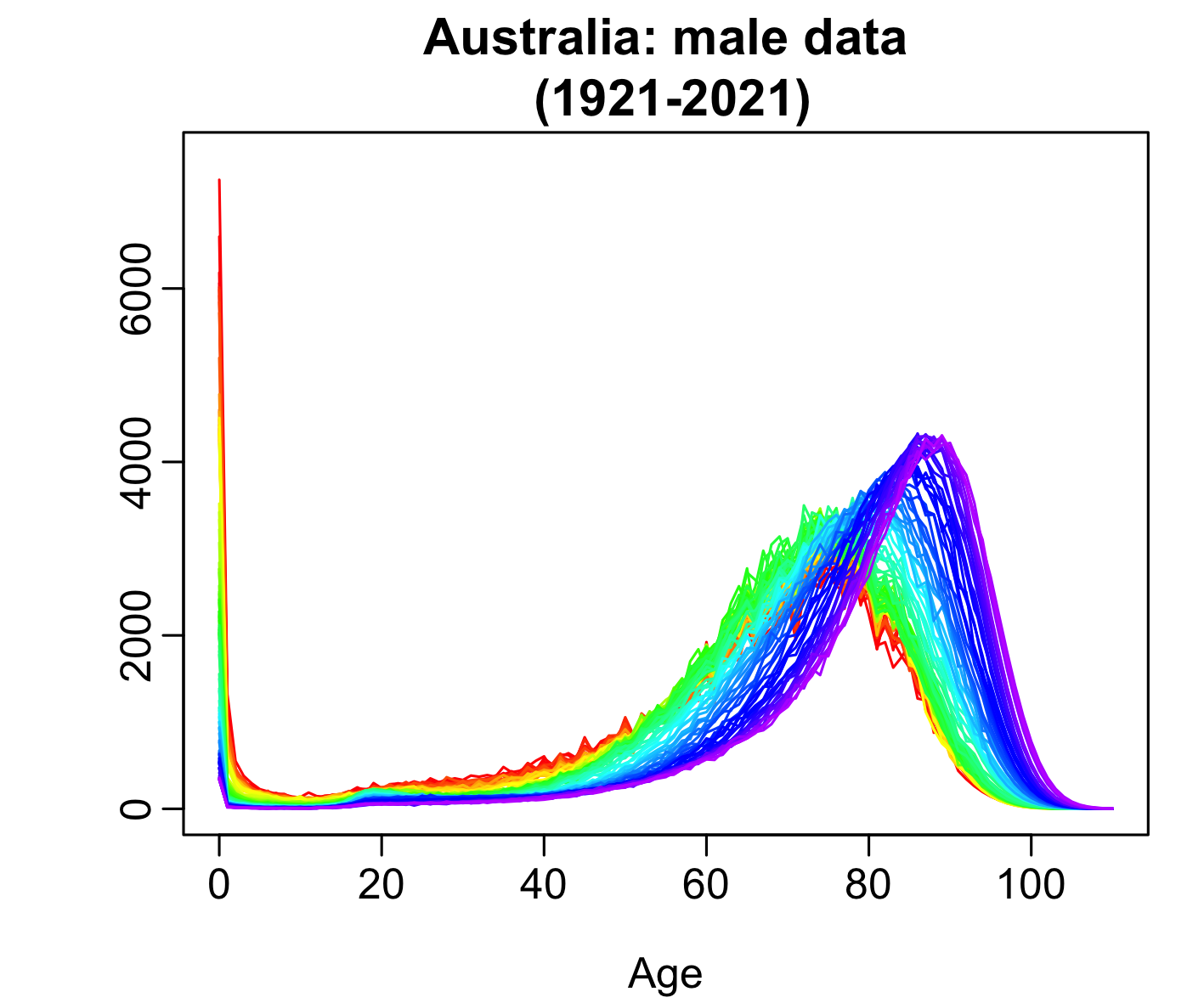}
\caption{Rainbow plots of Australian age-specific life-table death counts from 1921 to 2021 in a single-year group. The oldest years are shown in red, with the most recent years in violet. Curves are ordered chronologically according to the colors of the rainbow.}\label{fig:1}
\end{figure}

To understand the main features of the data, Figure~\ref{fig:1} displays rainbow plots of the age-specific life-table death counts for females and males in Australia, grouped by single year, from 1921 to 2021. The time order of the curves follows the color order of a rainbow, where data from the distant past are shown in red, while the most recent data are shown in purple \citep[see][for other examples]{HS10}. Both figures demonstrate a decreasing trend in infant death counts and a typical negative skew, with the peaks gradually shifting to higher ages for both sexes. This shift is a source of longevity risk, creating a major issue for insurers and government pension funds, especially in the sale and risk management of annuity products \citep[see][for a discussion]{DDG07}. Moreover, the spread of the distribution indicates lifespan variability. A decrease in variability over time can be directly observed and quantified using the interquartile range or the Gini coefficient \citep[see, e.g.,][]{WH99, VC13, DCH+17}. With the decrease in infant deaths and the increase in mortality at older ages, the spread of the distribution has narrowed.

While we present the Australian age distribution of deaths in Figure~\ref{fig:1}, in Section~\ref{sec:3_4}, we introduce new visualization plots for pairwise comparisons of the age distribution of death counts across sexes and countries.

\section{Transformation}\label{sec:3}

Let age-specific life-table death counts be denoted by $d^s_{t, i}$ or $d^s_{t, x}$, where $t$ denotes a year, $s$ represents females or males, and $i$ or $x$ denotes an age. For each year $t$, the life-table death counts sum to a radix~$10^5$ or one by normalization.

\subsection{Centered log-ratio transformation}\label{sec:3.1}

The CLR transformation maps the simplex to a hyperplane in Euclidean space, allowing principal component analysis to be applied. The transformation consists of the following steps:
\begin{asparaenum}[1)]
\item Compute the geometric mean function, estimated by a simple average
\begin{equation*}
\alpha^s_{n, x} = \exp\Big(\frac{1}{n}\sum^n_{t=1}\ln d^s_{t,x}\Big),
\end{equation*}
where $\ln(\cdot)$ denotes the natural logarithm. We compute the centered data by dividing each death count by its geometric mean
\begin{equation*}
\zeta^s_{t,x} = \frac{d^s_{t,x}}{\alpha_{n,x}^s}.
\end{equation*}
\item By taking the natural logarithm, we obtain
\begin{align*}
\beta_{t,x}^s = \ln \zeta^s_{t,x} &= \ln d_{t,x}^s - \ln \alpha_{n,x}^s \\
&= \ln d_{t,x}^s - \frac{1}{n}\sum^n_{t=1}\ln d_{t,x}^s,
\end{align*}
where $\bm{\beta}_{x}^s = (\beta_{1, x}^s,\dots,\beta_{n, x}^s)^{\top}$ denotes a set of unconstrained functional time series.
\item Applying eigendecomposition to the sample variance of $\bm{\beta}_{x}^s$, we can express
\begin{equation*}
\beta_{t, x}^{s} = \sum^{L_s}_{\ell=1}\gamma^{s}_{t, \ell}\phi_{\ell, x}^{s} + \omega_{t, x}^{s},
\end{equation*}
where $\phi_{\ell, x}^{s}$ is the $\ell$\textsuperscript{th} estimated principal component for age $x$, $\gamma^s_{t,\ell} = \langle \beta_{t, x}^{s}, \phi_{\ell, x}^{s}\rangle$ is the $\ell$\textsuperscript{th} estimated principal component score at time $t$, $\omega_{t, x}^{s}$ denotes the residuals of the model for age $x$ in year $t$, and $L_{s}$ denotes the number of principal components retained. To select the number of components $L_s$, we implement an eigenvalue ratio (EVR) criterion of \cite{LRS20}, as well as $L_{s}=6$ recommended in \cite{HBY13}.
 
\item \textit{Forecasting step:} By conditioning on the estimated principal components and observed data, the $h$-step-ahead forecast of $\beta_{n+h, x}^{s}$ can be estimated by
\begin{equation*}
\widehat{\beta}^{s}_{n+h|n, x} = \sum^{L_s}_{\ell=1}\widehat{\gamma}^s_{n+h|n, \ell}\phi_{\ell, x}^{s},
\end{equation*}
where $\widehat{\gamma}_{n+h|n,\ell}$ denotes the $h$-step-ahead forecast of the principal component scores. These forecasts can be obtained via a univariate time-series forecasting method, such as the exponential smoothing method of \cite{HKO+08}. The automatic algorithm of \cite{HKS+02} can be used to select the optimal exponential smoothing model among 15 possible candidates \citep[see Table~1][]{HK08}, based on the corrected Akaike information criterion.
\item By taking the inverse CLR transformation, we traverse back to the original simplex
\begin{equation*}
\widehat{\zeta}_{n+h|n,x}^s = \exp^{\widehat{\beta}^s_{n+h|n,x}}.
\end{equation*}
\item Then, we add back the geometric mean to obtain the $h$-step-ahead forecasts of the life-table death count $d_{n+h, x}^{s}$:
\begin{equation*}
\widehat{d}_{n+h|n, x}^{s} = \widehat{\zeta}_{n+h|n, x}^{s}\times \alpha_{n, x}^{s}.
\end{equation*}
\end{asparaenum}

\subsection{Cumulative distribution function transformation}\label{sec:3.2}

For a time series of CDFs, it is essential to scale the life-table death counts so that they sum to one. We describe the transformation in the following steps:
\begin{asparaenum}[1)]
\item Compute the empirical cumulative distribution function by the cumulative sum,
\begin{equation*}
D^s_{t,x} = \sum^x_{i=1}d^s_{t,i}, \quad x=1,\dots,111,\quad t=1,\dots,n,
\end{equation*}
where $D^s_{t,111} = 1$. Here, we relabel ages as $1,\dots,111$ to represent actual ages $0,\dots,109, 110+$.
\item Perform the logit transformation of $D_{t,x}^s$ for all ages except the last one,
\begin{equation*}
Z^s_{t,x} = \text{logit}(D^s_{t,x}) = \ln\left(\frac{D^s_{t,x}}{1-D^s_{t,x}}\right).
\end{equation*}
\item With a set of unconstrained data $\bm{Z}^{s}_{x} = (Z^{s}_{1,x}, \dots, Z^{s}_{n,x})^{\top}$, we implement a univariate functional time-series forecasting method. By computing the sample variance of $\bm{Z}^{s}_{x}$, we express a stochastic process $Z^{s}_{t,x}$ as
\begin{equation*}
Z^{s}_{t,x} = \sum^{K_{s}}_{k=1}\eta^{s}_{t, k}\psi^{s}_{k, x}+\varepsilon^{s}_{t, x},
\end{equation*}
where $\psi^{s}_{k, x}$ denotes the $k$\textsuperscript{th} principal component for age $x$ and sex $s$, $\eta^{s}_{t, k} = \langle Z^{s}_{t, x}, \psi^{s}_{k, x}\rangle$ is the estimated principal component score at time $t$, $\varepsilon^{s}_{t, x}$ denotes the model residual function for sex $s$ in year $t$, and $K_{s}$ denotes the number of principal components retained. We implement an eigenvalue ratio criterion of \cite{LRS20}, as well as $K_{s}=6$ recommended in \cite{HBY13}.
\item \textit{Forecasting step:} By conditioning on the estimated functional principal components $\bm{\Psi}^{s}_{x} = (\psi^{s}_{1, x},\dots,\psi^{s}_{K_{s}, x})$ and observed data $\bm{Z}_{x}^{s}=(Z_{1, x}^{s}, \dots, Z_{n, x}^{s})$, the $h$-step-ahead forecast of $Z_{n+h, x}^{s}$ can be obtained
\begin{equation*}
\widehat{Z}_{n+h|n, x}^{s} = \text{E}[Z_{n+h, x}^{s}|\bm{\Psi}^{s}_{x}, \bm{Z}_{x}^{s}] = \sum^{K_{s}}_{k=1}\widehat{\eta}_{n+h|n, k}^{s}\psi_{k, x}^{s},
\end{equation*}
where $\widehat{\eta}_{n+h|n, k}^{s}$ denotes the $h$-step-ahead univariate time-series forecast of the principal component scores. Among the univariate time-series methods, we consider the exponential smoothing method of \cite{HKO+08} to model and forecast each set of scores.
\item By taking the inverse logit transformation, we obtain
\begin{equation*}
\widehat{D}_{n+h|n, x}^{s} = \frac{\exp^{\widehat{Z}_{n+h|n, x}^{s}}}{1+\exp^{\widehat{Z}_{n+h|n, x}^{s}}}.
\end{equation*}
\item By taking the first-order differencing, we obtain
\begin{equation*}
\widehat{d}^{s}_{n+h|n, x} =
\left\{
\begin{array}{ll}
\widehat{D}^{s}_{n+h|n, 1} & x = 1 \\
\Delta^{x}_{i=1}\widehat{D}^{s}_{n+h|n, i}=\widehat{D}^{s}_{n+h|n, x} - \widehat{D}^{s}_{n+h|n, x-1}  & 2 \leq x \leq 111
\end{array}
\right.
\end{equation*}
where $\Delta$ represents the first-order differencing.
\end{asparaenum}

In Sections~\ref{sec:3.1} and~\ref{sec:3.2}, the forecasting steps are designed to model and forecast each series individually. This method does not account for the correlations between female and male data within the same country. Although explicitly modeling the correlation between multiple series can improve forecast accuracy, this is not the focus of this paper, as a similar study has been presented in \cite{SJ25}.

\section{Visualization plots of the gap in the age distribution of deaths}\label{sec:3_4}

\begin{figure}[!htb]
\centering
\includegraphics[width=13.1cm]{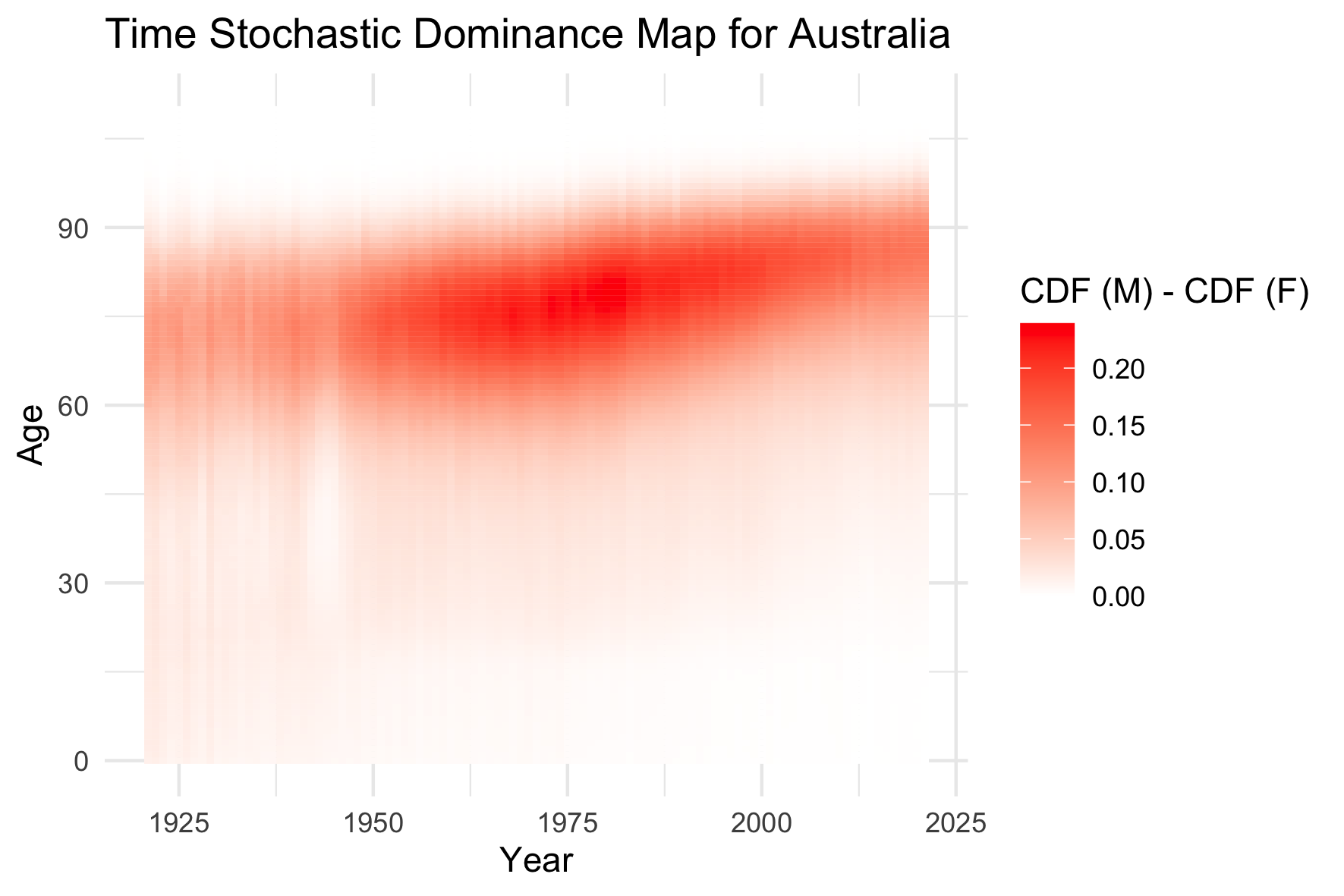}
\caption{Image plot showing time stochastic dominance map between the probabilities of dying between Australian males and females from 1921 to 2021.}\label{fig:image}
\end{figure}

While the two transformations work with life-table death counts, the CDF transformation is advantageous for visualizing differences between sexes and countries. In the first step of the CDF transformation, the age distribution of deaths for a given year is converted into a CDF, providing a scale-free means of comparing different populations in terms of their probabilities of dying, commonly denoted $q_x$. By computing the differences, represented by 
\begin{equation*}
G_t(u) = CDF^{\text{M}}_{t}(u) - \text{CDF}^{\text{F}}_t(u), 
\end{equation*}
between the two populations, we can visualize the gap in the probability of dying for Australia, shown in Figure~\ref{fig:image}. The gap between the Australian female and male populations is largest between 1950 and 2000, particularly among individuals aged 60 to 90. Since the lower bound of the gap is 0, it indicates that the male population has a higher probability of dying than the female population.

We can summarize domain strength at each time $t$ using an integral measure, such as
\begin{equation}
S_{t} = \int_{u=0}^{110+} \left[\text{CDF}^{\text{M}}_t(u) - \text{CDF}^{\text{F}}_t(u)\right]du,\label{eq:integral}
\end{equation}
where the integral can be approximated via the trapezoidal rule. When $S_t > 0$, it implies that females tend to live longer. The integral measure is closely related to the difference between life expectancy at birth. Recall that the life expectancy at birth can be expressed as
\begin{equation*}
e_0 = \int_{0}^{\infty} S(u)du
\end{equation*}
where $S(u)$ is the survival function to the last age considered. Since the survival function is the complement of the CDF, it can be shown that
\begin{align*}
S_{t} &= \int_{u=0}^{110+} \left[\text{CDF}^{\text{M}}_t(u) - \text{CDF}^{\text{F}}_t(u)\right]du \\
&= \int^{110+}_{u=0}[S_t^{\text{F}}(u) - S_t^{\text{M}}(u)]du \\
&=(e_{0}^{\text{F}}-e_0^{\text{M}})_t.
\end{align*}

\begin{figure}[!htb]
\centering
\includegraphics[width=12cm]{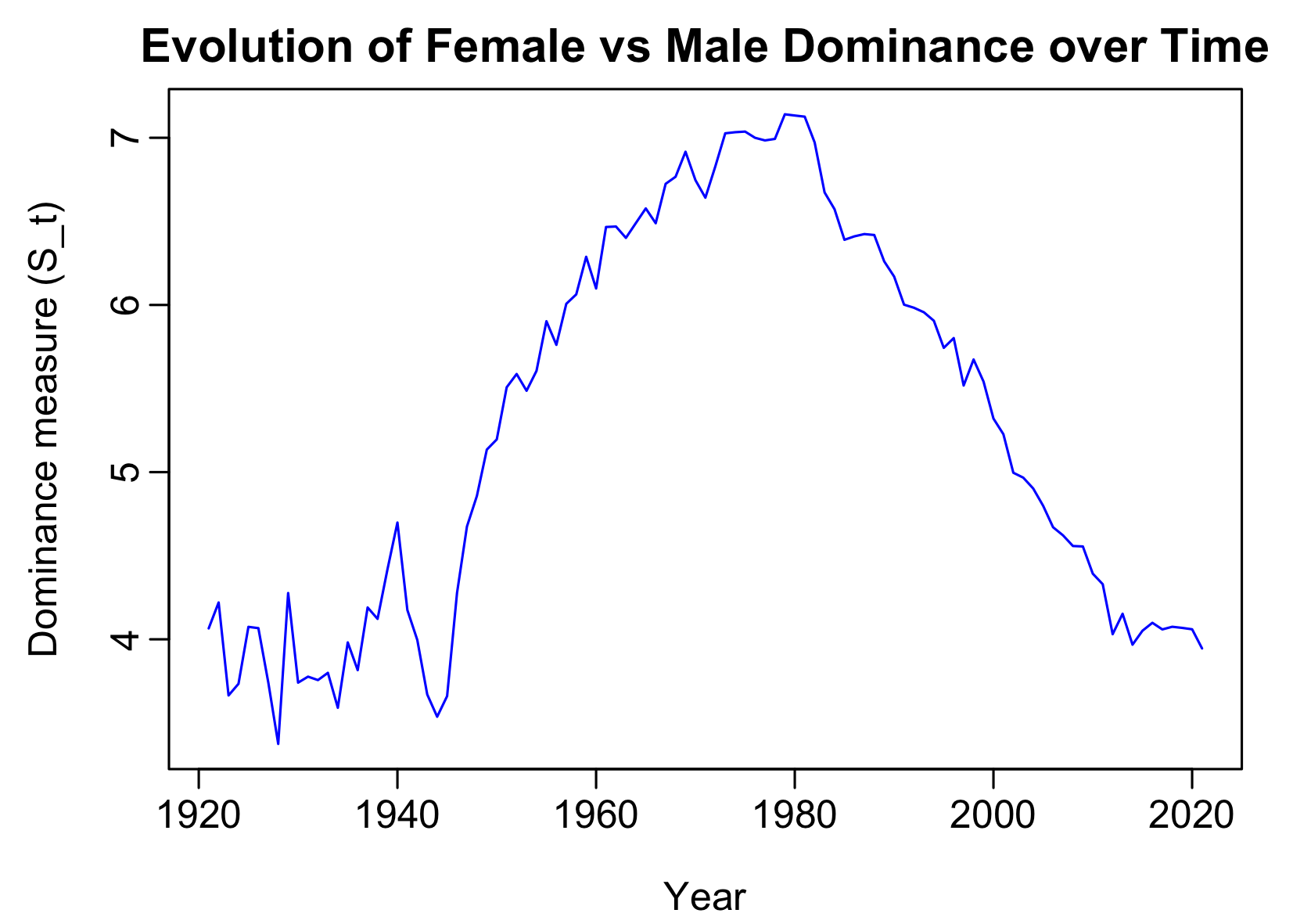}
\caption{A time series plot of the integral measure of the probabilities of dying between the Australian males and females from 1921 to 2021.}\label{fig:plot}
\end{figure}

In Figure~\ref{fig:plot}, we present a time series plot of $S_t$ to show the gap-integral measure between Australian females and males, aggregated by age.

\begin{figure}[!htb]
\centering
\includegraphics[width=8.5cm]{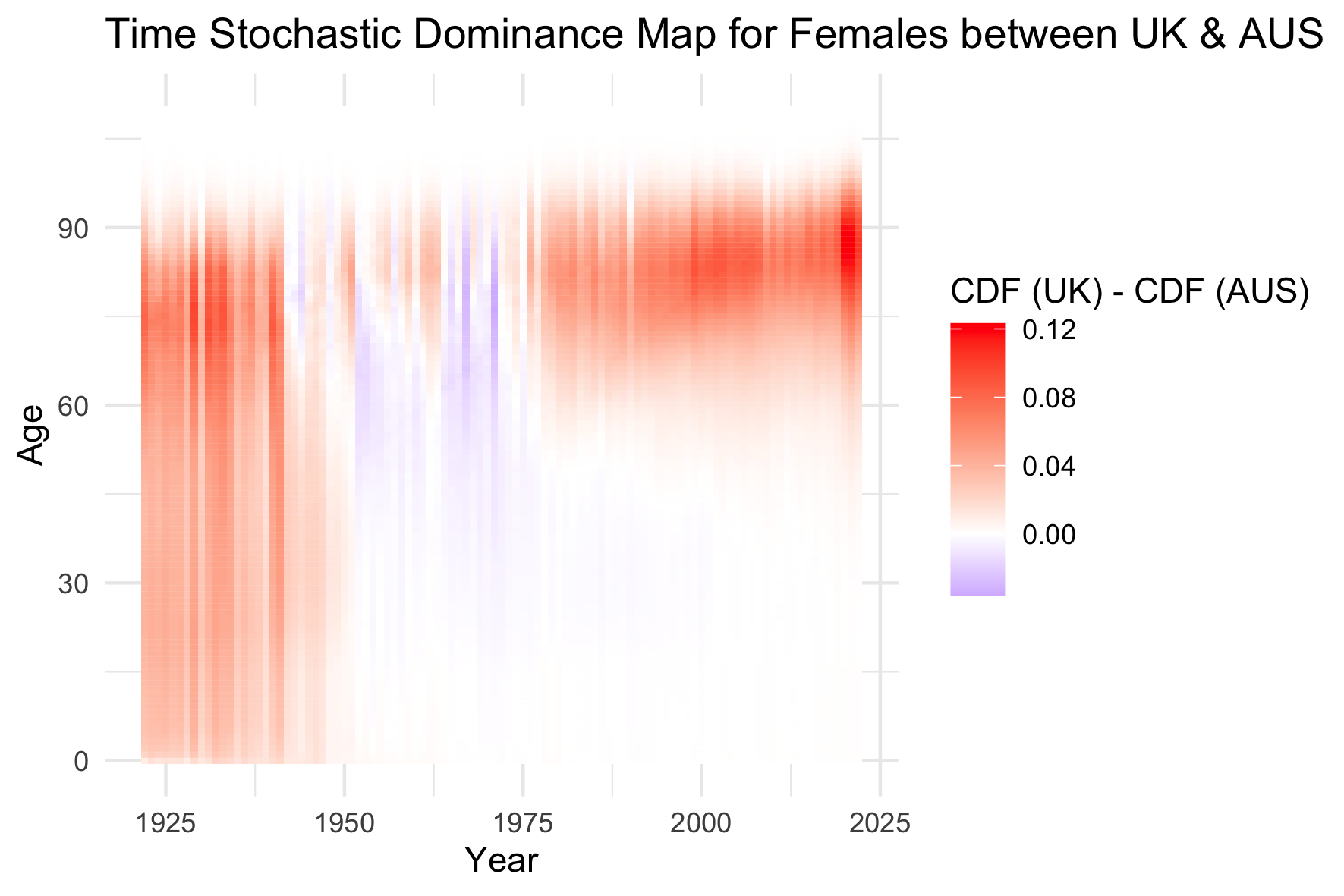}
\quad
\includegraphics[width=8.5cm]{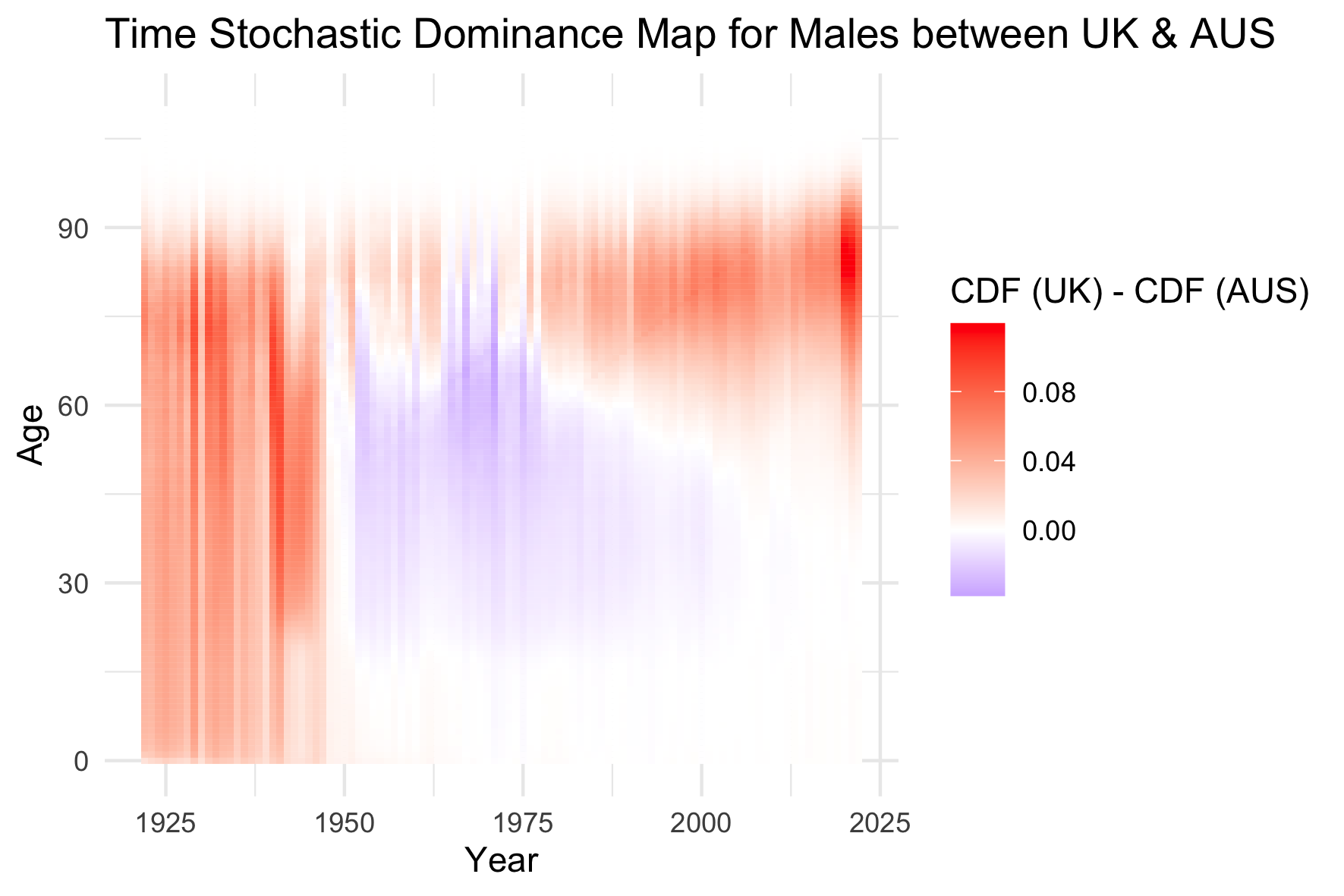}
\caption{Image plot showing time stochastic dominance map between the probabilities of dying between the UK and Australian males and females from 1921 to 2021.}\label{fig:UK_AUS_image}
\end{figure}

While Figures~\ref{fig:image} and~\ref{fig:plot} compare Australian females and males, the CDF, being a scale-free measure, can also be used to compare the probability of dying across countries. In Figure~\ref{fig:UK_AUS_image}, we present image plots comparing the probability of dying of females and males between the UK and Australia. For comparison, UK females generally have a higher probability of dying than Australian females, except between 1950 and 1970, especially between ages 30 and 80. In addition, UK males have a higher probability of dying than Australian males, except between 1950 and 2000, especially between ages 20 and 60.

Using the integral measure in~\eqref{eq:integral}, we present two time series plots showing the gap integral measure between the UK and Australian females and males. Except for 1950 and 1975, the UK population generally has a higher probability of dying than the Australian population.
\begin{figure}[!htb]
\centering
\subfloat[Female data]
{\includegraphics[width=8.5cm]{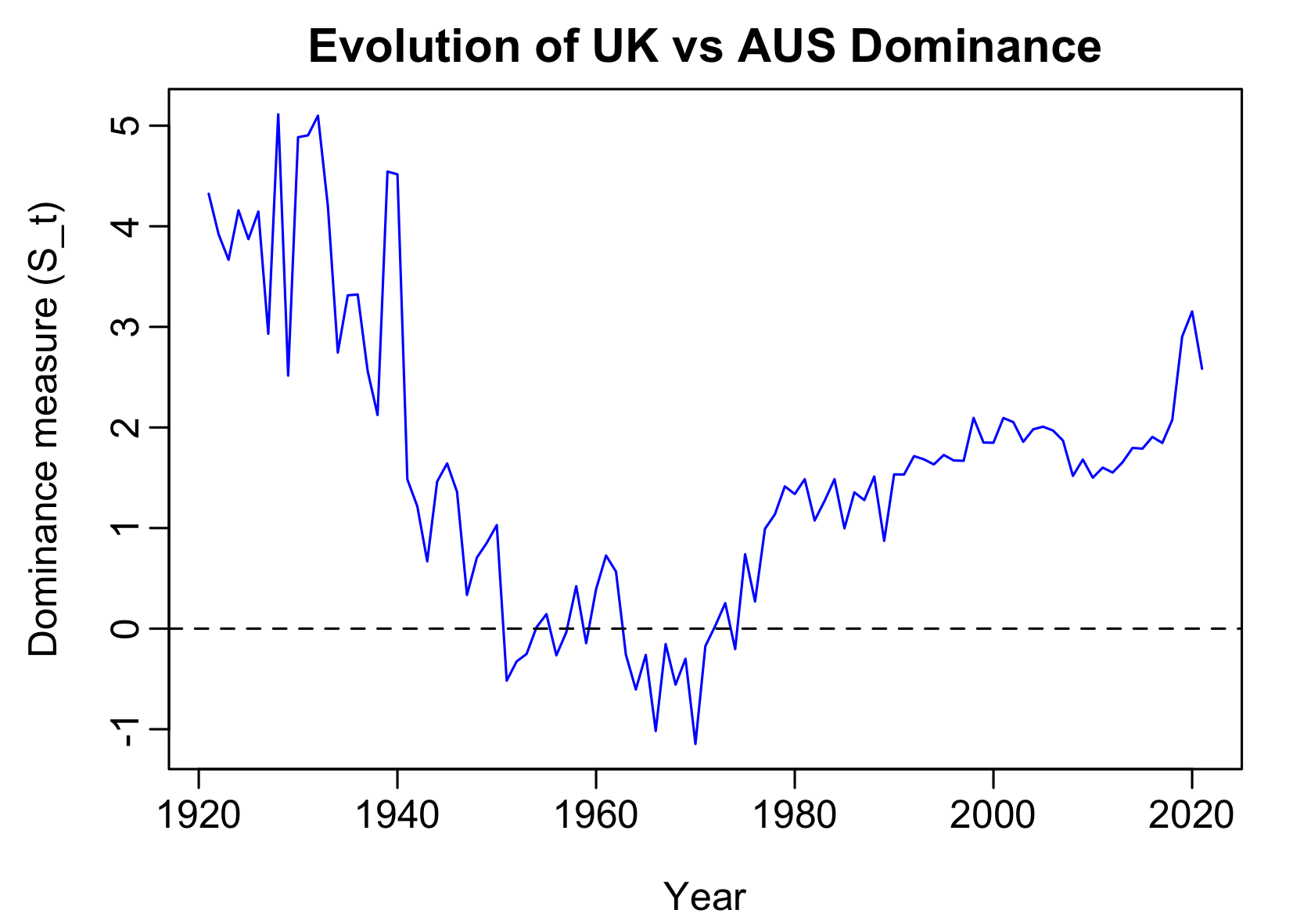}}
\quad
\subfloat[Male data]
{\includegraphics[width=8.5cm]{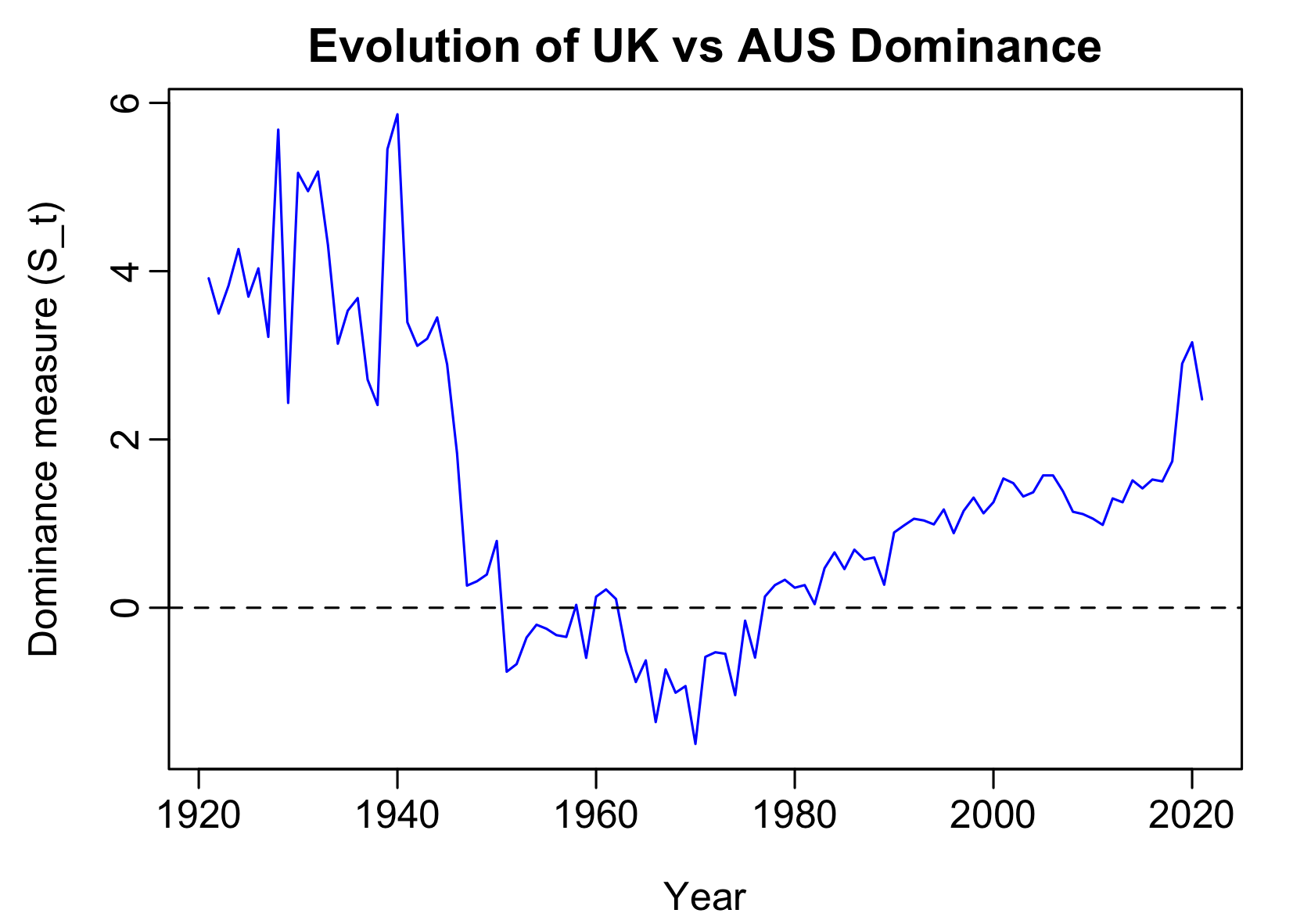}}
\caption{Time series plots of the integral measure of the probabilities of dying between the UK and Australian males and females from 1921 to 2021.}\label{fig:UK_AUS_plot}
\end{figure}

\section{Forecast accuracy comparison}\label{sec:4}

\subsection{Expanding window}\label{sec:4.1}

We implement an expanding-window scheme, which is commonly used to evaluate model and parameter stability, as well as prediction accuracy over time. The expanding-window analysis assesses the stability of a model's parameter estimates and forecasts over an expanding window across the sample. Since each data set has a different sample period, we keep the last 20 years of data for evaluation and the remaining data for training. For the 20 years of data in the testing sample, we consider one-step-ahead to 20-step-ahead forecasts to assess the model's performance at short to medium horizons. Through the expanding-window scheme, we produce 20 one-step-ahead forecasts, 19 two-step-ahead forecasts, $\dots$, and one 20-step-ahead forecast. We compare these forecasts with the holdout data in the testing sample to determine the accuracy of the out-of-sample forecast. In Figure~\ref{fig:tikz}, we present a concept diagram.
\begin{figure}[!htb]
\begin{center}
\begin{tikzpicture}
\draw[->] (0,0) -- (10,0) node[right] {Time};    
\draw[fill=blue!20] (0,-0.5) rectangle (3,0.5) node[midway] {Train};
\draw[fill=red!20] (3,-0.5) rectangle (3.5,0.5) node[midway] {F};    
\draw[fill=blue!20] (0,-1.5) rectangle (5,-0.5) node[midway] {Train};
\draw[fill=red!20] (5,-1.5) rectangle (5.5,-0.5) node[midway] {F};    
\draw[fill=blue!20] (0,-2.5) rectangle (7,-1.5) node[midway] {Train};
\draw[fill=red!20] (7,-2.5) rectangle (7.5,-1.5) node[midway] {F};    
\draw[fill=blue!20] (0,-3.5) rectangle (9,-2.5) node[midway] {Train};
\draw[fill=red!20] (9,-3.5) rectangle (9.5,-2.5) node[midway] {F};    
\node[left] at (0,0) {};
\node[left] at (0,-1) {};
\node[left] at (0,-2) {\hspace{-0.8in}{$\vdots$}};
\node[left] at (0,-3) {};
\draw[fill=blue!20] (6.5,1) rectangle (7,1.5);
\node[right] at (7,1.25) {Training Window};
\draw[fill=red!20] (6.5,0.5) rectangle (7,1);
\node[right] at (7,0.75) {Forecast (F) when $h=1$};
\end{tikzpicture}
\end{center}
\caption{A diagram of the expanding-window forecast scheme.}\label{fig:tikz}
\end{figure}

\vspace{-.2in}

\subsection{Point forecast evaluation metrics}\label{sec:4.2}

Since the age distribution of death counts can be considered a probability density function, we consider two density evaluation metrics: the symmetric discrete Kullback-Leibler divergence (KLD) \citep{KL51} and the Jensen-Shannon divergence (JSD) \citep{Shannon48}. 

The KLD measures information loss by approximating an unknown density with its approximation. For two probability density functions, denoted by $d_{n+\xi}(u)$ and $\widehat{d}_{n+\xi|n}(u)$, the symmetric discrete KLD is defined as
\begin{align*}
\text{KLD}(h) =\ & D_{\text{KL}}(d_{n+\xi,x}||\widehat{d}_{n+\xi|n,x}) + D_{\text{KL}}(\widehat{d}_{n+\xi|n,x}||d_{n+\xi,x}) \\
=\ & \frac{1}{111\times (21-h)}\sum^{20}_{\xi=h}\sum^{111}_{x=1}[d_{n+\xi,x} (\ln d_{n+\xi,x} - \ln \widehat{d}_{n+\xi|n,x})  +  \widehat{d}_{n+\xi|n,x} (\ln \widehat{d}_{n+\xi|n,x} - \ln d_{n+\xi,x})],
\end{align*}
where $\xi$ is the forecasting period. A feature of the symmetric discrete KLD is its non-negativity.

Even though we have a sample of observed densities in the forecasting period, they may not be the actual densities. An alternative is the JSD, which can be viewed as a symmetric and smoothed version of the KLD. The JSD is defined by
\begin{equation*}
\text{JSD}(h) = \frac{1}{2}D_{\text{KL}}(d_{n+\xi, x}||\delta_{n+\xi, x}) + \frac{1}{2}D_{\text{KL}}(\widehat{d}_{n+\xi|n, x}||\delta_{n+\xi, x}),
\end{equation*}
where $\delta_{n+\xi, x}$ measures a common quantity between $d_{n+\xi, x}$ and $\widehat{d}_{n+\xi, x}$. An example of $\delta_{n+\xi, x}$ can be its geometric mean $\delta_{n+\xi, x} = \sqrt{d_{n+\xi, x}\widehat{d}_{n+\xi, x}}$.

\subsection{Comparison of point forecast accuracy}\label{sec:4.3}

\begin{figure}[!htb]
\centering
{\includegraphics[width=4.42cm]{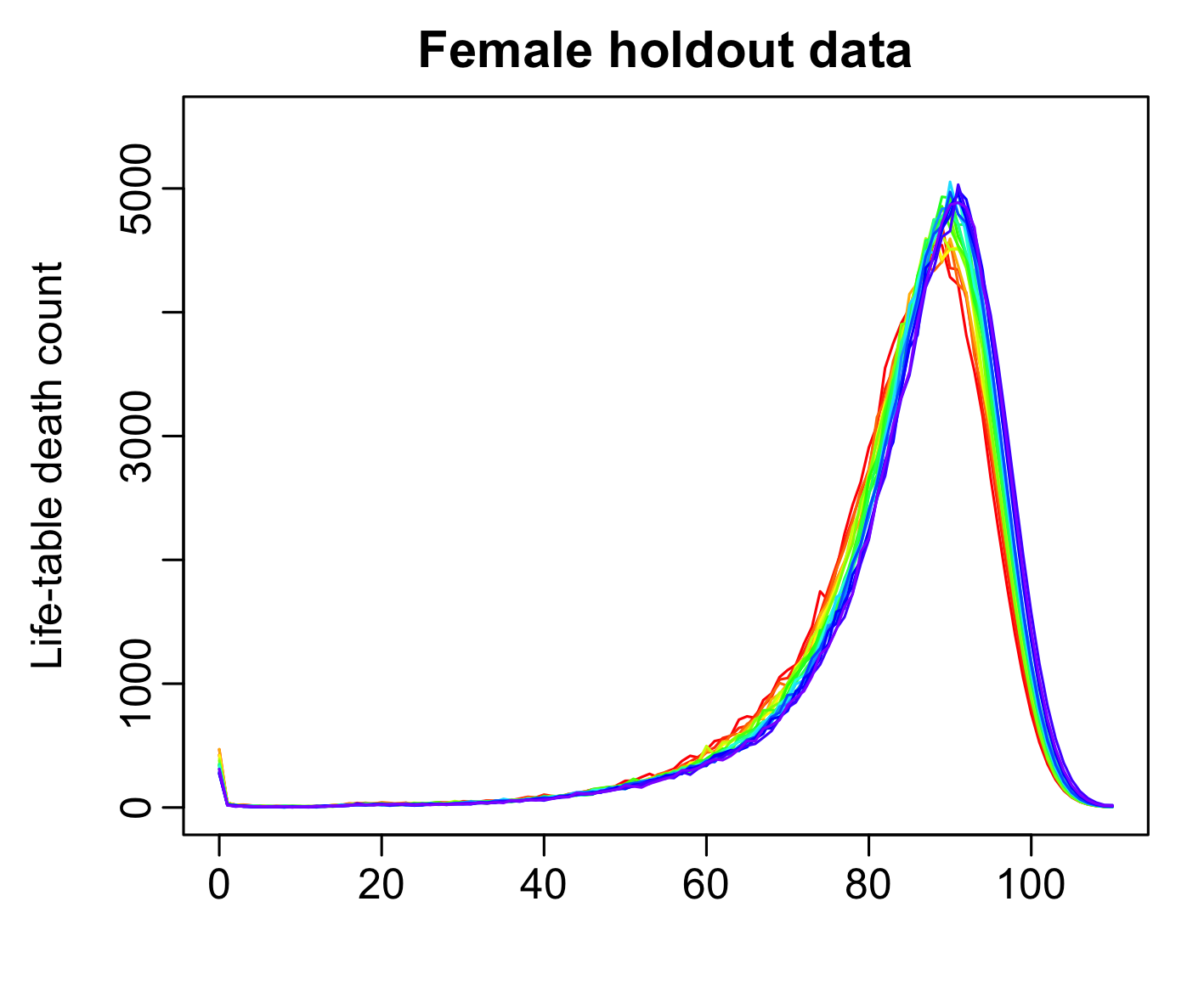}}
{\includegraphics[width=4.42cm]{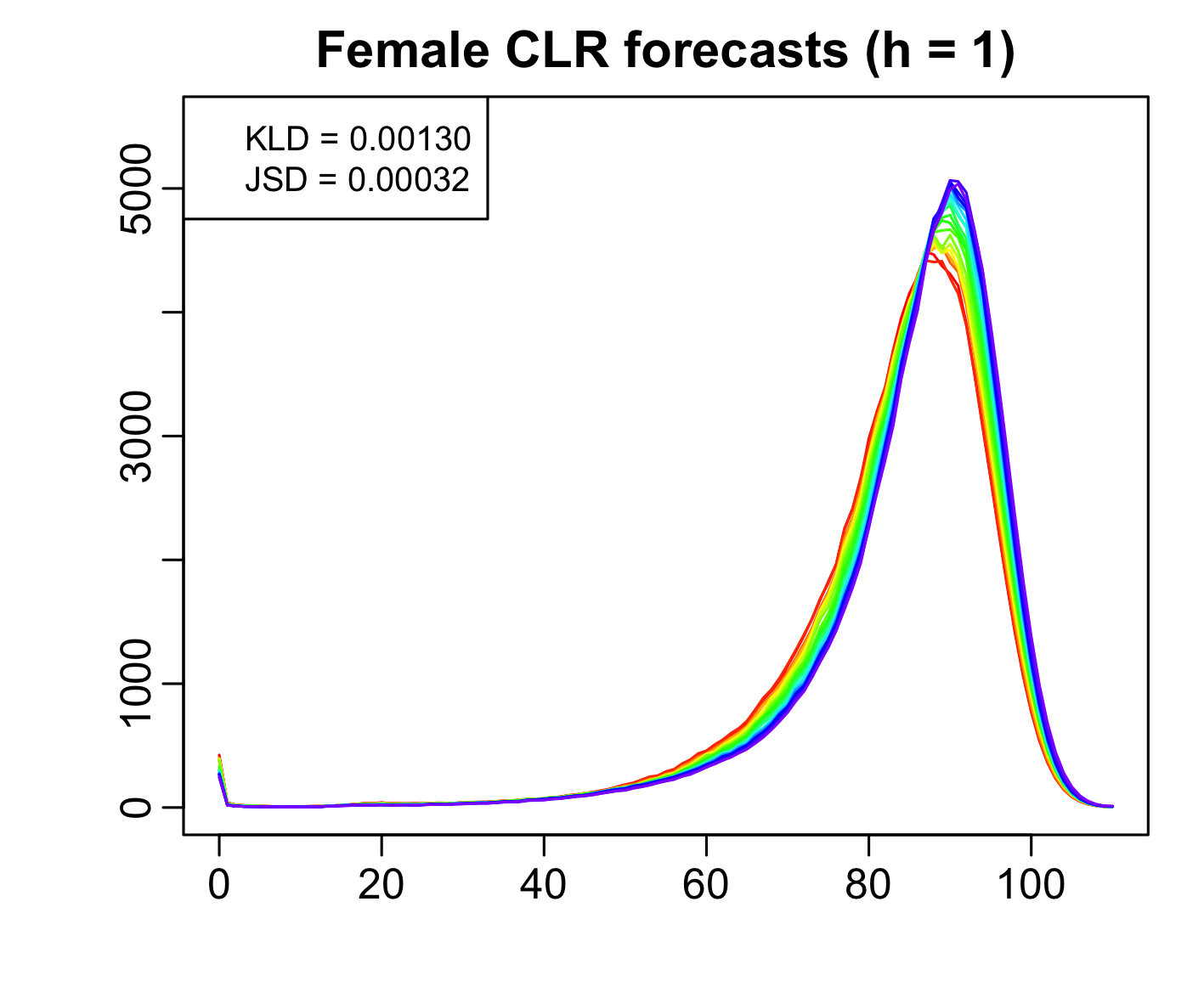}}
{\includegraphics[width=4.42cm]{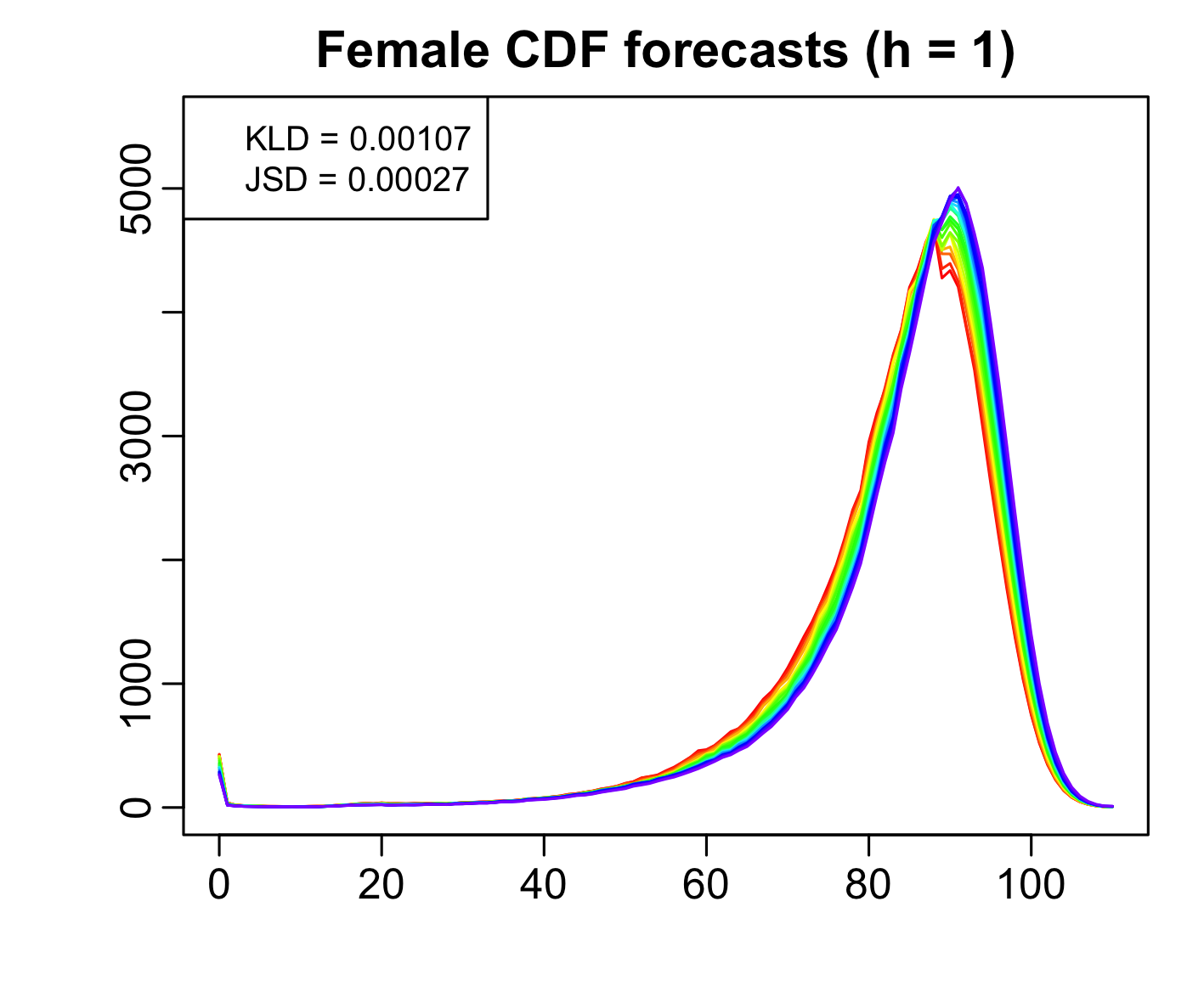}}
{\includegraphics[width=4.42cm]{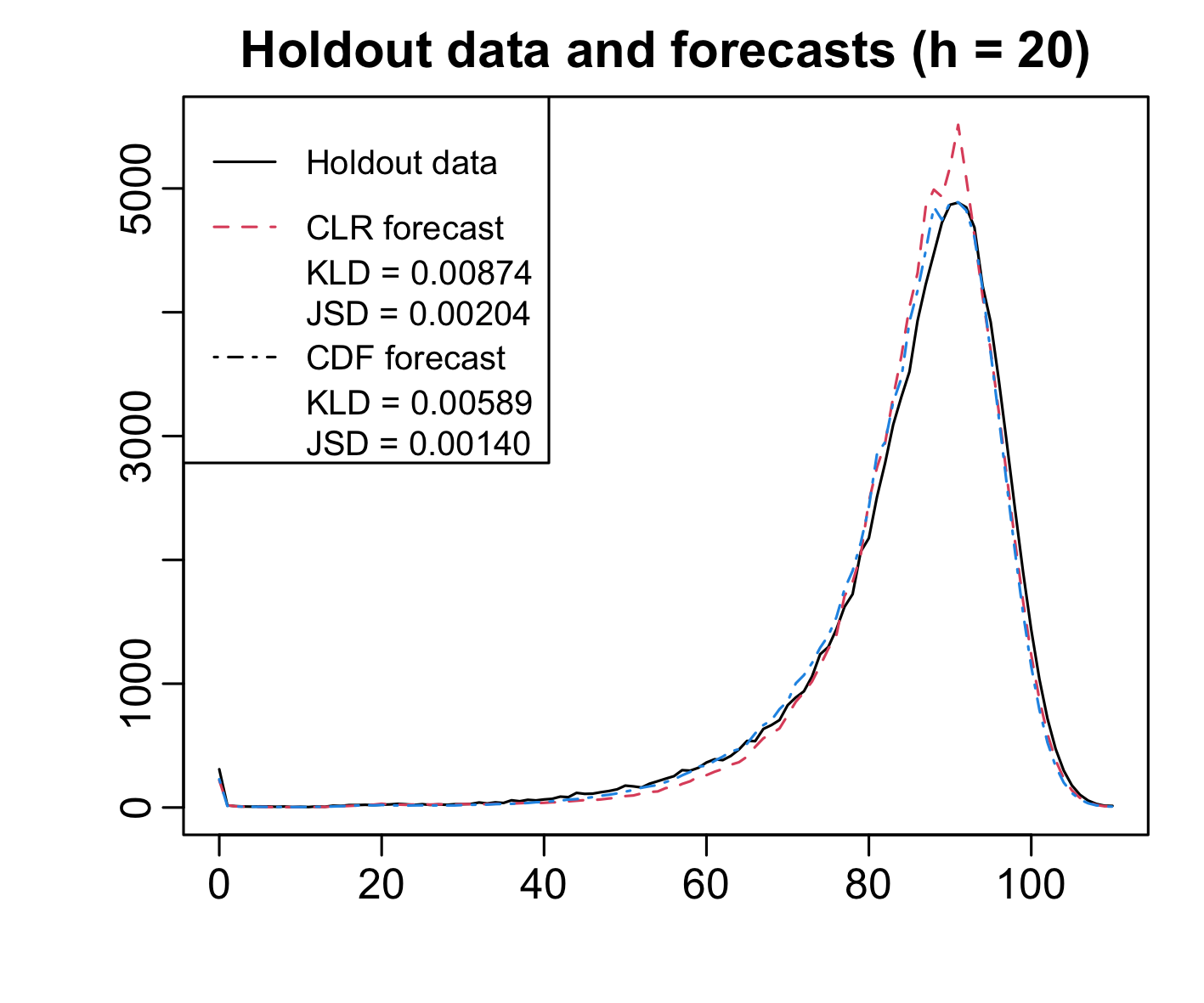}}
\\
{\includegraphics[width=4.42cm]{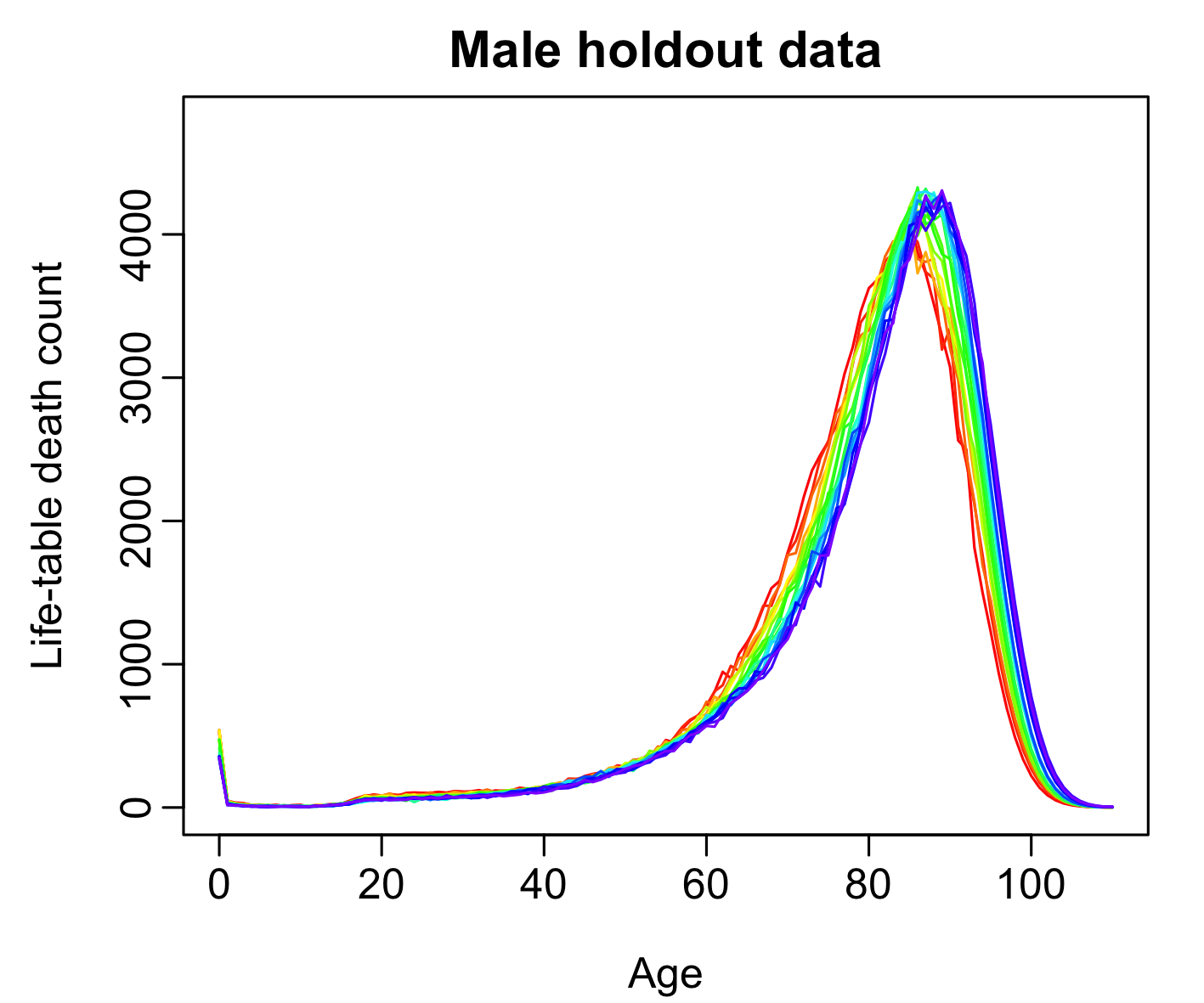}}
{\includegraphics[width=4.42cm]{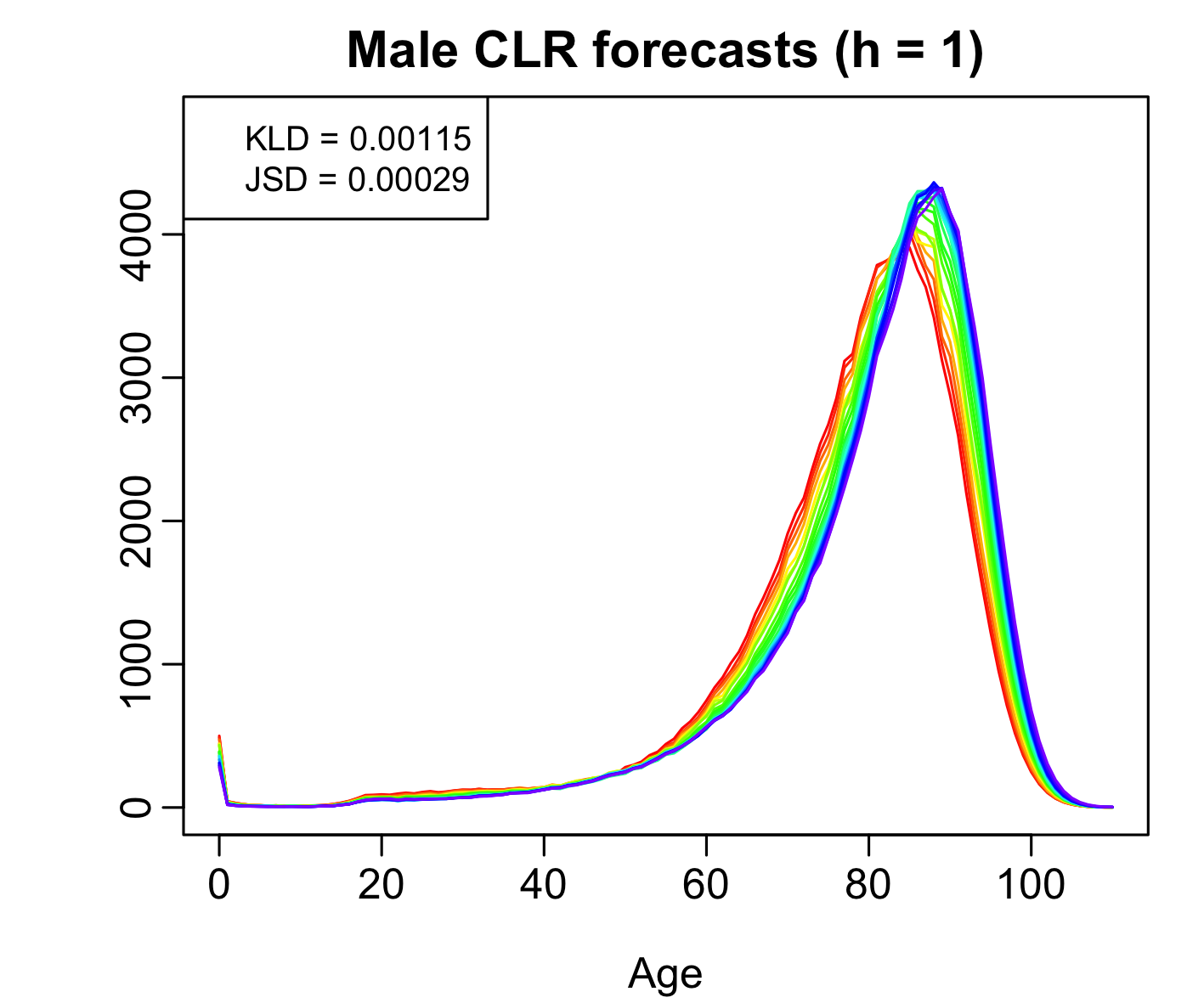}}
{\includegraphics[width=4.42cm]{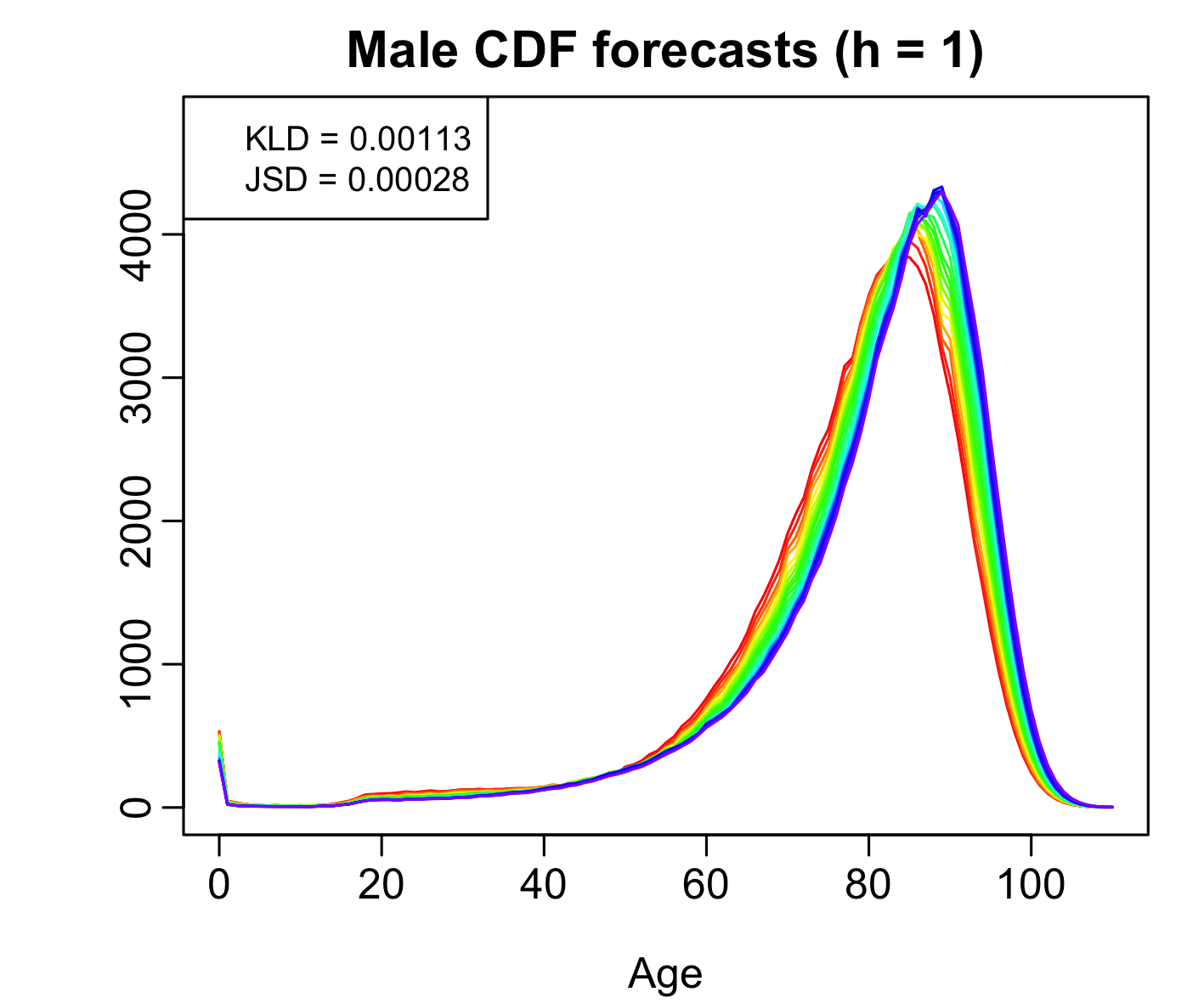}}
{\includegraphics[width=4.42cm]{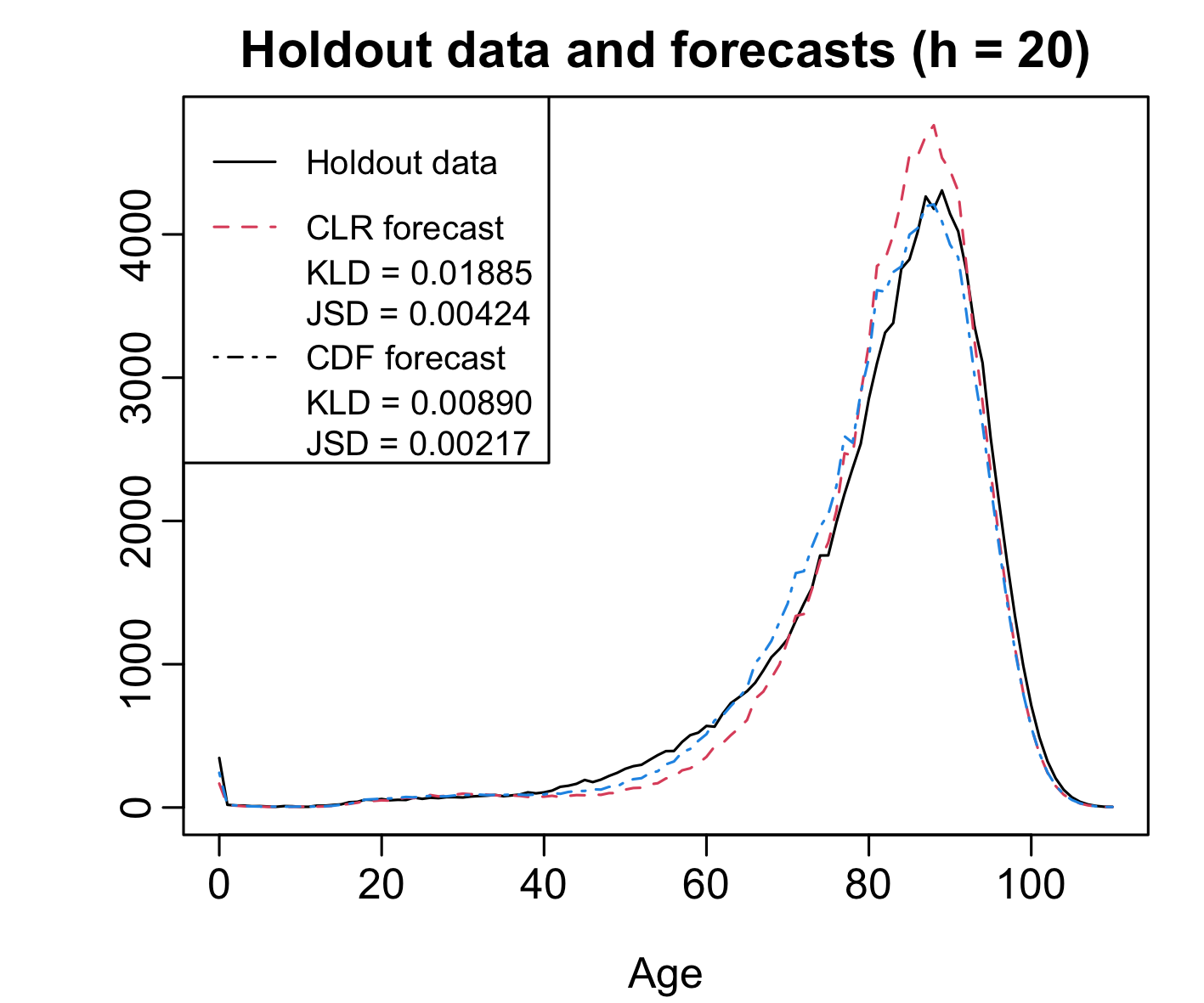}}
\caption{\small One-step-ahead and 20-step-ahead forecasts of life-table death counts for Australian women (first row) and men (second row). As measured by the KLD and JSD, we compute these errors obtained from the CLR and CDF transformations.}\label{fig:2}
\end{figure}

In Figure~\ref{fig:2}, we demonstrate the forecasts of the life-table death counts obtained from the CLR and CDF transformations. For one-step-ahead forecasts, the differences between the two transformations are marginal. As the forecast horizon increases to $h=20$, the difference is more visible. As measured by KLD and JSD, the CDF transformation produces smaller forecast errors than those of the CLR transformation for the $h=1$ and $h=20$ considered.

In Table~\ref{tab:KLD_tab}, we compare the KLD across the 24 countries, averaged over 20 forecast horizons. Countries such as Iceland and Ireland are challenging to predict accurately using the CLR transformation. In contrast, the CDF transformation shows moderate errors. Between the two ways of selecting the number of components, it is advantageous to use $K=6$ for forecasting.
\begin{center}
\tabcolsep 0.17in
\renewcommand{\arraystretch}{0.84}
\begin{longtable}{@{}lrrrrrrrr@{}}
\caption{Averaged over 20 forecast horizons, a comparison of the female and male KLD errors among the 24 countries. Heterogeneity in forecast accuracy highlights differences in prediction accuracy across the two transformations using our chosen forecasting method. Eigenvalue ratio (EVR) is a criterion for selecting the number of components.} \label{tab:KLD_tab} \\
\toprule
& \multicolumn{4}{c}{Female} & \multicolumn{4}{c}{Male} \\
\cmidrule(lr){2-5}\cmidrule(lr){6-9}
		& \multicolumn{2}{c}{EVR} & \multicolumn{2}{c}{$K=6$} & \multicolumn{2}{c}{EVR} & \multicolumn{2}{c}{$K=6$} \\
\cmidrule(lr){2-3}\cmidrule(lr){4-5}\cmidrule(lr){6-7}\cmidrule(lr){8-9}
Country 		& CLR & CDF & CLR & CDF   & CLR & CDF & CLR & CDF  \\
\midrule
\endfirsthead\\
\toprule
& \multicolumn{4}{c}{Female} & \multicolumn{4}{c}{Male} \\
\cmidrule(lr){2-5}\cmidrule(lr){6-9}
		& \multicolumn{2}{c}{EVR} & \multicolumn{2}{c}{$K=6$} & \multicolumn{2}{c}{EVR} & \multicolumn{2}{c}{$K=6$} \\
\cmidrule(lr){2-3}\cmidrule(lr){4-5}\cmidrule(lr){6-7}\cmidrule(lr){8-9}
Country 				& CLR & CDF & CLR & CDF   & CLR & CDF & CLR & CDF  \\
\midrule
\endhead
\bottomrule
\endlastfoot
AUS & 0.0117 & 0.0091 & 0.0040 & 0.0029 & 0.0208 & 0.0195 & 0.0096 & 0.0047 \\ 
AUT & 0.0092 & 0.0056 & 0.0101 & 0.0053 & 0.0186 & 0.0074 & 0.0334 & 0.0071 \\ 
BEL & 0.0144 & 0.0059 & 0.0117 & 0.0051 & 0.0509 & 0.0305 & 0.0209 & 0.0085 \\ 
BGR & 0.0771 & 0.0335 & 0.0220 & 0.0242 & 0.0527 & 0.0220 & 0.0190 & 0.0159 \\ 
CAN & 0.0066 & 0.0052 & 0.0051 & 0.0035 & 0.0355 & 0.0340 & 0.0147 & 0.0102 \\ 
CZE & 0.0406 & 0.0123 & 0.0106 & 0.0137 & 0.1508 & 0.0244 & 0.0115 & 0.0142 \\ 
DEN & 0.0176 & 0.0052 & 0.0142 & 0.0067 & 0.0380 & 0.0179 & 0.0285 & 0.0184 \\ 
FIN & 0.0481 & 0.0177 & 0.0191 & 0.0053 & 0.0991 & 0.0468 & 0.0351 & 0.0109 \\ 
FRA & 0.0250 & 0.0081 & 0.0162 & 0.0074 & 0.0460 & 0.0333 & 0.0242 & 0.0247 \\ 
HUN & 0.0750 & 0.0060 & 0.0074 & 0.0068 & 0.1154 & 0.0618 & 0.0655 & 0.0434 \\ 
ICE & 0.3189 & 0.0696 & 0.3708 & 0.0567 & 0.1629 & 0.0465 & 0.2282 & 0.0475 \\ 
IRE & 0.1731 & 0.0125 & 0.0121 & 0.0090 & 0.2546 & 0.0650 & 0.0455 & 0.0230 \\ 
ITA & 0.0104 & 0.0223 & 0.0067 & 0.0143 & 0.0442 & 0.0280 & 0.0201 & 0.0205 \\ 
JPN & 0.0954 & 0.0846 & 0.1072 & 0.0805 & 0.0075 & 0.0129 & 0.0096 & 0.0120 \\ 
NLD & 0.0303 & 0.0134 & 0.0134 & 0.0101 & 0.0602 & 0.0334 & 0.0392 & 0.0306 \\ 
NOR & 0.0323 & 0.0160 & 0.0170 & 0.0068 & 0.0798 & 0.0439 & 0.0362 & 0.0208 \\ 
NZ & 0.0209 & 0.0075 & 0.0091 & 0.0057 & 0.2086 & 0.0290 & 0.0176 & 0.0095 \\ 
PRT & 0.0090 & 0.0063 & 0.0076 & 0.0053 & 0.0180 & 0.0095 & 0.0284 & 0.0091 \\ 
SLO & 0.1185 & 0.0257 & 0.0214 & 0.0234 & 0.0794 & 0.0537 & 0.0430 & 0.0297 \\ 
SPA & 0.0217 & 0.0171 & 0.0124 & 0.0089 & 0.0304 & 0.0112 & 0.0302 & 0.0058 \\ 
SWE & 0.0295 & 0.0101 & 0.0154 & 0.0166 & 0.0351 & 0.0040 & 0.0268 & 0.0047 \\ 
SWI & 0.0083 & 0.0068 & 0.0062 & 0.0077 & 0.0190 & 0.0108 & 0.0069 & 0.0058 \\ 
UK & 0.0175 & 0.0068 & 0.0100 & 0.0045 & 0.0342 & 0.0293 & 0.0082 & 0.0055 \\ 
USA & 0.0086 & 0.0073 & 0.0082 & 0.0061 & 0.0172 & 0.0170 & 0.0158 & 0.0192 \\ 
\midrule
Mean & 0.0508 & 0.0173 & 0.0308 & \textBF{0.0140} & 0.0700 & 0.0288 & 0.0341 & \textBF{0.0167} \\
\end{longtable}
\end{center}

\vspace{-.3in}

\begin{figure}[!htb]
\centering
\includegraphics[width=8.72cm]{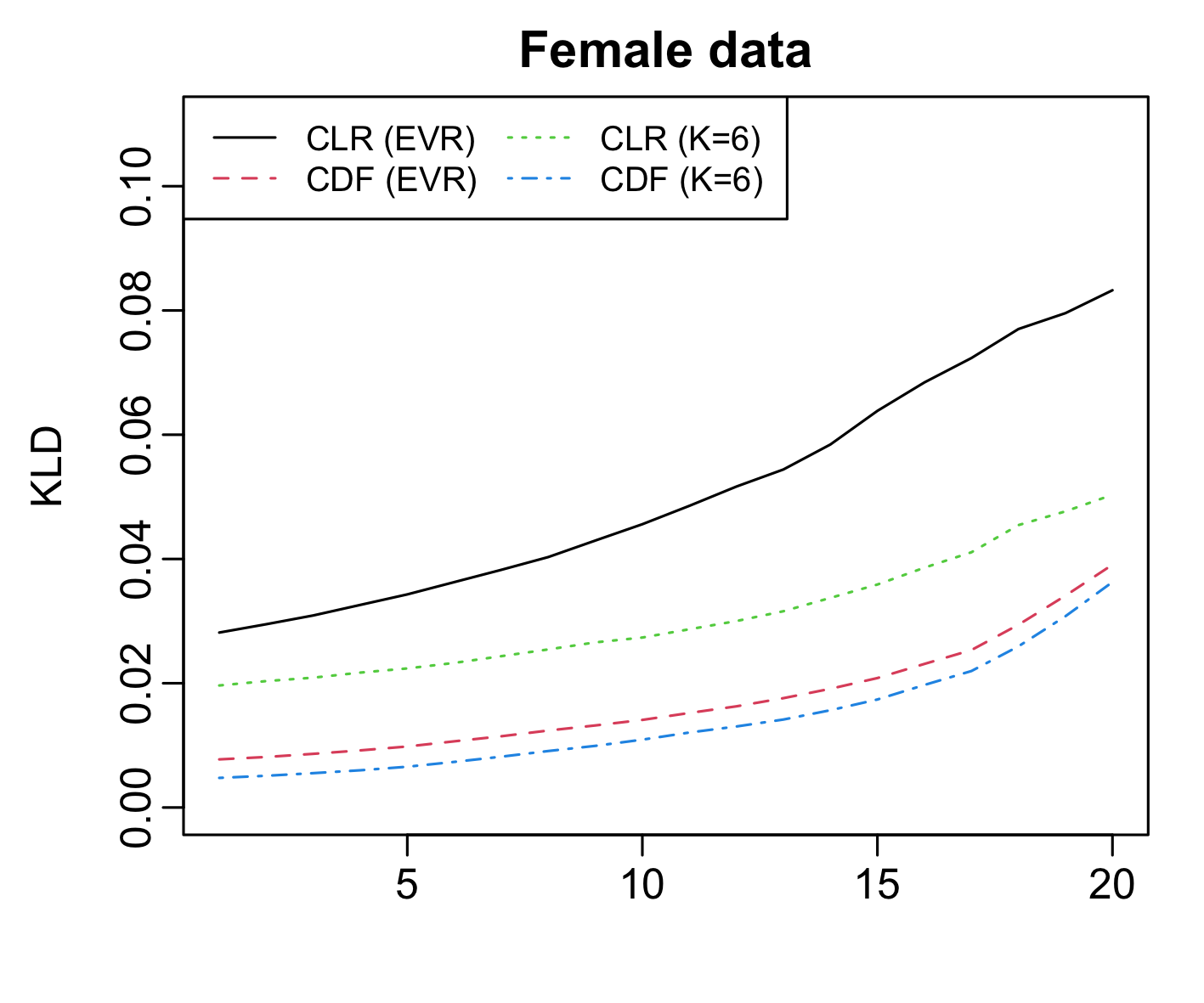}
\quad
\includegraphics[width=8.72cm]{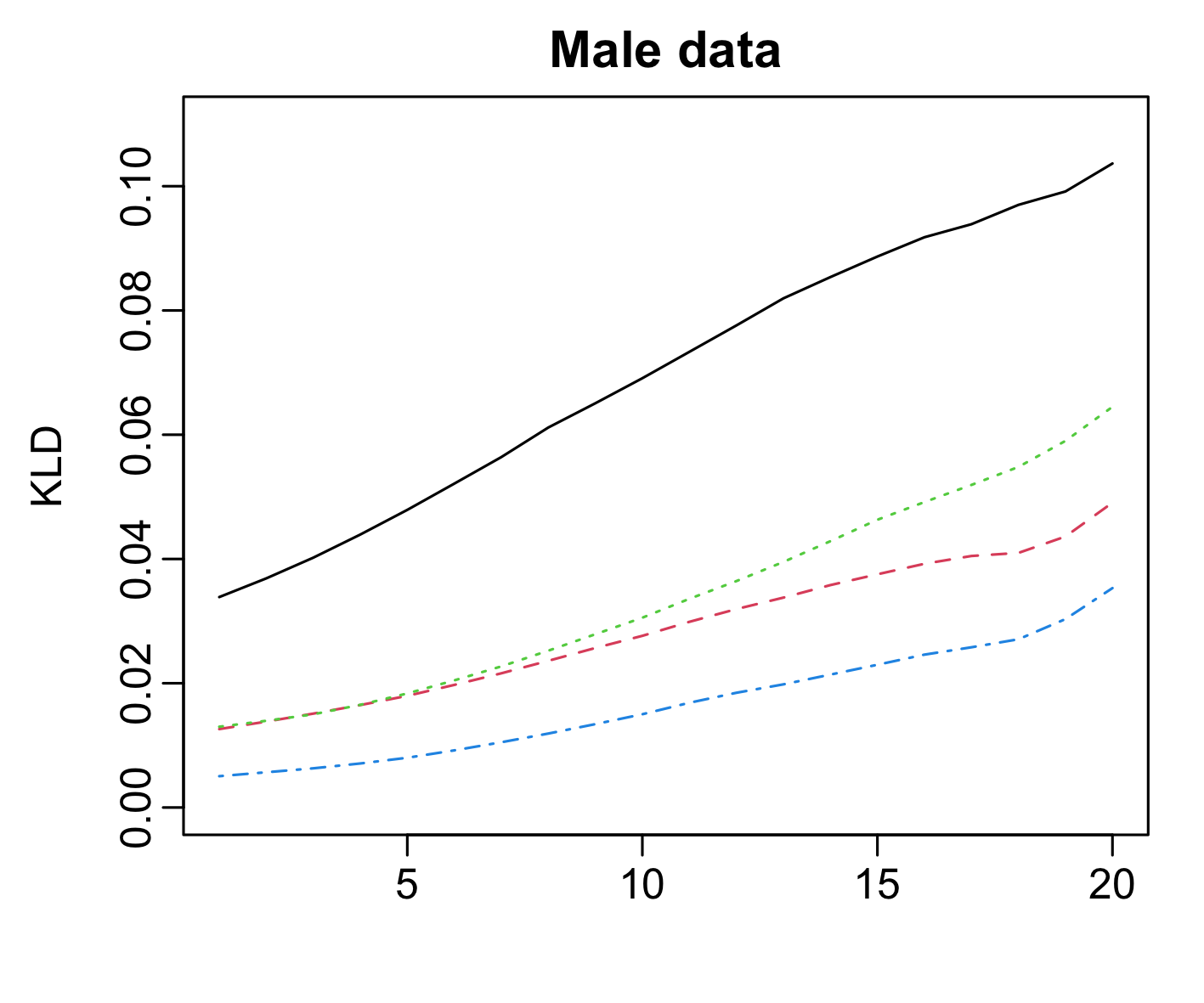}
\\
\includegraphics[width=8.72cm]{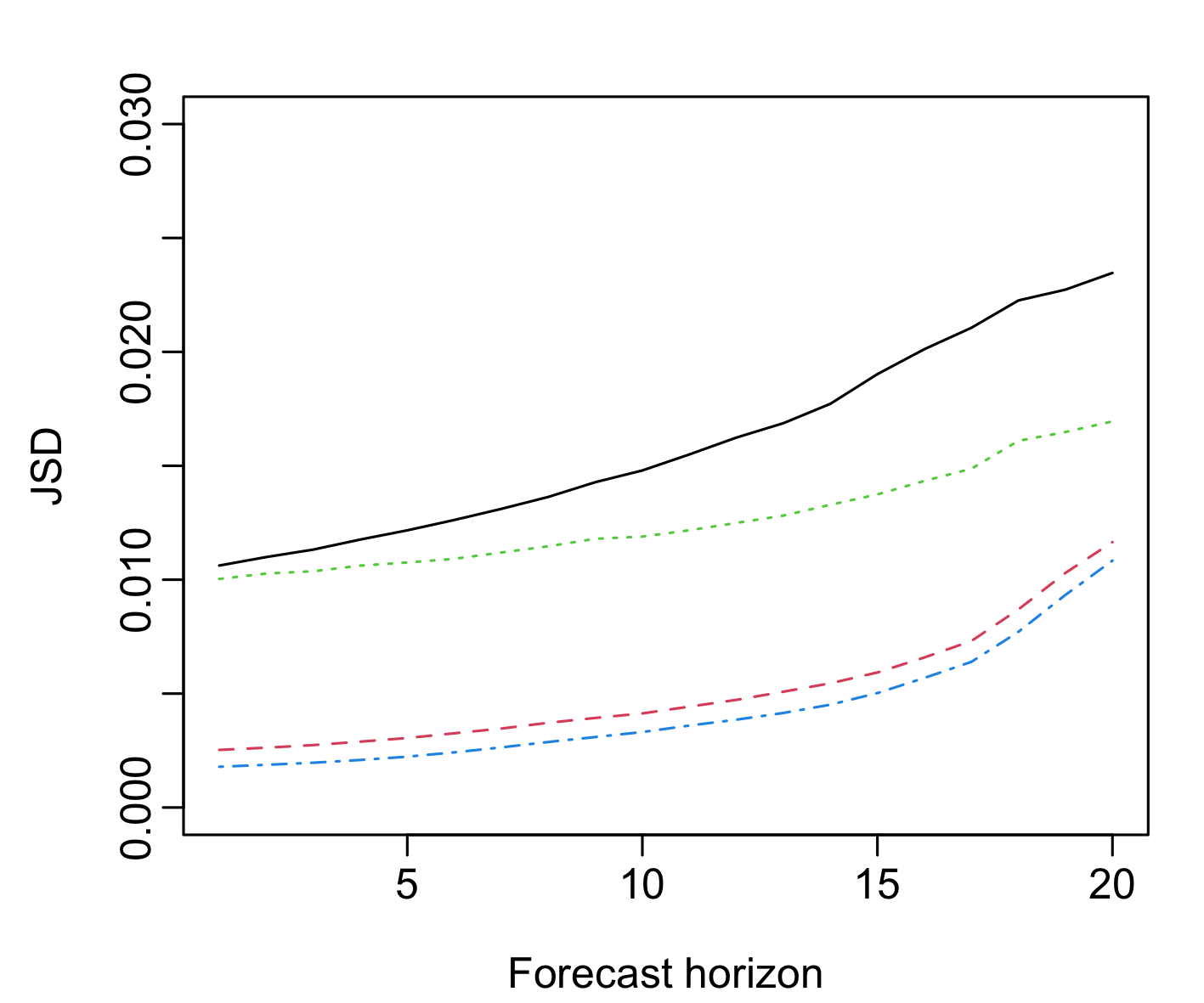}
\quad
\includegraphics[width=8.72cm]{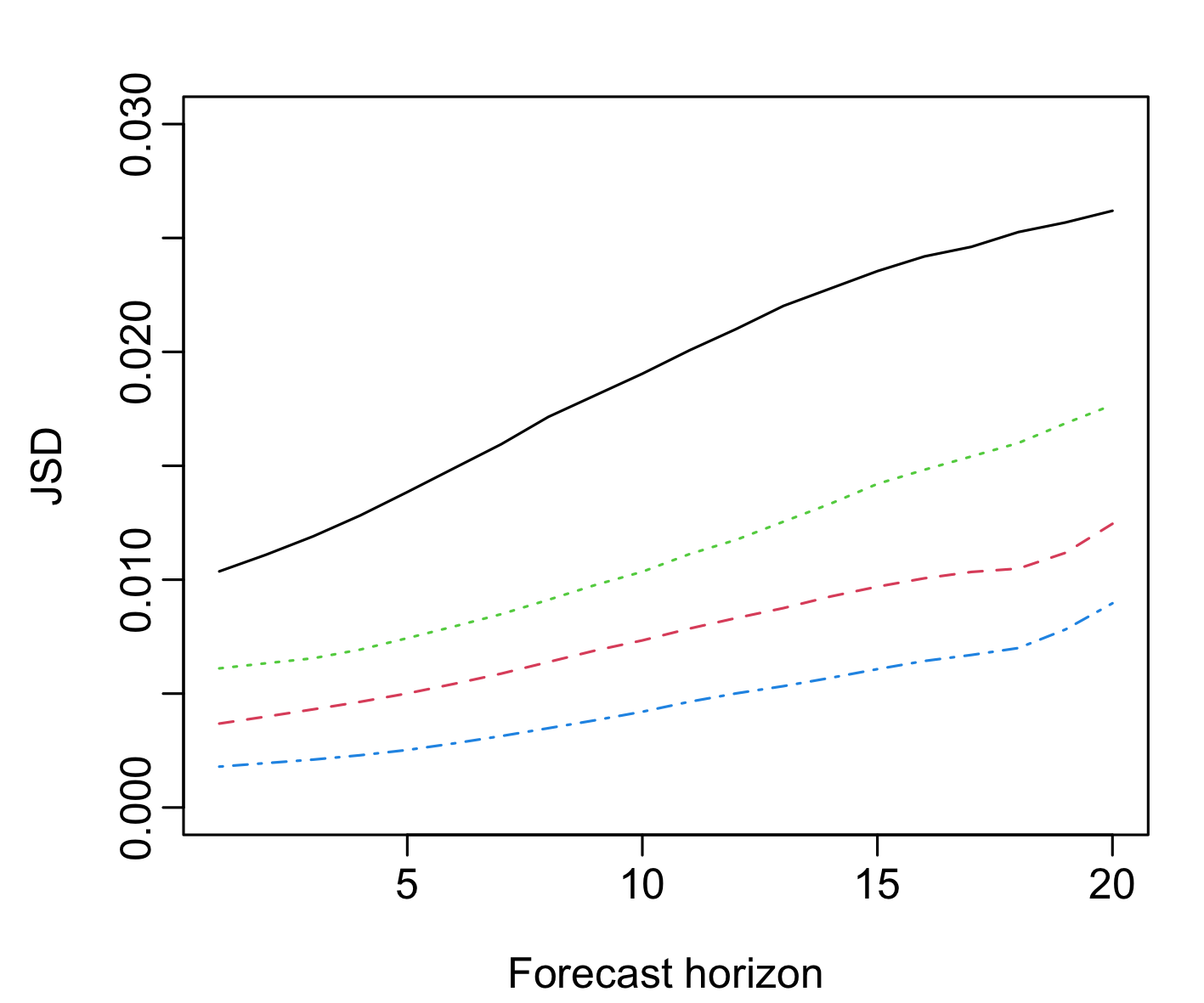}
\caption{Comparison of the point forecast accuracy, measured by the KLD and JSD, between the CLR and CDF transformations. For each transformation, we perform a principal component analysis, selecting the number of components using the EVR criterion or setting $K=6$.}\label{fig:3}
\end{figure}

Averaging over point forecast errors obtained from the 24 countries, Figure~\ref{fig:3} presents horizon-specific plots to compare the accuracy of the two transformations, as measured by the KLD and JSD. Between the CLR and CDF transformations, the CDF transformation \textit{consistently} outperforms the CLR transformation. This superiority is credited to the monotonicity constraint enjoyed by the CDF transformation. For modeling the unconstrained data in Euclidean space, it is advantageous to use $K=6$ for forecasting. In contrast, the number of components selected by the EVR tends to be fewer than optimal. Choosing a larger number of components results in only a minor loss, whereas selecting too few components leads to a more substantial loss.

\subsection{Construction of pointwise prediction intervals}

Using the data in the training sample, we applied an expanding-window approach to obtain $h$-step-ahead density forecasts in the validation set for $h=1,\dots,19$. For different forecast horizons, we have different numbers of curves in the validation set. For $h=1$, we have 20 years to evaluate forecast errors; for $h=19$, we have two years to evaluate the residual functions between the holdout data in the validation set and their forecasts. Based on these residuals, we can compute the functional standard deviation, which requires at least two years of data. The forecast errors can be denoted by $\widehat{\epsilon}_{\nu, x} = d_{\nu,x}-\widehat{d}_{\nu,x}$, where $\widehat{d}_{\nu,x}$ denotes the forecasts obtained from the two transformations, and $\nu=1,\dots, (21-h)$ denotes the number of observations in the validation set for $h=1,\dots,19$. When $h=1$, we have 20 observations in the validation set; when $h=19$, we have two observations.

Let us denote $\Gamma_{\nu, x} = \text{sd}(\widehat{\epsilon}_{\nu, x})$, where $\text{sd}(\cdot)$ can be pointwise standard deviation. We denote $\bm{\Gamma}_{\nu}=(\Gamma_{\nu,1},\dots,\Gamma_{\nu,111})^{\top}$ and $\bm{\widehat{\epsilon}}_{\nu}=(\widehat{\epsilon}_{\nu, 1},\dots,\widehat{\epsilon}_{\nu, 111})^{\top}$. For a level of significance $\alpha$, commonly $\alpha=0.2$ or~0.05, our aim is to find $\theta_{h, \alpha}$ such that approximately $100(1-\alpha)\%$ of the residuals satisfy
\begin{equation*}
-\theta_{h, \alpha}\bm{\Gamma}_{\nu} \leq \bm{\widehat{\epsilon}}_{\nu} \leq \theta_{h, \alpha}\bm{\Gamma}_{\nu},\qquad v=1,\dots,(21-h).
\end{equation*}
According to the law of large numbers, one can achieve the following
\begin{equation*}
\text{Pr}(-\theta_{h, \alpha}\Gamma_{\nu, x} \leq d_{n+h,x}-\widehat{d}_{n+h|n,x} \leq \theta_{h, \alpha}\Gamma_{\nu, x}) \approx \frac{1}{(21-h)\times 111}\sum^{21-h}_{\nu=1}\sum^{111}_{x=1}\mathds{1}(-\theta_{h, \alpha}\Gamma_{\nu, x}\leq \widehat{\epsilon}_{\nu, x}\leq \theta_{h, \alpha}\Gamma_{\nu,x}),
\end{equation*}
where $\mathds{1}(\cdot)$ represents the binary indicator function.

To determine the optimal $\theta_{h, \alpha}$, the samples in the validation set are used to calibrate a prediction interval so that its empirical coverage probability is close to its nominal coverage probability \citep[see also][]{SH25b}. With the estimated $\theta_{h, \alpha}$, the $100(1-\alpha)\%$ prediction interval of $d_{n+h,x}$ can be obtained as
\begin{equation*}
\Big[\widehat{d}_{n+h|n,x} - \theta_{h, \alpha}\Gamma_{\nu, x}, \quad \widehat{d}_{n+h|n,x} + \theta_{h, \alpha}\Gamma_{\nu, x}\Big],
\end{equation*}
where $v$ denotes the number of observations in the validation set. As the forecast horizon $h$ increases, the value of $\theta_{h, \alpha}$ generally increases, reflecting an increasing degree of forecast uncertainty.

\subsection{Interval forecast evaluation metrics}\label{sec:4.4}

For each year in the forecasting period, the $h$-step-ahead prediction intervals were computed at the $(1-\alpha)$ nominal coverage probability, where $\alpha$ denotes a level of significance. For a chosen~$\alpha$, let us denote $\widehat{d}_{n+\xi|n,x}^{\text{lb}}$ and $\widehat{d}_{n+\xi|n,x}^{\text{ub}}$, as the lower and upper bounds, respectively. We compute the empirical coverage probability, defined as the proportion of holdout data points that fall within the lower and upper bounds for a given significance level. It can be expressed as
\begin{equation*}
\text{ECP}_{\alpha}(h) = 1-\frac{\sum^{20}_{\xi=h}\sum^{111}_{x=1}\left[\mathds{1}(d_{n+\xi,x}<\widehat{d}_{n+\xi|n,x}^{\text{lb}})+\mathds{1}(d_{n+\xi,x}>\widehat{d}_{n+\xi|n,x}^{\text{ub}})\right]}{111\times (21 - h)},
\end{equation*}
where the denominator is the number of curves in the forecasting period, which depends on the forecast horizon.

From the ECP, we compute the coverage probability difference (CPD), the absolute difference between the empirical and nominal coverage probabilities, and use it to assess the accuracy of the interval forecast. As an absolute measure, the CPD cannot inform us about undercoverage or overcoverage. However, it eliminates the possibility of the canceling effect, in which undercoverage and overcoverage may occur concurrently. The CPD can be expressed as
\begin{equation*}
\text{CPD}_{\alpha}(h) = |\text{ECP}_{\alpha}(h) - (1 - \alpha)|.
\end{equation*}
The smaller the value of CPD, the more accurate the prediction interval provided by one method.

The ECP and CPD measure the accuracy of prediction intervals, and neither considers the sharpness of the prediction intervals, i.e., the distance between lower and upper bounds. We also consider the interval score of \cite{GR07}, defined as
\begin{align*}
S_{\alpha}(\widehat{d}_{n+\xi|n, x}^{\text{lb}}, \widehat{d}_{n+\xi|n, x}^{\text{ub}}, d_{n+\xi,x}) = (\widehat{d}_{n+\xi|n, x}^{\text{ub}} - \widehat{d}_{n+\xi|n, x}^{\text{lb}}) &+ \frac{2}{\alpha}(\widehat{d}_{n+\xi|n, x}^{\text{lb}} - d_{n+\xi, x})\mathds{1}(d_{n+\xi, x}<\widehat{d}_{n+\xi|n, x}^{\text{lb}}) \\
& + \frac{2}{\alpha}(d_{n+\xi, x} - \widehat{d}_{n+\xi|n, x}^{\text{ub}})\mathds{1}(d_{n+\xi, x}>\widehat{d}_{n+\xi|n, x}^{\text{ub}}).
\end{align*}
The interval score rewards a narrow prediction interval if and only if the holdout observation lies within the prediction interval. The optimal interval score is achieved when $d_{n+\xi, x}$ is not only between $\widehat{d}_{n+\xi|n, x}^{\text{lb}}$ and $\widehat{d}_{n+\xi|n, x}^{\text{ub}}$, but the distance between $\widehat{d}_{n+\xi|n, x}^{\text{lb}}$ and $\widehat{d}_{n+\xi|n, x}^{\text{ub}}$ is minimal for a given age $x$.

For different ages and years in the forecasting period, the mean interval score is defined by
\begin{equation*}
\overline{S}_{\alpha}(h) = \frac{\sum^{20}_{\xi=h}\sum^{111}_{x=1}S_{\alpha}(\widehat{d}_{n+\xi|n,x}^{\text{lb}}, \widehat{d}_{n+\xi|n,x}^{\text{ub}}; d_{n+\xi,x})}{111\times (21-h)},
\end{equation*}
where $S_{\alpha}(\widehat{d}_{n+\xi|n, x}^{\text{lb}}, \widehat{d}_{n+\xi|n, x}^{\text{ub}}, d_{n+\xi, x})$ denotes the interval score at the $\xi$\textsuperscript{th} curve in the forecasting period. Since we require at least two curves to compute the standard deviation, the forecast horizon is $h=1, 2,\dots,19$.

\subsection{Comparison of interval forecast accuracy}\label{sec:4.5}

Averaging over the interval forecasts from the 24 countries, Table~\ref{tab:2} presents horizon-specific forecast errors for comparing the accuracy of the 80\% prediction intervals constructed by the two transformations. Differing from point forecast accuracy, the CDF transformation does not consistently outperform the CLR transformation; however, it generally provides superior accuracy with smaller CPD and mean interval scores. Furthermore, we compare two methods for selecting the number of components in principal component analysis for modeling unconstrained data in the transformed space. Due to limited space, we do not report the results for the 95\% nominal coverage probability; however, these can be obtained upon request from the corresponding author.
\begin{center}
\tabcolsep 0.172in
\renewcommand{\arraystretch}{0.88}
\begin{longtable}{@{}llrrrrrrrr@{}}
\caption{Comparison of the interval forecast accuracy, as measured by the CPD$_{\alpha}$ and mean interval score $\overline{S}_{\alpha}$, between the CLR and CDF transformations at the level of significance $\alpha=0.2$. For each transformation, we compare two methods for selecting the number of components.}\label{tab:2} \\
\toprule
&	& \multicolumn{4}{c}{Female} & \multicolumn{4}{c}{Male} \\
&  & \multicolumn{2}{c}{EVR}   & \multicolumn{2}{c}{$K=6$}   & \multicolumn{2}{c}{EVR}   & \multicolumn{2}{c}{$K=6$} \\
\cmidrule(lr){3-4}  \cmidrule(lr){5-6}   \cmidrule(lr){7-8}   \cmidrule(lr){9-10}
Metric & $h$ & CLR & CDF & CLR & CDF  & CLR  & CDF & CLR  & CDF  \\ 
\midrule
\endfirsthead
\toprule
&  & \multicolumn{2}{c}{EVR}   & \multicolumn{2}{c}{$K=6$}   & \multicolumn{2}{c}{EVR}   & \multicolumn{2}{c}{$K=6$} \\
\cmidrule(lr){3-4}  \cmidrule(lr){5-6}   \cmidrule(lr){7-8}   \cmidrule(lr){9-10}
Metric & $h$ & CLR & CDF & CLR & CDF  & CLR  & CDF & CLR  & CDF  \\ 
\midrule
\endhead
\midrule
\multicolumn{10}{r}{\textit{Continued on next page}} \\
\endfoot
\endlastfoot
CPD$_{\alpha}$ 	&  1 & 0.027 & 0.023 & 0.089 & 0.100 & 0.152 & 0.047 & 0.082 & 0.102 \\ 
		 		&  2 & 0.134 & 0.040 & 0.037 & 0.015 & 0.143 & 0.002 & 0.040 & 0.010 \\ 
 				&  3 & 0.006 & 0.049 & 0.062 & 0.042 & 0.039 & 0.053 & 0.005 & 0.004 \\ 
 				&  4 & 0.131 & 0.089 & 0.097 & 0.091 & 0.135 & 0.116 & 0.086 & 0.012 \\ 
 				&  5 & 0.103 & 0.070 & 0.064 & 0.040 & 0.052 & 0.146 & 0.147 & 0.000 \\ 
 				&  6 & 0.103 & 0.005 & 0.048 & 0.047 & 0.676 & 0.022 & 0.014 & 0.018 \\ 
 				&  7 & 0.073 & 0.103 & 0.045 & 0.017 & 0.160 & 0.025 & 0.079 & 0.088 \\ 
 				&  8 & 0.027 & 0.030 & 0.093 & 0.091 & 0.027 & 0.155 & 0.011 & 0.008 \\ 
 				&  9 & 0.040 & 0.102 & 0.062 & 0.093 & 0.197 & 0.090 & 0.008 & 0.033 \\ 
 				&  10 & 0.083 & 0.022 & 0.002 & 0.077 & 0.404 & 0.185 & 0.242 & 0.259 \\ 
 				&  11 & 0.162 & 0.029 & 0.098 & 0.019 & 0.178 & 0.069 & 0.032 & 0.080 \\ 
 				&  12 & 0.179 & 0.031 & 0.038 & 0.043 & 0.574 & 0.144 & 0.043 & 0.033 \\ 
 				&  13 & 0.130 & 0.106 & 0.153 & 0.172 & 0.018 & 0.026 & 0.119 & 0.098 \\ 
 				&  14 & 0.226 & 0.115 & 0.027 & 0.144 & 0.096 & 0.121 & 0.082 & 0.134 \\ 
 				&  15 & 0.117 & 0.024 & 0.219 & 0.064 & 0.350 & 0.354 & 0.159 & 0.073 \\ 
 				&  16 & 0.119 & 0.070 & 0.040 & 0.018 & 0.283 & 0.187 & 0.027 & 0.056 \\ 
 				&  17 & 0.200 & 0.117 & 0.099 & 0.155 & 0.160 & 0.123 & 0.110 & 0.186 \\ 
 				&  18 & 0.197 & 0.110 & 0.004 & 0.083 & 0.035 & 0.101 & 0.052 & 0.092 \\ 
 				&  19 & 0.196 & 0.138 & 0.133 & 0.102 & 0.241 & 0.151 & 0.192 & 0.196 \\ 
\cmidrule{2-10}
 				& Mean   & 0.119 & \textBF{0.067} & 0.074 & 0.074 & 0.206 & 0.112 & 0.081 & \textBF{0.078} \\ 
\midrule
$\overline{S}_{\alpha}$ 	&  1 & 347 & 326 & 149 & 138 & 729 & 415 & 144 & 150 \\ 
					&  2 & 432 & 286 & 241 & 197 & 634 & 339 & 353 & 243 \\ 
					&  3 & 480 & 309 & 333 & 211 & 703 & 699 & 345 & 273 \\ 
					&  4 & 1301 & 787 & 441 & 404 & 722 & 514 & 392 & 374 \\ 
					&  5 & 263 & 259 & 190 & 151 & 599 & 707 & 337 & 213 \\ 
					&  6 & 621 & 343 & 294 & 305 & 1846 & 530 & 420 & 428 \\ 
					&  7 & 432 & 321 & 335 & 333 & 999 & 608 & 811 & 578 \\ 
					&  8 & 709 & 575 & 485 & 301 & 2743 & 1044 & 706 & 443 \\ 
					&  9 & 770 & 366 & 578 & 355 & 2461 & 947 & 776 & 717 \\ 
					&  10 & 1015 & 360 & 395 & 440 & 1254 & 1048 & 1207 & 1097 \\ 
					&  11 & 1409 & 1075 & 975 & 781 & 2029 & 664 & 979 & 662 \\ 
					&  12 & 2188 & 527 & 521 & 468 & 2879 & 1471 & 1179 & 610 \\ 
					&  13 & 504 & 645 & 439 & 635 & 2614 & 918 & 784 & 728 \\ 
					&  14 & 1432 & 1803 & 1514 & 1702 & 355 & 449 & 471 & 450 \\ 
					&  15 & 1340 & 601 & 837 & 509 & 2250 & 1556 & 1347 & 1108 \\ 
					&  16 & 1313 & 869 & 823 & 479 & 1999 & 1641 & 1137 & 924 \\ 
					&  17 & 1208 & 472 & 433 & 501 & 2672 & 1213 & 1168 & 975 \\ 
					&  18 & 1088 & 602 & 518 & 462 & 1476 & 1021 & 1100 & 814 \\ 
					&  19 & 2824 & 1500 & 1429 & 1556 & 1985 & 1884 & 1789 & 1516 \\ 
\cmidrule{2-10}
					& Mean   & 1036 & 633 & 575 & \textBF{523} & 1629 & 930 & 813 & \textBF{648} \\ 
\bottomrule
\end{longtable}
\end{center}

\vspace{-.2in}

\begin{figure}[!htb]
\centering
\subfloat[CPD$_{\alpha=0.2}$ for female data]
{\includegraphics[width=8.56cm]{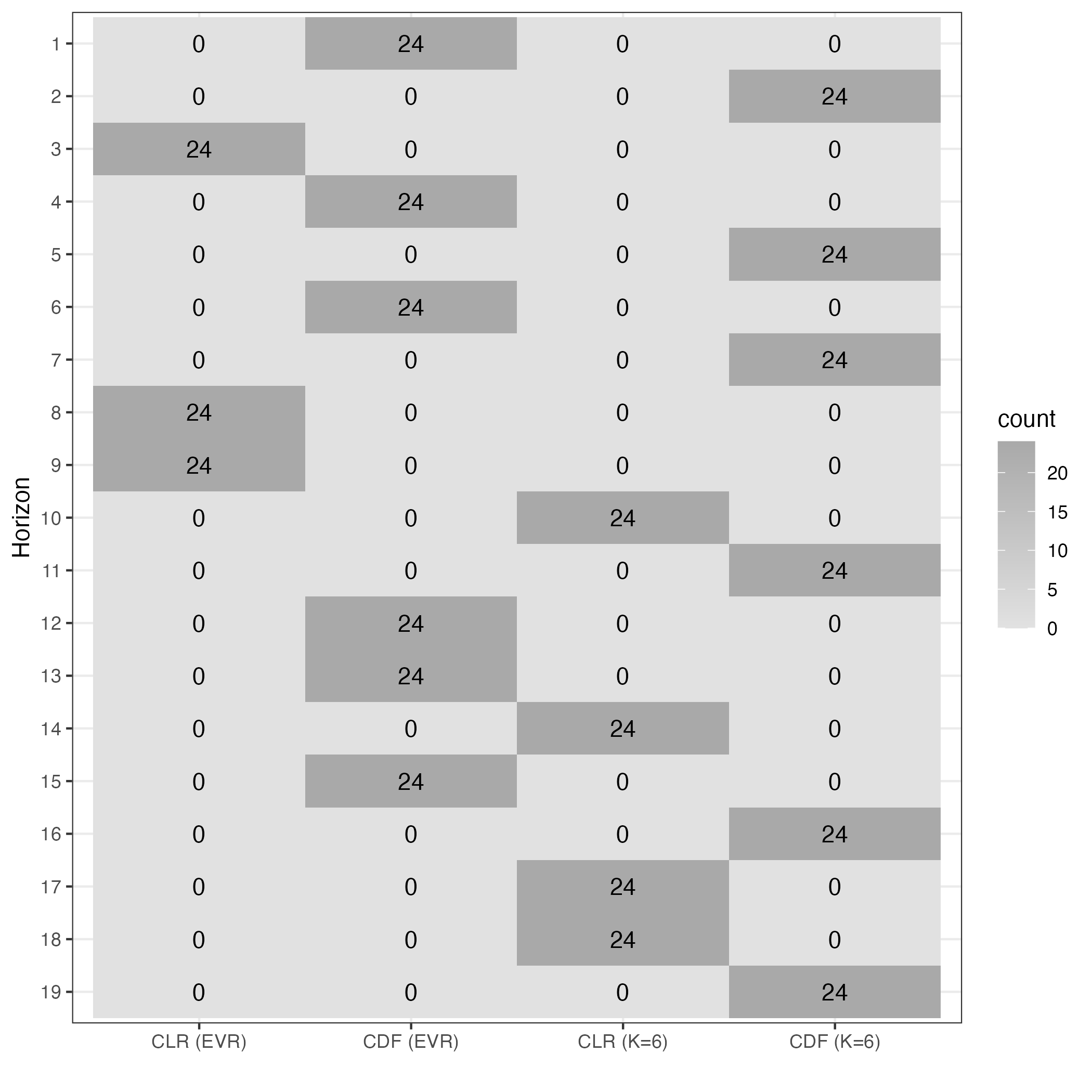}}
\qquad
\subfloat[CPD$_{\alpha=0.2}$ for male data]
{\includegraphics[width=8.56cm]{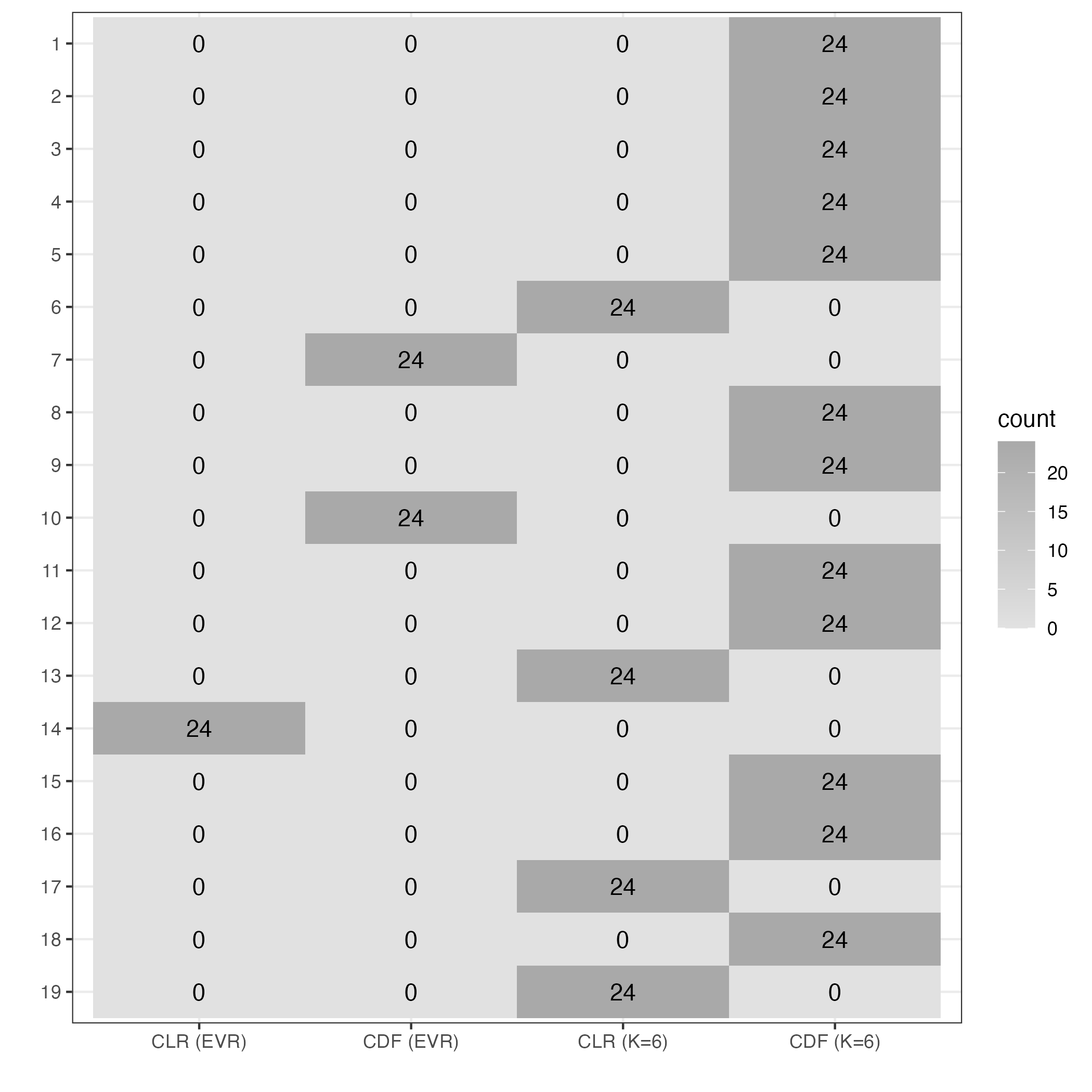}}
\\
\subfloat[$\overline{S}_{\alpha=0.2}$ for female data]
{\includegraphics[width=8.56cm]{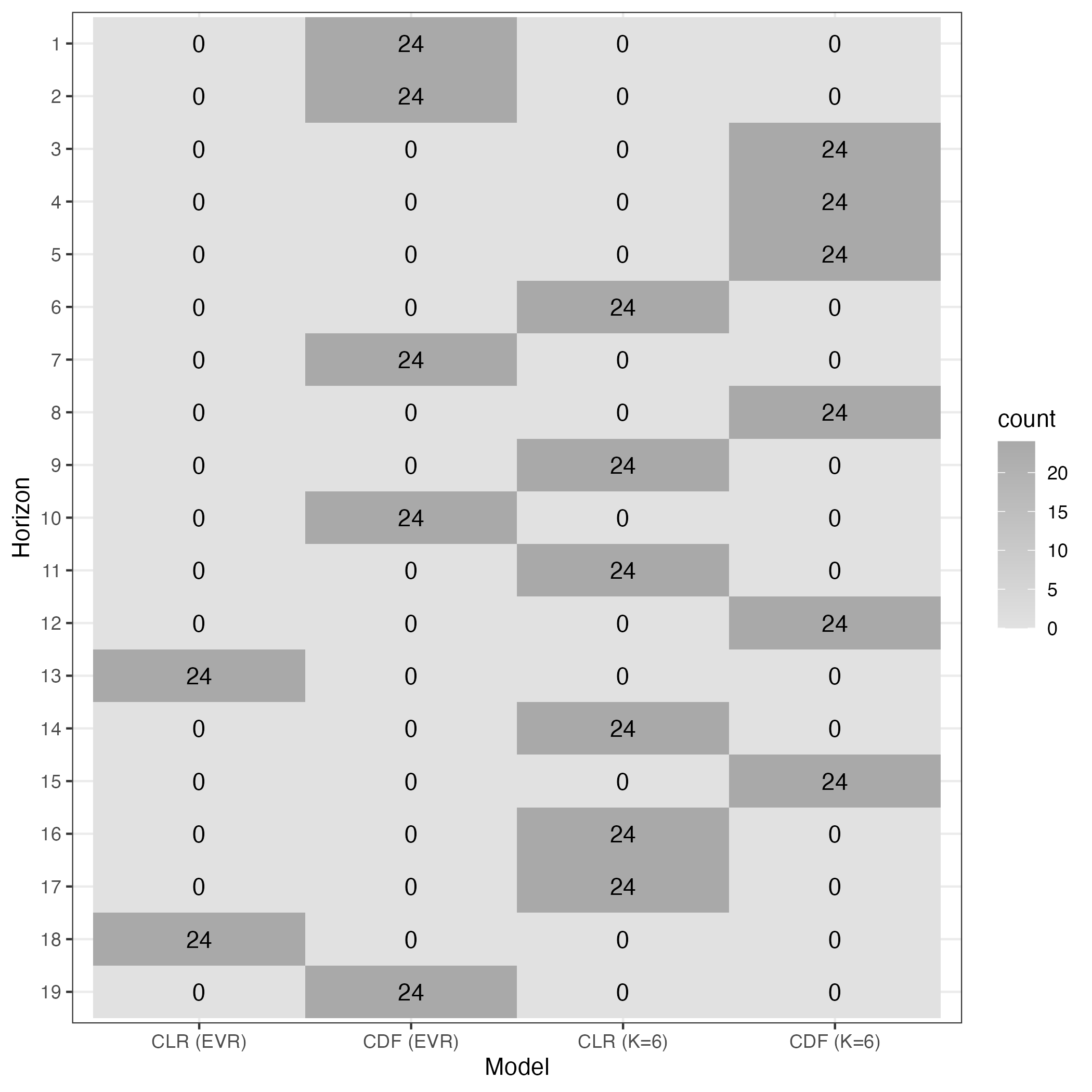}}
\qquad
\subfloat[$\overline{S}_{\alpha=0.2}$ for male data]
{\includegraphics[width=8.56cm]{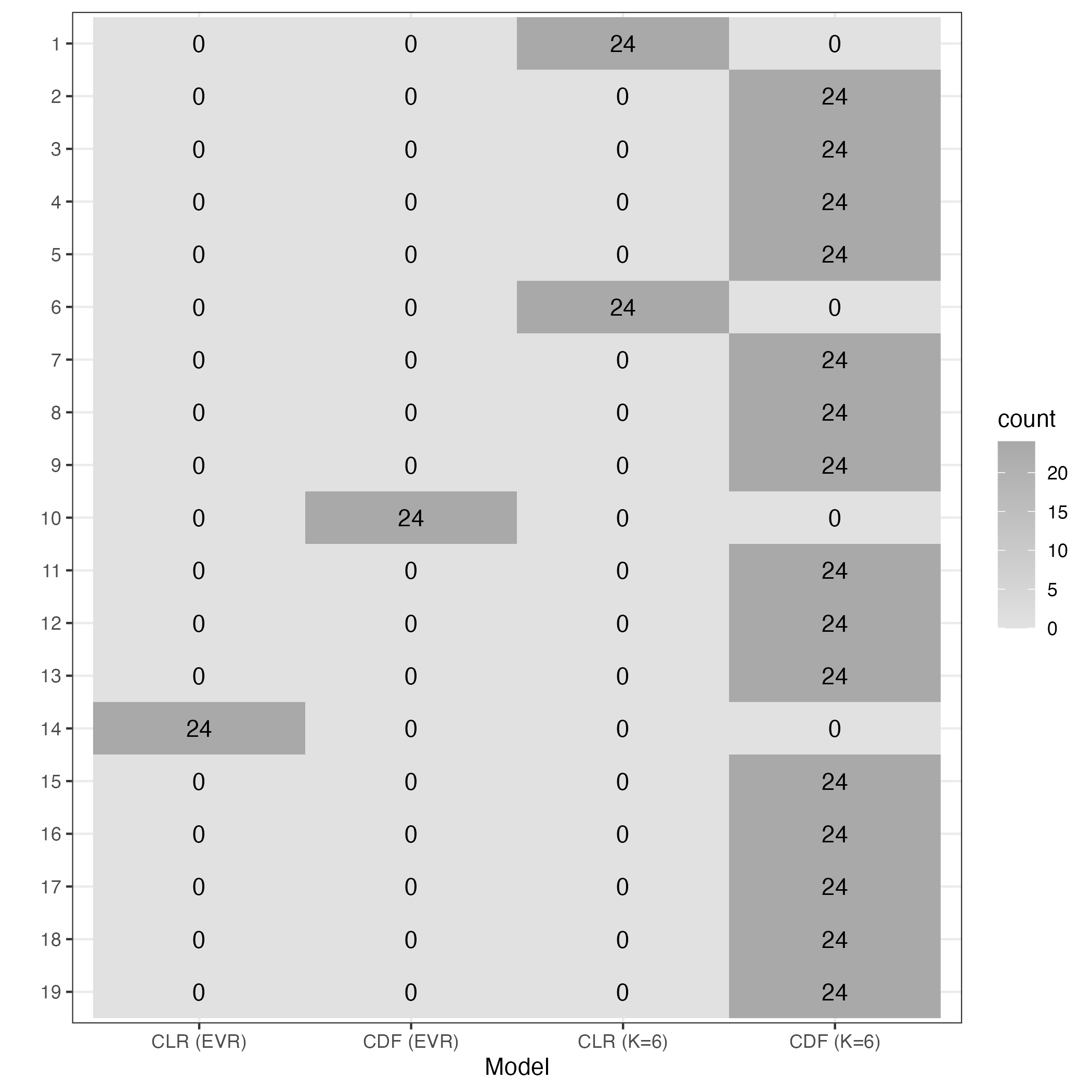}}
\caption{Heatmaps showing the frequency of the most accurate transformation, as measured by the CPD and mean interval score, at the nominal coverage probability of 80\%. Each row sums up to 24 countries.}\label{fig:4}
\end{figure}

In Table~\ref{tab:2}, we compare the interval forecast accuracy using the CPD and mean interval scores, aggregated across 24 countries. Based on the averaged CPD$_{\alpha}$ and $\overline{S}_{\alpha}$ across various forecast horizons, the CDF transformation generally provides smaller interval forecast errors than those obtained from the CLR transformation. In Figure~\ref{fig:4}, we present several heatmaps illustrating the frequency with which each method yields the smallest errors. For modeling the female data, the difference between the CLR and CDF transformations is marginal. Across forecast horizons, the inconsistency between the two transformations may stem from the optimal estimation of the parameters $\theta_{h,\alpha}$ in both. For modeling male data, the CDF transformation with $K=6$ is recommended, as it yields the smallest CPD and the smallest mean interval score.

\section{Multi-country comparisons}\label{sec:5}

\subsection{Comparison of life expectancy at birth} \label{sec:5.1}

As a widely used summary measure for both demographers and actuaries, we evaluate and compare forecast life expectancies at birth across countries \citep[see also][]{SBH11}. Using the \verb|LifeTable|  function of the MortalityLaws package \citep{Pascariu25} in \Rlogo \ , we convert the forecast age distribution of deaths into life expectancy at birth.
\begin{figure}[!htb]
\centering
\includegraphics[width=8.74cm]{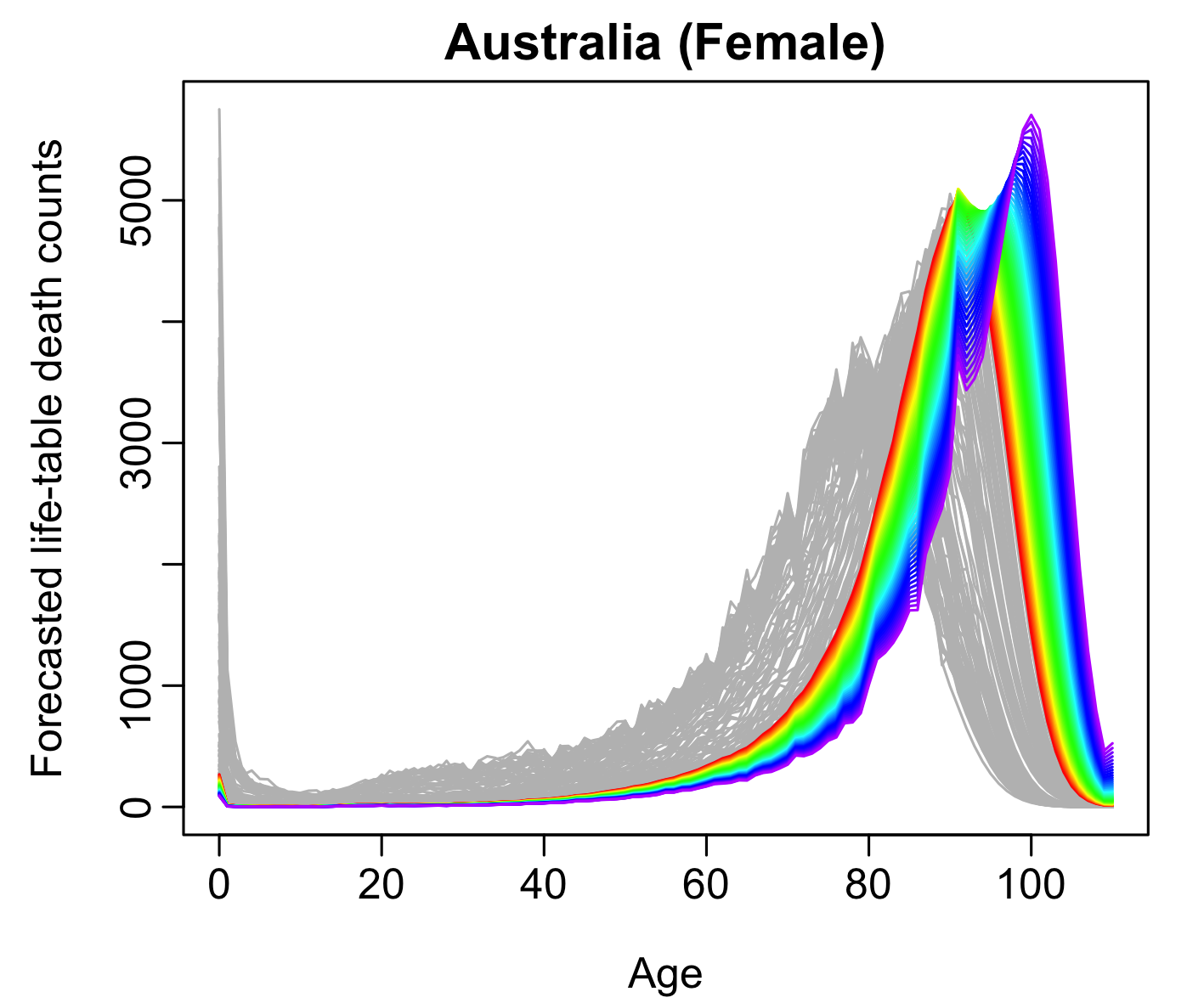}
\quad
\includegraphics[width=8.74cm]{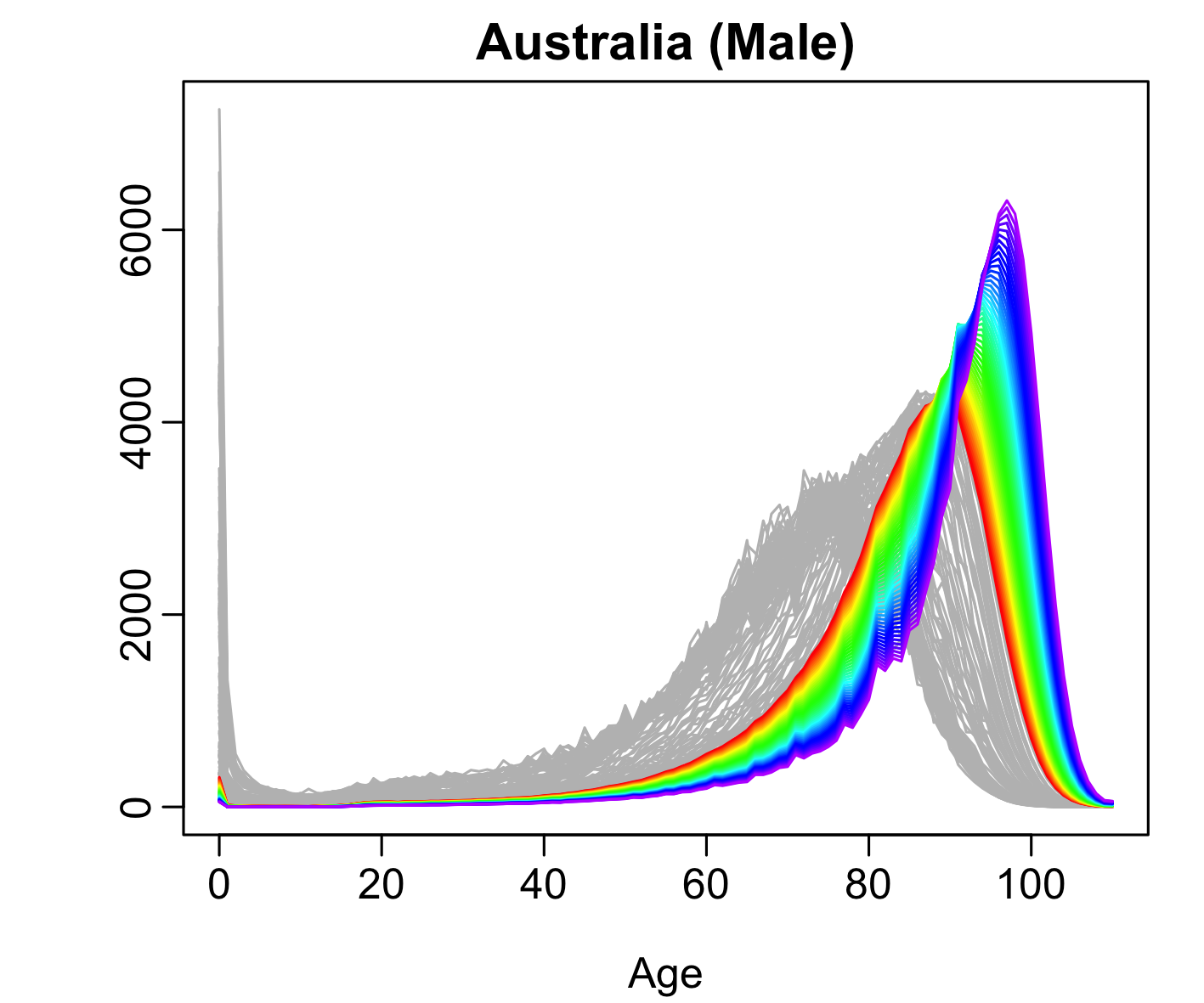}
\caption{Using the CDF transformation, we obtain age-specific life-table death count forecasts from 2022 to 2071 for Australian females and males.}\label{fig:5}
\end{figure}

The CDF transformation generally yields smaller forecast errors than the CLR transformation. Using the CDF transformation, we obtain 50-years-ahead forecasts of life-table death counts for all countries. As a demonstration, we display the forecasts for Australia in Figure~\ref{fig:5} and observe a continuing trend of greater longevity. These forecasts are time-series extrapolations of past trends and should be used with caution. Any extreme shocks, such as the COVID-19 pandemic, are likely to have a substantial impact on the mortality trend. 

\begin{figure}[!htb]
\centering
\subfloat[Forecasts for 2022]
{\includegraphics[width=8.78cm]{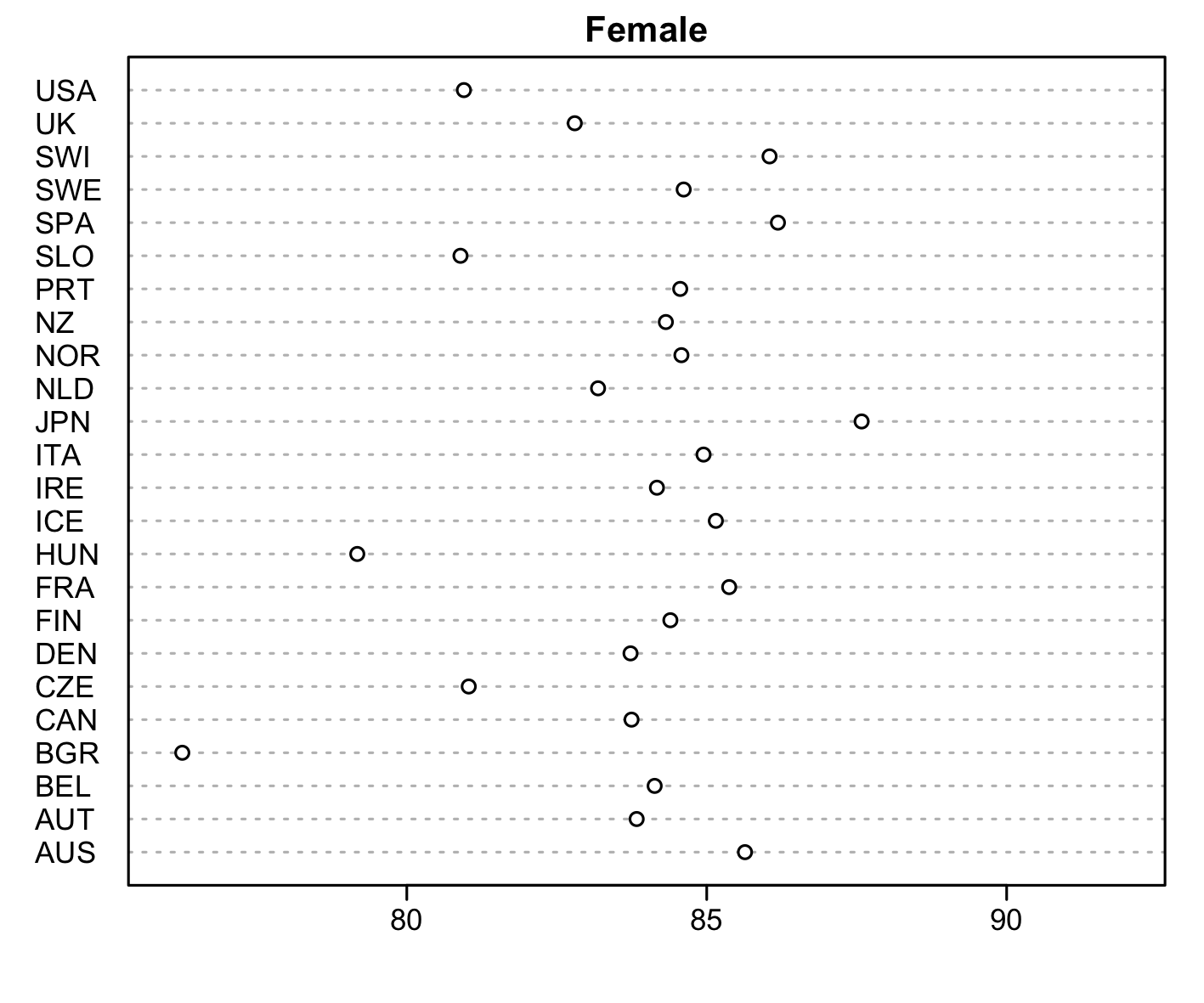}}
\quad
\subfloat[Forecasts for 2022]
{\includegraphics[width=8.78cm]{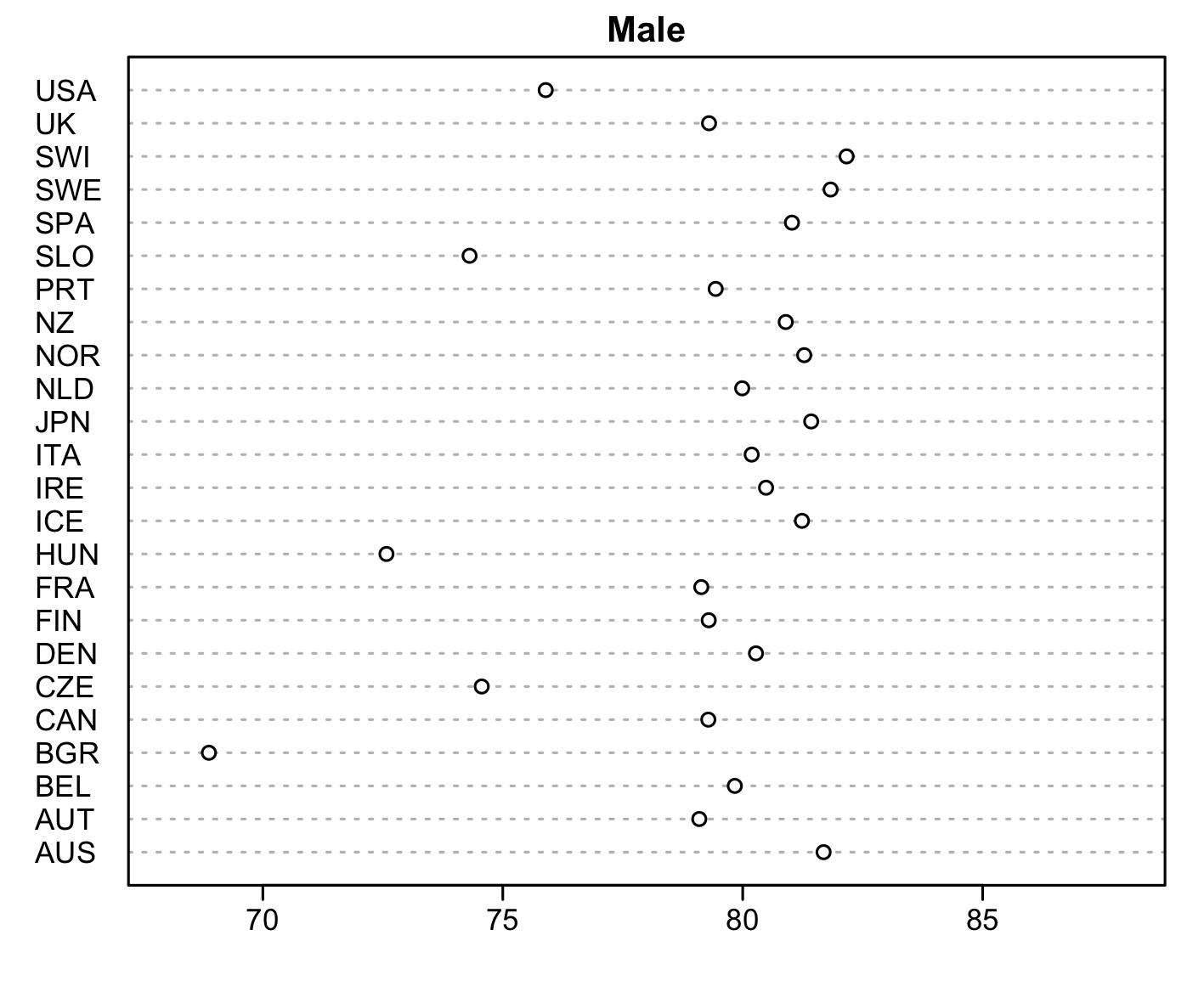}}
\\
\subfloat[Forecasts for 2050]
{\includegraphics[width=8.78cm]{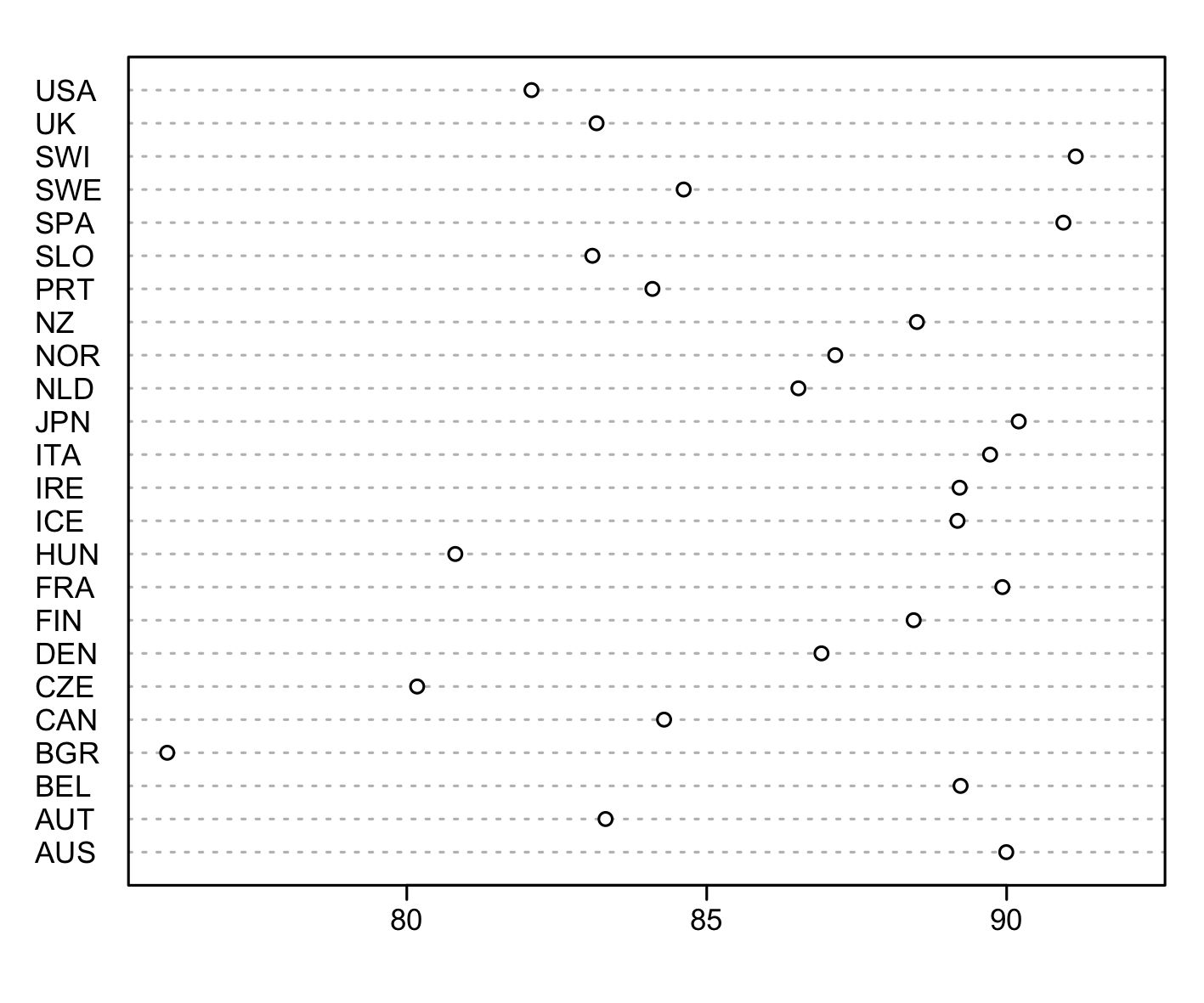}}
\quad
\subfloat[Forecasts for 2050]
{\includegraphics[width=8.78cm]{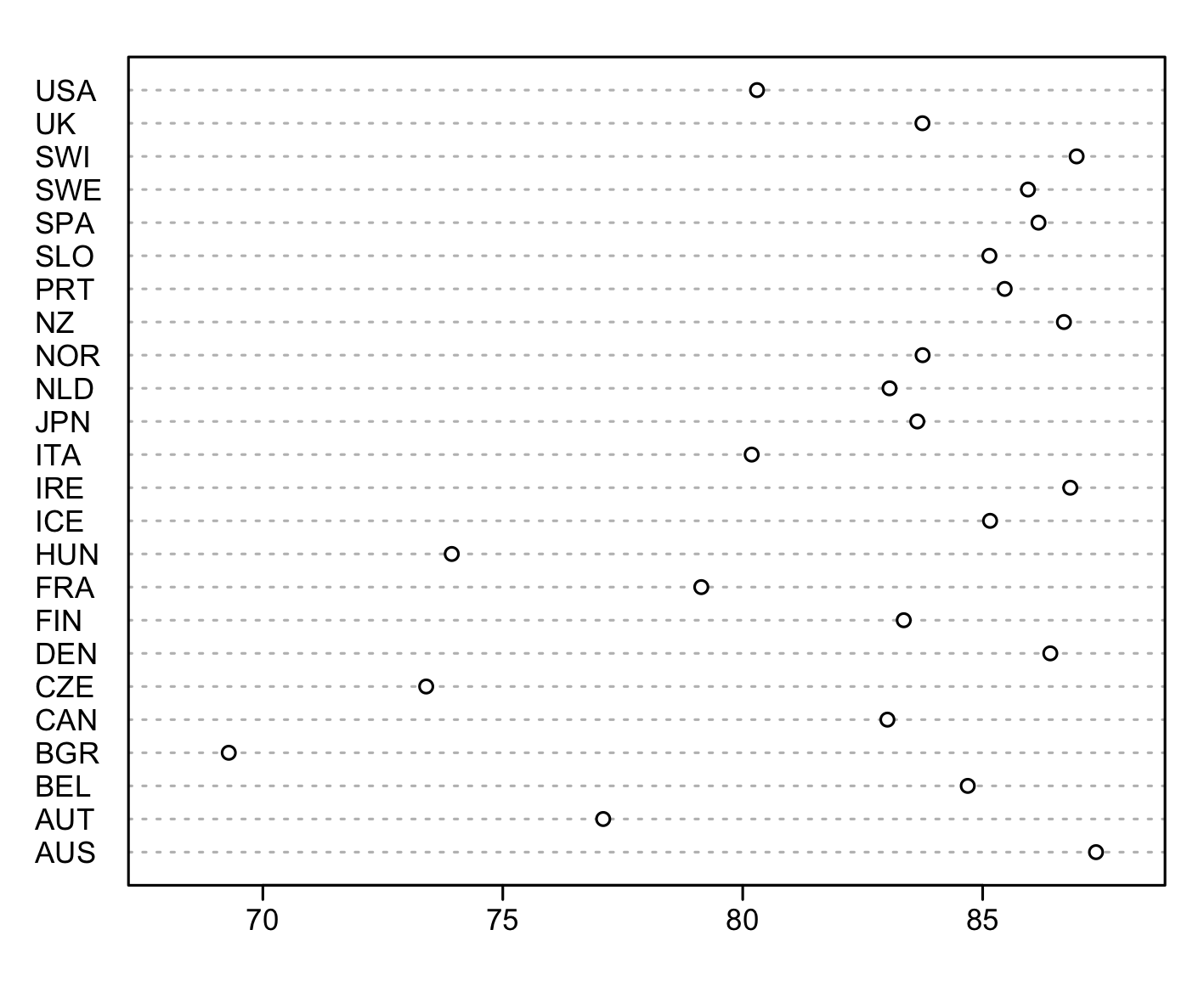}}
\caption{Estimated period life expectancy at birth of the female and male populations in the 24 countries in 2022 and 2050.}\label{fig:6}
\end{figure}

Via a standard life-table calculation, we convert the forecasts of life-table death counts $(d_{x})$ (from 2022 to 2071) to their corresponding \textit{period} life expectancies ($e_x$). In Figure~\ref{fig:6}, we display the forecasted life expectancy at birth of the female and male populations in the 24 countries. In 2022, Japan had the highest life expectancy for females, but by 2050, the highest may be in Switzerland. In 2022, Switzerland had the highest life expectancy among men, but by 2050, Australia may have the highest. Among all countries, Bulgaria has the lowest life expectancy at birth, due to factors such as public health, socioeconomic status, and healthcare.

\begin{figure}[!htb]
\centering
\subfloat[Forecasts for 2022]
{\includegraphics[width=8.78cm]{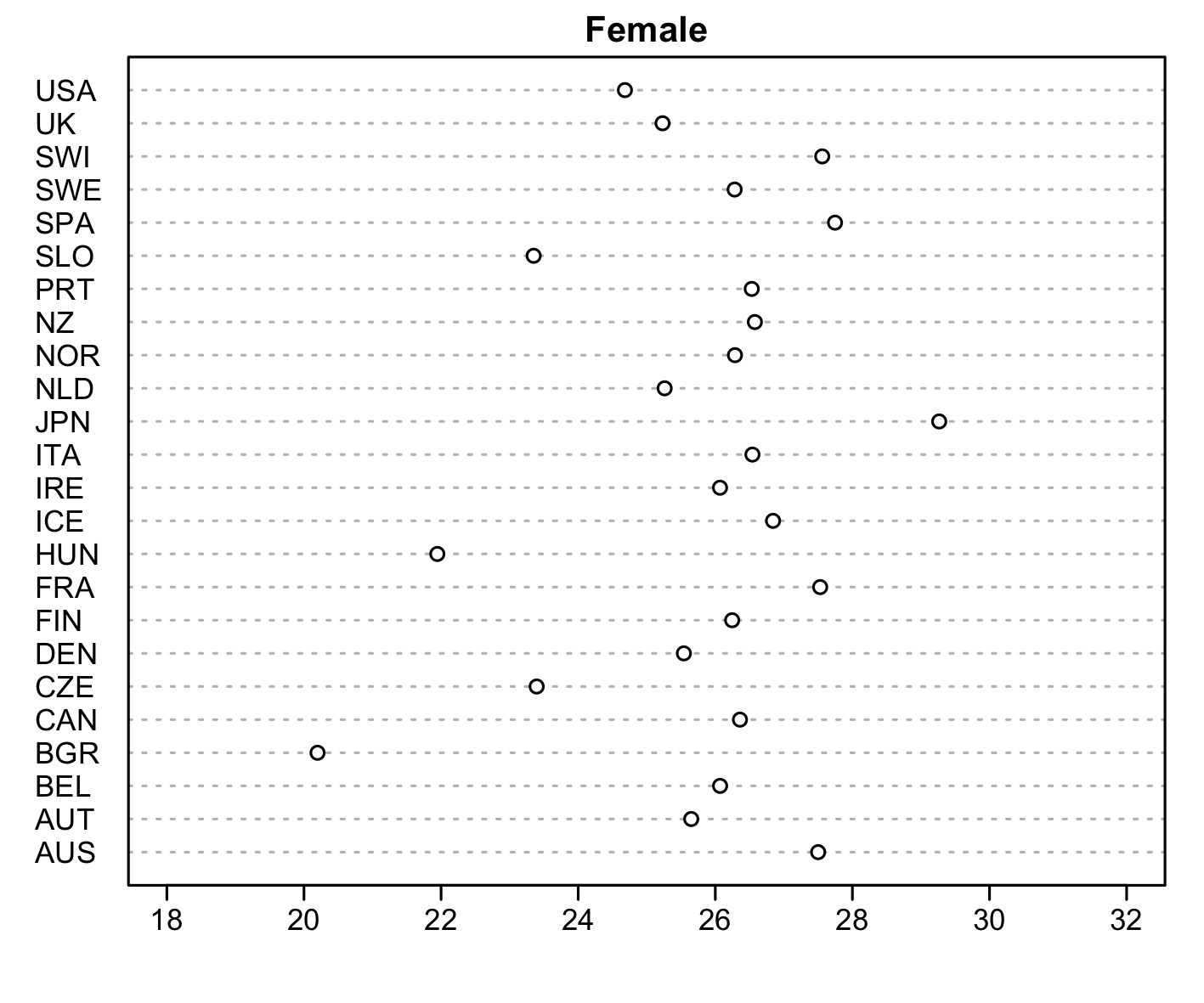}}
\quad
\subfloat[Forecasts for 2022]
{\includegraphics[width=8.78cm]{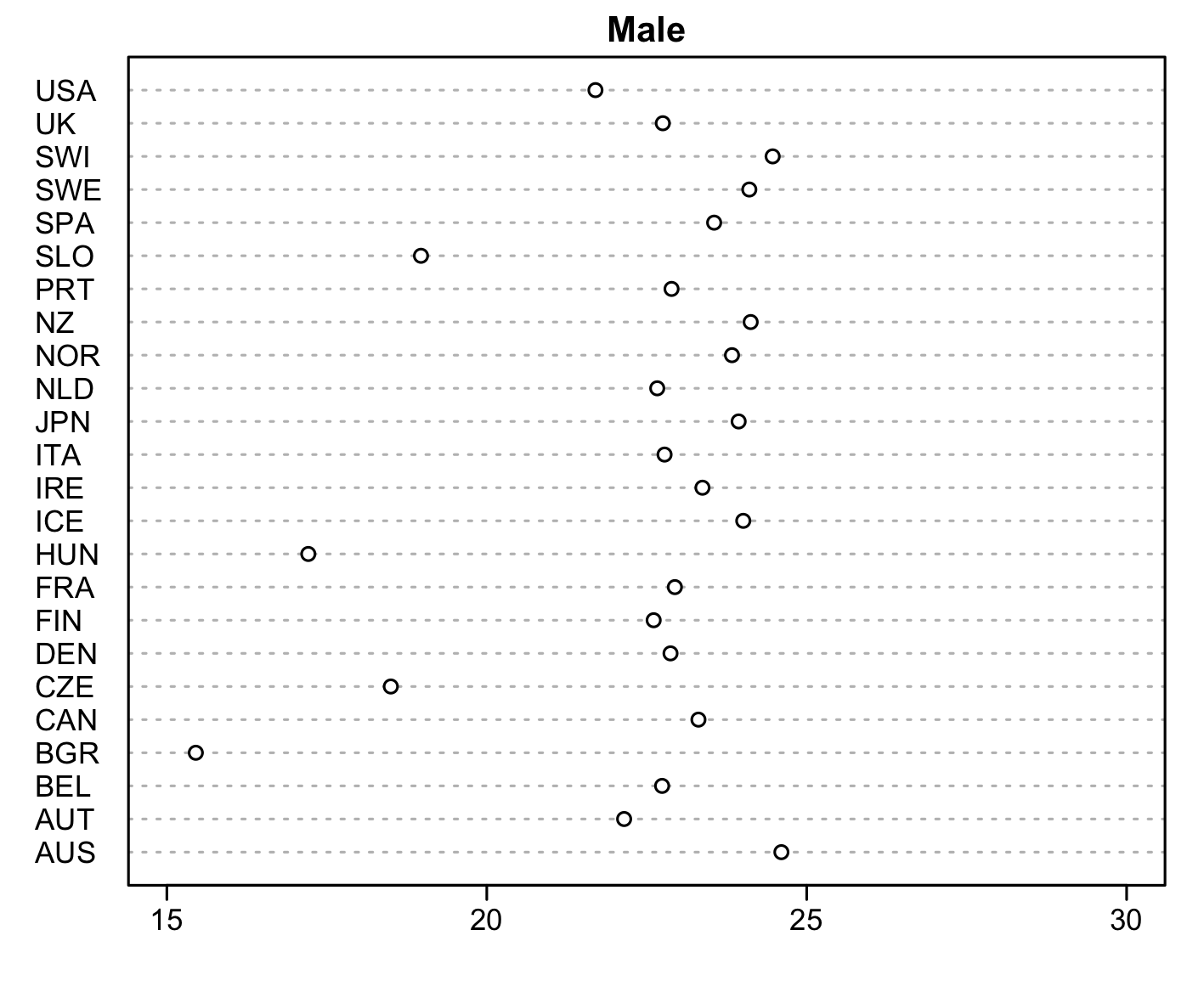}}
\\
\subfloat[Forecasts for 2050]
{\includegraphics[width=8.78cm]{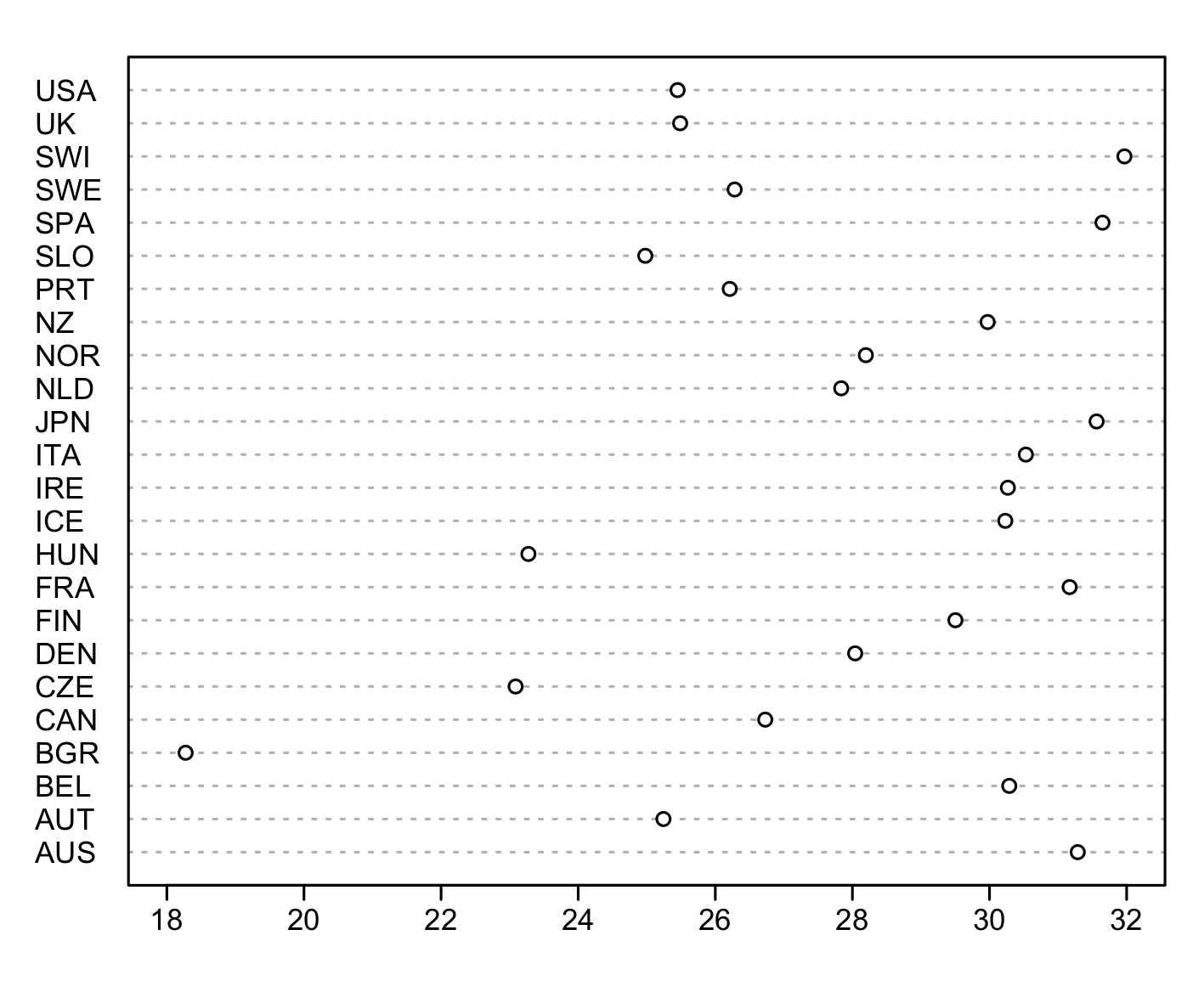}}
\quad
\subfloat[Forecasts for 2050]
{\includegraphics[width=8.78cm]{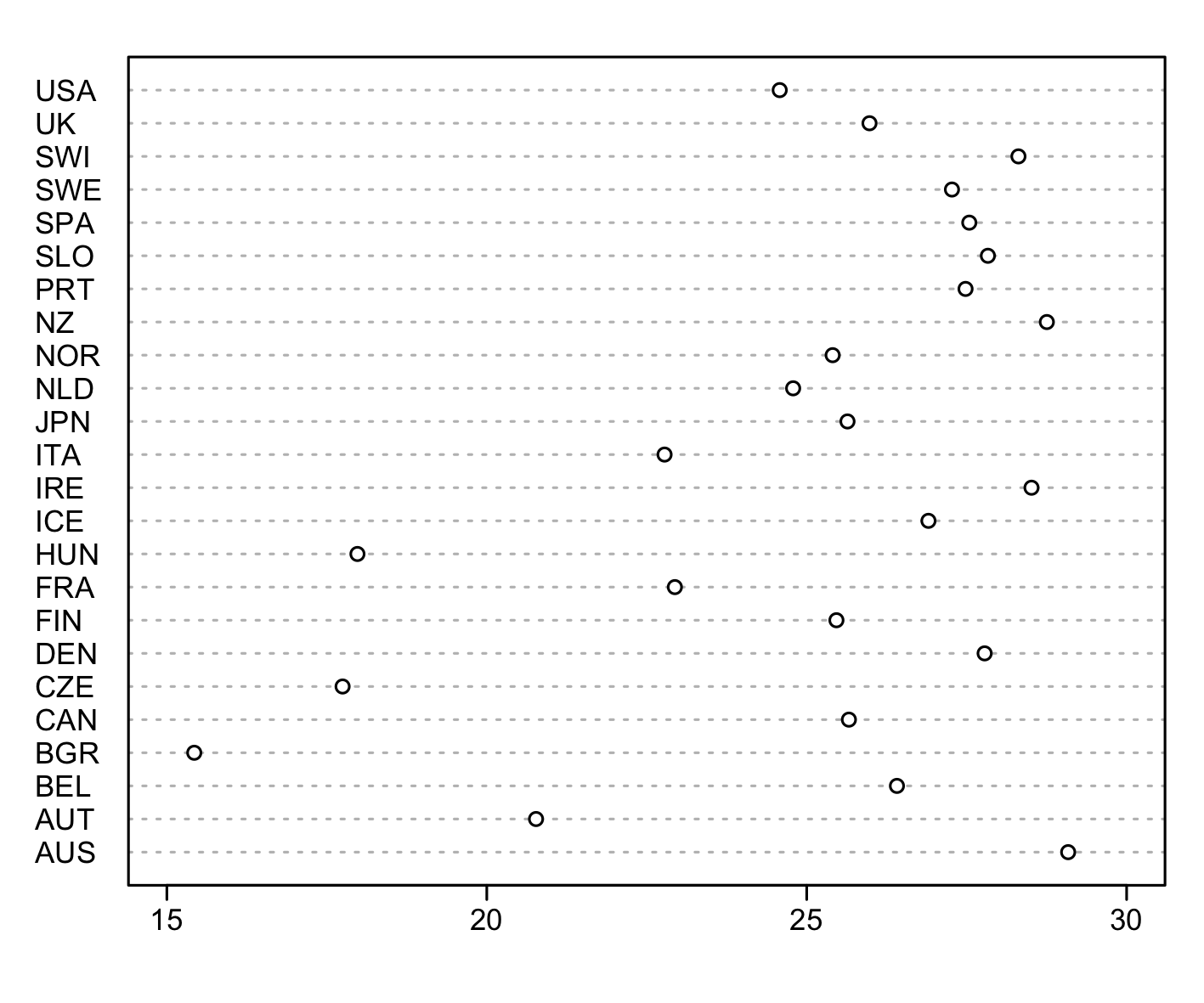}}
\caption{Estimated period life expectancy at age 60 of the female and male populations in the 24 countries in 2022 and 2050.}\label{fig:10}
\end{figure}

In Figure~\ref{fig:10}, we display the forecasted life expectancy at \textit{age 60} of the female and male populations in the 24 countries. In 2022, Japan had the highest life expectancy for females, but by 2050, the highest may be in Switzerland. Among males, Australia had the highest life expectancy in 2022 and may still maintain its lead by 2050. Among all countries, Bulgaria has the lowest life expectancy at 60 years of age.

\vspace{-.2in}

\subsection{Comparison of annuity price}\label{sec:5.2}

An immediate application of life-table death count forecasts for individuals aged 60 and over is in the superannuation industry, whose profitability and solvency crucially depend on accurate mortality forecasts to appropriately hedge longevity risks. When a person retires, a reliable way to ensure a steady financial income is to purchase an annuity. Such an annuity is a financial contract offered by insurers that guarantees a steady stream of payments for either a temporary or the lifetime of the annuitant in exchange for an initial premium.

Among many annuity products, we study temporary annuities, which have grown in popularity in a number of countries, including Australia and the USA, because lifetime immediate annuities, where rates are locked in for life, have been demonstrated to deliver poor value for money \citep[see, e.g.,][Chapter 6]{CT08}. Temporary annuities pay a predetermined, guaranteed income level higher than that provided by a lifetime annuity for a similar premium amount. Fixed-term annuities also offer an attractive alternative to lifetime annuities, allowing the purchaser to purchase a deferred annuity later.

Via a standard life-table calculation, we convert the forecasts of life-table death counts $(\bm{d}_x)$ to their corresponding survival probabilities ($\bm{p}_x$) from 2022 to 2071. We plug the age-specific survival probabilities into the calculation of single-premium term immediate annuities, and we adopt a \textit{cohort} approach to calculating the survival probabilities, as is the practice of insurance companies, that is, survival probability $p_{60}$ is computed from the forecast life-table death counts in 2022, $p_{61}$ is computed from the forecast life-table death counts in 2023, $\dots$, $p_{109}$ is computed from the forecast life-table death counts in 2071. The $\tau$ year survival probability of a person with entry age~$x$ at the contract time is determined by
\begin{align*}
_{\tau}p_x = \prod^{\tau}_{j=1}p_{x+j-1} &= \prod^{\tau}_{j=1}(1-q_{x+j-1}) \\
&= \prod^{\tau}_{j=1}(1-\frac{d_{x+j-1}}{l_{x+j-1}}),
\end{align*}
where $d_{x+j-1}$ denotes the number of death counts between ages $x+j-1$ and $x+j$, and $l_{x+j-1}$ represents the number of lives alive at age $x+j-1$.

Once we compute the survival probabilities for ages 60 and beyond, the calculation of annuity prices also depends on the zero-coupon bond price. For an $x$-year-old with a benefit, a monetary amount per year is given by
\begin{equation*}
a_{x:\overline{T}|}  = \sum^{T}_{\tau=1}B(0,\tau)_{\tau}p_{x},
\end{equation*}
where $B(0,\tau)$ is the price of the $\tau$-year bond and $_{\tau}p_x$ denotes the survival probability.

Table~\ref{tab:3} provides an example of annuity calculations using Australian data. We calculate the optimal estimate of annuity prices for various entry ages and maturities for female and male policyholders residing in Australia. We assume a constant interest rate at the value of $\lambda=4.1\%$ as of 18 May 2025, and hence the zero-coupon bond is given as $B(0,\tau) = \exp^{-\lambda\tau}$.
\begin{table}[!htb]
\centering
\caption{Australian annuity price estimates for various maturities and initial contract ages. The estimates are based on the 50-years-ahead mortality forecasts and the current interest rate of $4.1\%$ as of 18 May 2025.}\label{tab:3}
\begin{small}
\begin{tabular}{@{}lrrrrrrrrrrrr@{}}
  \toprule
  	& \multicolumn{6}{c}{Female}   	& \multicolumn{6}{c}{Male} \\
	\cmidrule(lr){2-7}	\cmidrule(ll){8-13}
Age & $T=5$ & 10 & 15 & 20 & 25 & 30 & 5 & 10 & 15 & 20 & 25 & 30 \\ 
  \midrule
  60 & 4.379 & 7.859 & 10.589 & 12.688 & 14.234 & 15.289 & 4.347 & 7.746 & 10.355 & 12.303 & 13.681 & 14.573 \\ 
  65 & 4.359 & 7.780 & 10.409 & 12.345 & 13.667 & 14.443 & 4.314 & 7.627 & 10.100 & 11.849 & 12.980 & 13.579 \\ 
  70 & 4.322 & 7.644 & 10.091 & 11.761 & 12.741 & 13.187 & 4.264 & 7.447 & 9.698 & 11.154 & 11.925 & 12.190 \\ 
  75 & 4.263 & 7.403 & 9.547 & 10.805 & 11.377 & 11.527 & 4.187 & 7.147 & 9.062 & 10.076 & 10.425 & 10.479 \\ 
  80 & 4.135 & 6.958 & 8.615 & 9.369 & 9.566 & 9.586 & 4.031 & 6.640 & 8.020 & 8.496 & 8.569 & 8.572 \\ 
  85 & 3.939 & 6.252 & 7.304 & 7.579 & 7.607 &  & 3.807 & 5.822 & 6.516 & 6.622 & 6.628 & \\ 
  90 & 3.552 & 5.167 & 5.589 & 5.632 &  &  & 3.307 & 4.445 & 4.620 & 4.629 &  &  \\ 
  95 & 3.020 & 3.809 & 3.890 &  &  &  & 2.504 & 2.888 & 2.908 &  &  & \\ 
  100 & 2.113 & 2.328 &  &  &  &  & 1.581 & 1.664 &  &  &  &  \\ 
  105 & 1.312 &  &  &  &  &  & 1.003 &  & &  &  &  \\ 
   \bottomrule
\end{tabular}
\end{small}
\end{table}

\begin{figure}[!htb]
\centering
\includegraphics[width=8.65cm]{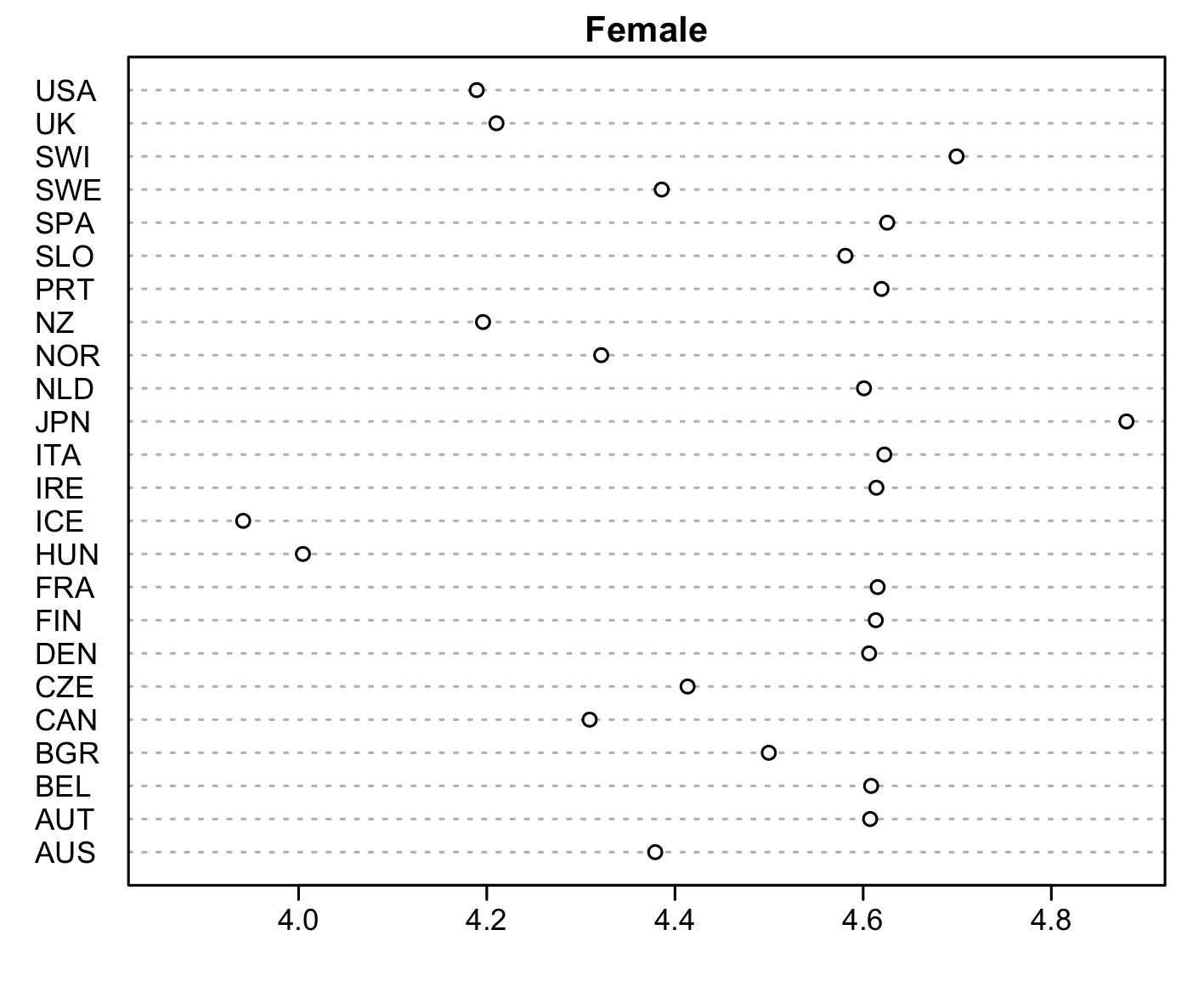}
\quad
\includegraphics[width=8.65cm]{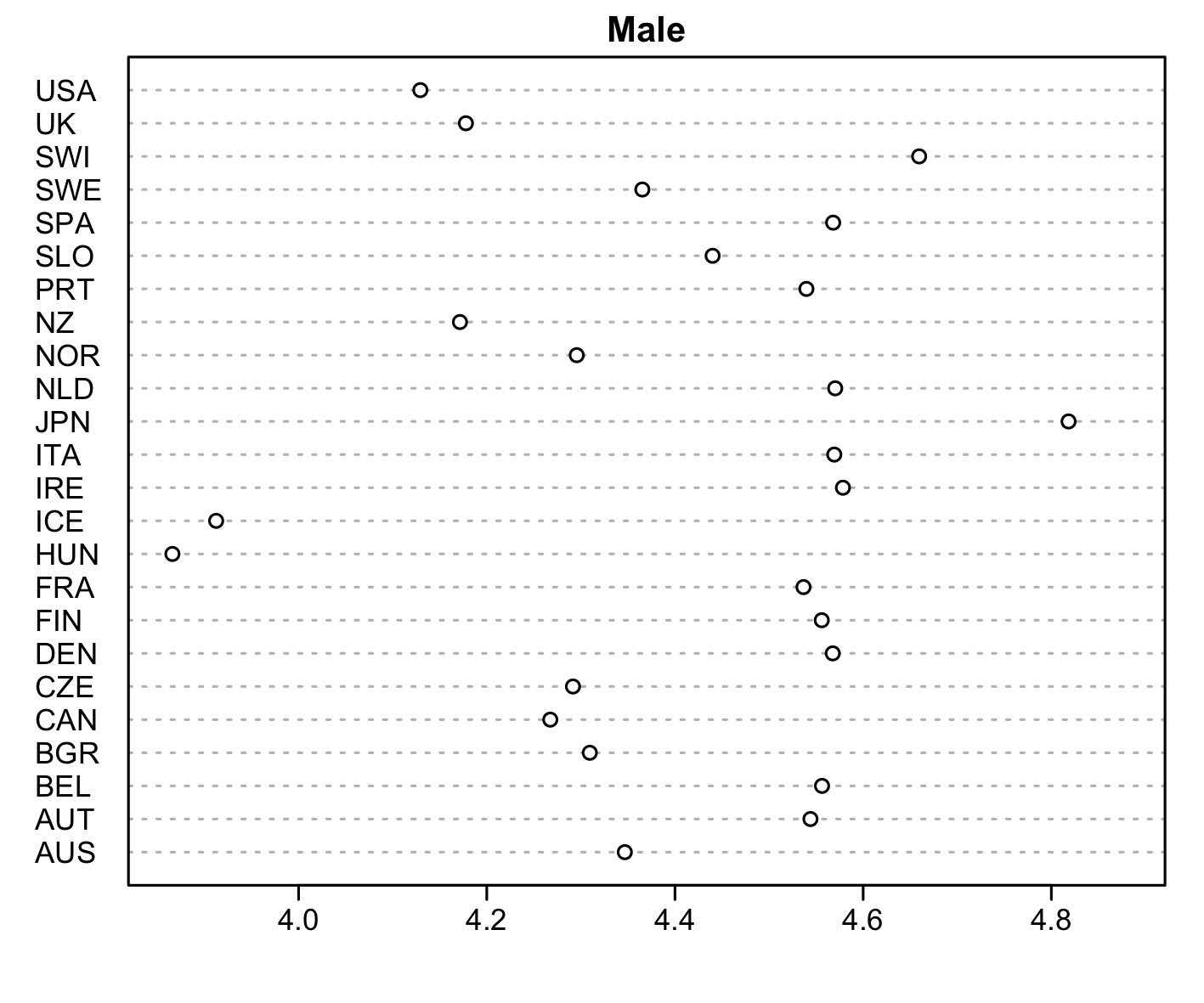}
\caption{With a return of one monetary amount annually, estimated annuity prices of the female and male populations in the 24 countries at entry age 60 with a maturity period of 5 years.}\label{fig:7}
\end{figure}

Figure~\ref{fig:7} shows dot charts for the estimated annuity prices of the 24 countries at entry age 60 with a maturity period of 5 years. With a return of one monetary amount, the estimated annuity price is the highest for Japan, while the two countries with the lowest values are \mbox{Iceland and Hungary.}

\begin{figure}[!htb]
\centering
\includegraphics[width=8.65cm]{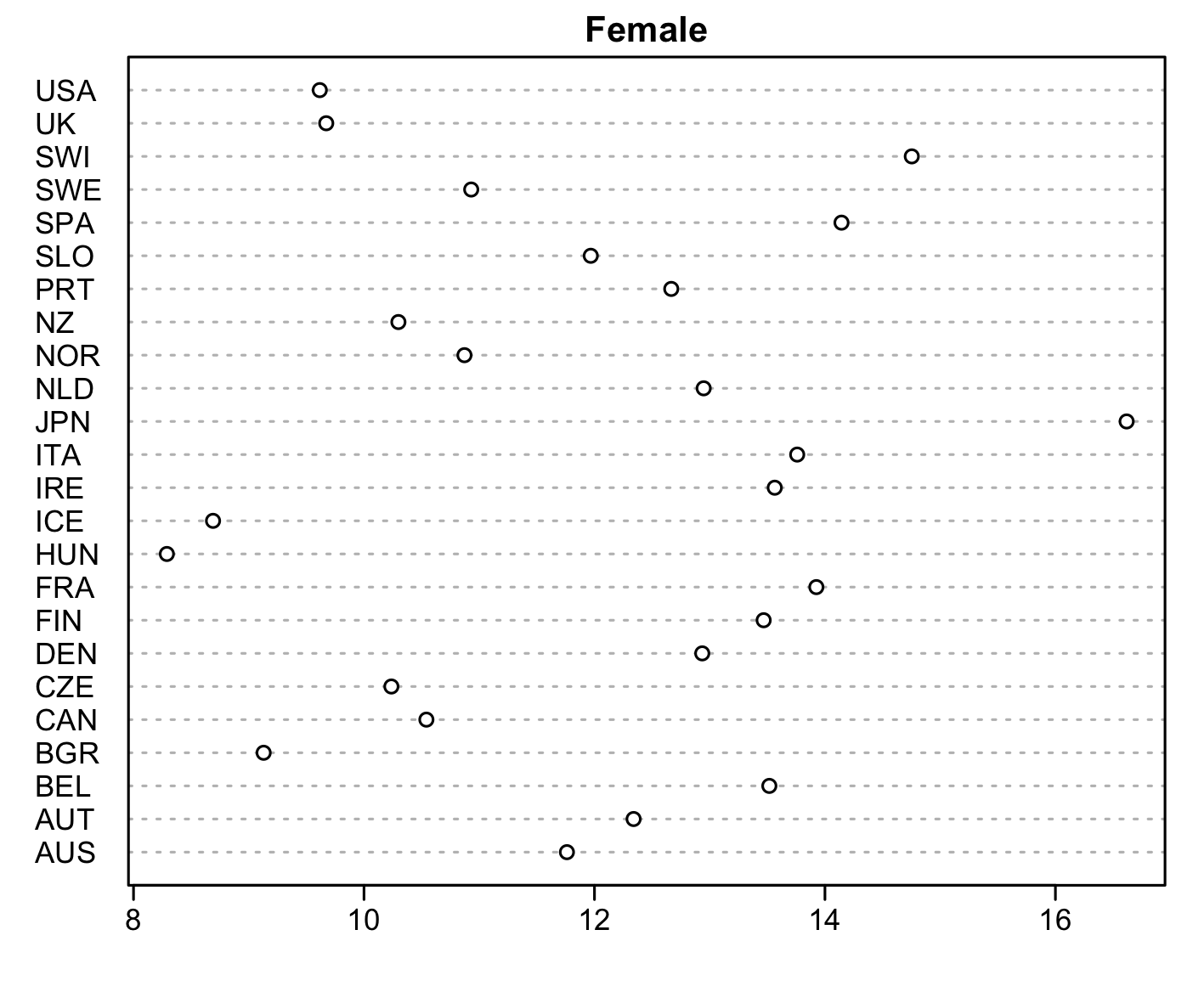}
\quad
\includegraphics[width=8.65cm]{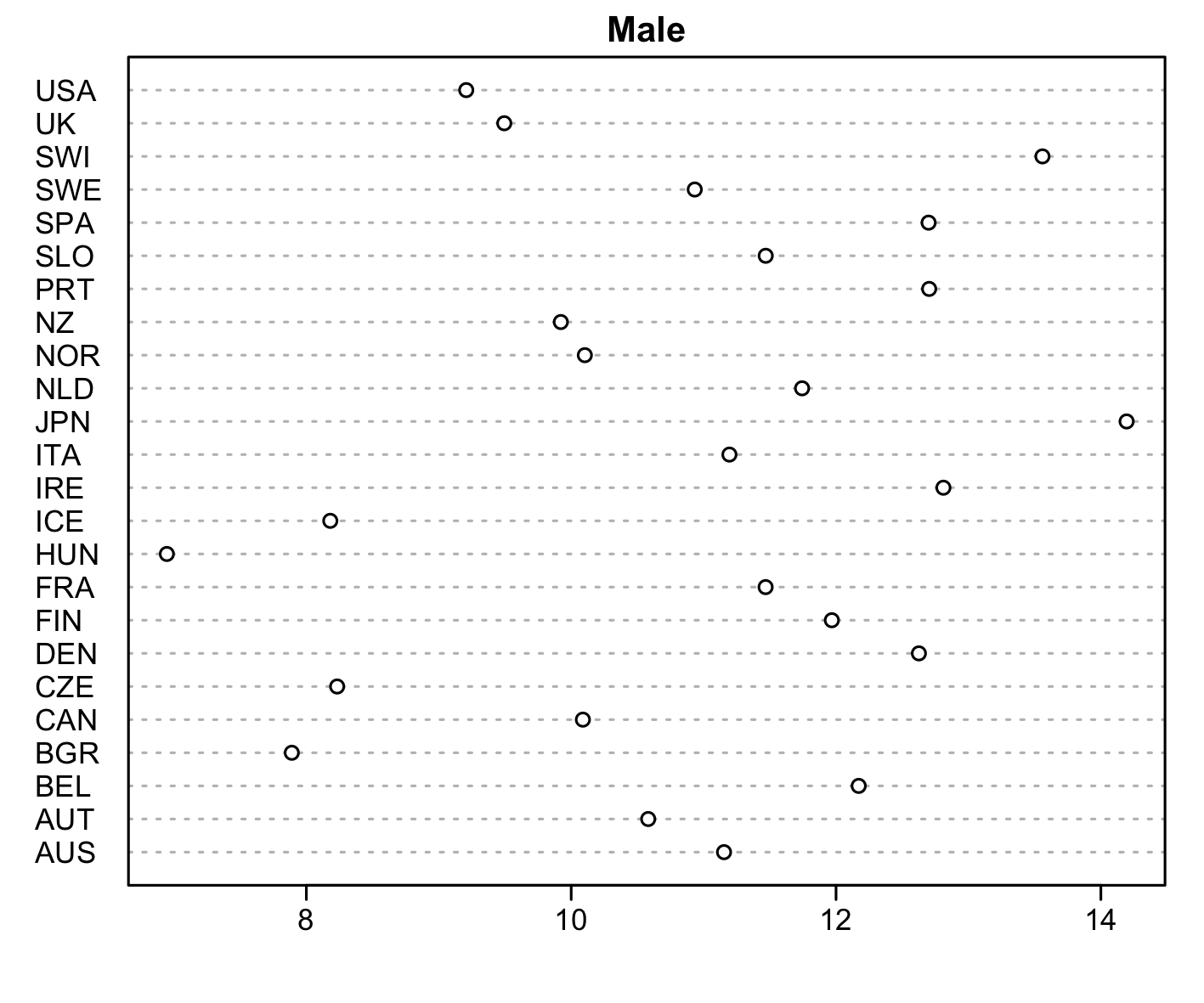}
\caption{With a return of one monetary amount annually, estimated annuity prices of the female and male populations in the 24 countries at entry age 70 with a maturity period of 20 years.}\label{fig:8}
\end{figure}

In Figure~\ref{fig:8}, we present dot charts of the estimated annuity prices for the 24 countries at entry age 70 with a 20-year maturity. With a return of one monetary amount, the estimated annuity price is highest in Japan and lowest in Hungary.

To measure forecast uncertainty, we construct the 50-years-ahead pointwise prediction intervals for the life-table death counts. For each of the 50 years, we search for the optimal tuning parameter that minimizes the CPD between the empirical coverage probability and the nominal coverage probabilities of 80\% and 95\%. In Table~\ref{tab:4}, we present the 80\% and 95\% pointwise prediction intervals of annuities for different entry ages and maturities, where age + maturity $\leq 110$.
\begin{small}
\begin{center}
\tabcolsep 0.05in
\renewcommand*{\arraystretch}{.92}
\begin{longtable}{@{}llcccccc@{}}
\caption{With a return of \$1 annually, the 80\% and 95\% pointwise prediction intervals of annuity prices with different entry ages and maturities $(T)$ for female and male populations residing in Australia.}\label{tab:4}\\
\toprule
Sex 	& Age & $T=5$ & 10 & 15 & 20 & 25 & 30 \\ 
\midrule
\endfirsthead
\toprule
Sex 	& Age & $T=5$ & 10 & 15 & 20 & 25 & 30 \\ 
\midrule
\endhead
\hline \multicolumn{8}{r}{{Continued on next page}} \\
\endfoot
\endlastfoot
\multicolumn{4}{l}{\hspace{-.05in}{80\% nominal coverage probability}} & \\
\cmidrule{1-4}
F 	& 60 & (4.373, 4.386) & (7.834, 7.887) & (10.530, 10.659)  & (12.577, 12.822) & (14.055, 14.457) & (15.044, 15.603) \\ 
	& 65 & (4.348, 4.372) & (7.734, 7.835) & (10.306, 10.536) & (12.162, 12.578) & (13.405, 14.009) & (14.118, 14.874) \\ 
  	& 70 & (4.301, 4.349) & (7.566, 7.742) & (9.923, 10.307) & (11.501, 12.105) & (12.407, 13.191) & (12.818, 13.687) \\ 
  	& 75 & (4.232, 4.304) & (7.286, 7.557) & (9.330, 9.837) & (10.504, 11.215) & (11.037, 11.844) & (11.185, 11.994) \\ 
  	& 80 & (4.083, 4.203) & (6.816, 7.149) & (8.386, 8.929) & (9.098, 9.742) & (9.296, 9.936) & (9.321, 9.946) \\ 
  	& 85 & (3.898, 3.996) & (6.137, 6.411) & (7.153, 7.513) & (7.435, 7.776) & (7.472, 7.791)  &  \\ 
  	& 90 & (3.506, 3.615) & (5.097, 5.264) & (5.537, 5.657) & (5.595, 5.679) &  &  \\ 
  	& 95 & (2.990, 3.041) & (3.703, 3.883) & (3.743, 3.993) &  & &  \\ 
  	& 100 & (1.975, 2.203) & (2.086, 2.490) &  &  &  &  \\ 
  	& 105 & (0.858, 1.543) &  &  &  &  &  \\ 
\midrule	
M 	& 60 & (4.336, 4.359) & (7.699, 7.799) & (10.253, 10.477) & (12.139, 12.505) & (13.452, 13.972) & (14.275, 14.965) \\ 
	& 65 & (4.292, 4.341) & (7.550, 7.722) & (9.956, 10.282) & (11.632, 12.133) & (12.682, 13.387) & (13.212, 14.097) \\ 
  	& 70 & (4.235, 4.300) & (7.363, 7.557) & (9.541, 9.912) & (10.907, 11.507) & (11.596, 12.410) & (11.8218, 12.741) \\ 
 	& 75 & (4.165, 4.216) & (7.065, 7.265) & (8.884, 9.330) & (9.801, 10.499) & (10.102, 10.927) & (10.147, 10.994) \\ 
 	& 80 & (3.987, 4.096) & (6.488, 6.869) & (7.748, 8.441) & (8.162, 9.015) & (8.225, 9.1051) & (8.228, 9.110) \\ 
 	& 85 & (3.730, 3.922) & (5.610, 6.144) & (6.228, 6.956) & (6.321, 7.083) & (6.326, 7.090) &  \\ 
 	& 90 & (3.209, 3.447) & (4.263, 4.707) & (4.422, 4.904) & (4.430, 4.914) &  &  \\ 
 	& 95 & (2.457, 2.565) & (2.828, 2.966) & (2.847, 2.987) &  &  &  \\ 
 	& 100 & (1.569, 1.596) & (1.650, 1.680) &  &  &  &  \\ 
 	& 105 & (1.000, 1.006) &  &  &  &  &  \\ 	
\midrule
\multicolumn{4}{l}{\hspace{-.05in}{95\% nominal coverage probability}} & \\
\cmidrule{1-4}
F 	& 60 & (4.369, 4.392) & (7.818, 7.911) & (10.494, 10.717) & (12.513, 12.931) & (13.957, 14.635) & (14.916, 15.849) \\ 
	& 65 & (4.341, 4.383) & (7.709, 7.878) & (10.250, 10.636) & (12.068, 12.758) & (13.274, 14.270) & (13.962, 15.199) \\ 
 	& 70 & (4.289, 4.370) & (7.525, 7.818) & (9.840, 10.472) & (11.377, 12.363) & (12.253, 13.524) & (12.650, 14.055) \\ 
 	& 75 & (4.216, 4.335) & (7.230, 7.671) & (9.232, 10.047) & (10.372, 11.507) & (10.890, 12.175) & (11.036, 12.326) \\ 
 	& 80 & (4.061, 4.250) & (6.757, 7.276) & (8.293, 9.136) & (8.990, 9.986) & (9.186, 10.178) & (9.214, 10.185) \\ 
 	& 85 & (3.882, 4.029) & (6.095, 6.504) & (7.098, 7.636) & (7.381, 7.892) & (7.421, 7.902) &  \\ 
 	& 90 & (3.490, 3.650) & (5.072, 5.318) & (5.518, 5.696)  & (5.581, 5.710) &  &  \\ 
 	& 95 & (2.975, 3.048) & (3.649, 3.908) & (3.674, 4.028)  &  &  &  \\ 
 	& 100 & (1.902, 2.233) & (1.971, 2.544)  &  &  &  &  \\ 
 	& 105 & (0.622, 1.610) &  &  &  &  &  \\ 	
\midrule
M 	& 60 & (4.328, 4.369) & (7.671, 7.842) & (10.193, 10.574) & (12.047, 12.666)  & (13.330, 14.203) & (14.124, 15.279) \\ 
 	& 65 & (4.279, 4.362) & (7.509, 7.794) & (9.882, 10.421) & (11.524, 12.352) & (12.541, 13.703)  & (13.043, 14.506) \\ 
 	& 70 & (4.221, 4.327) & (7.323, 7.640) & (9.470, 10.074) & (10.798, 11.779)  & (11.454, 12.791) & (11.666, 13.175) \\ 
 	& 75 & (4.156, 4.238) & (7.031, 7.353) & (8.811, 9.533)  & (9.689, 10.829) & (9.973, 11.320)  & (10.015, 11.397) \\ 
 	& 80 & (3.970, 4.143) & (6.428, 7.044) & (7.641, 8.767) & (8.033, 9.420) & (8.091, 9.523) & (8.094, 9.528) \\ 
 	& 85 & (3.699, 4.010)  & (5.525, 6.391) & (6.114, 7.294) & (6.203, 7.436) & (6.207, 7.443) &  \\ 
 	& 90 & (3.169, 3.544)  & (4.191, 4.886) & (4.345, 5.098) & (4.353, 5.109) &  &  \\ 
 	& 95 & (2.439, 2.599) & (2.806, 3.010) & (2.824, 3.031) &  &  &  \\ 
 	& 100 & (1.565, 1.602) & (1.645, 1.687)  &  &  &  &  \\ 
 	& 105 & (0.999, 1.007) &  & &  &  &  \\ 	
\bottomrule
\end{longtable}
\end{center}
\end{small}

\vspace{-.3in}
    
\begin{figure}[!htb]
\centering
\subfloat[Entry age 60 with 5-year maturity]
{\includegraphics[width=8.55cm]{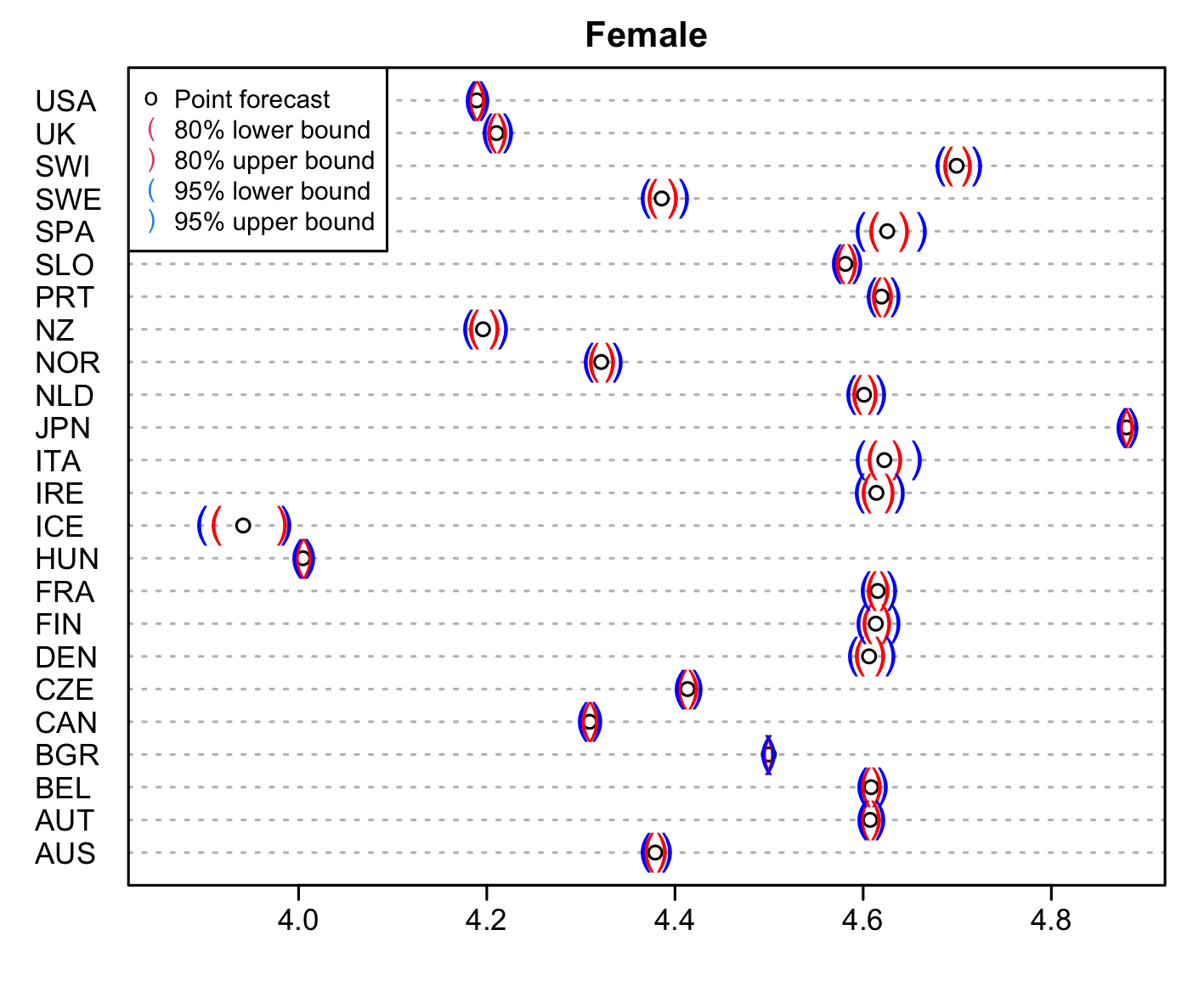}\label{fig:9a}}
\quad
\subfloat[Entry age 60 with 5-year maturity]
{\includegraphics[width=8.55cm]{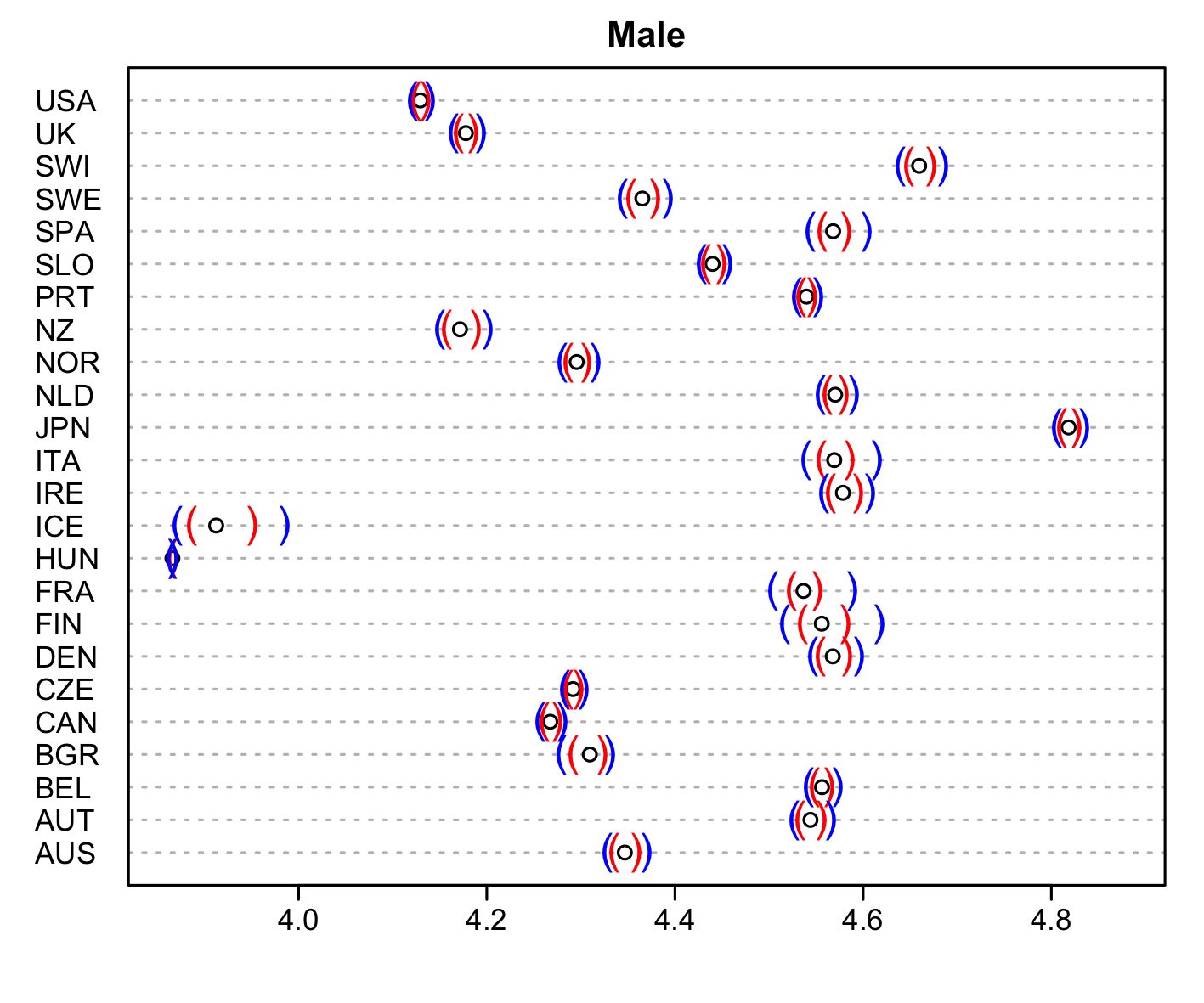}\label{fig:9b}}
\\
\subfloat[Entry age 70 with 20-year maturity]
{\includegraphics[width=8.55cm]{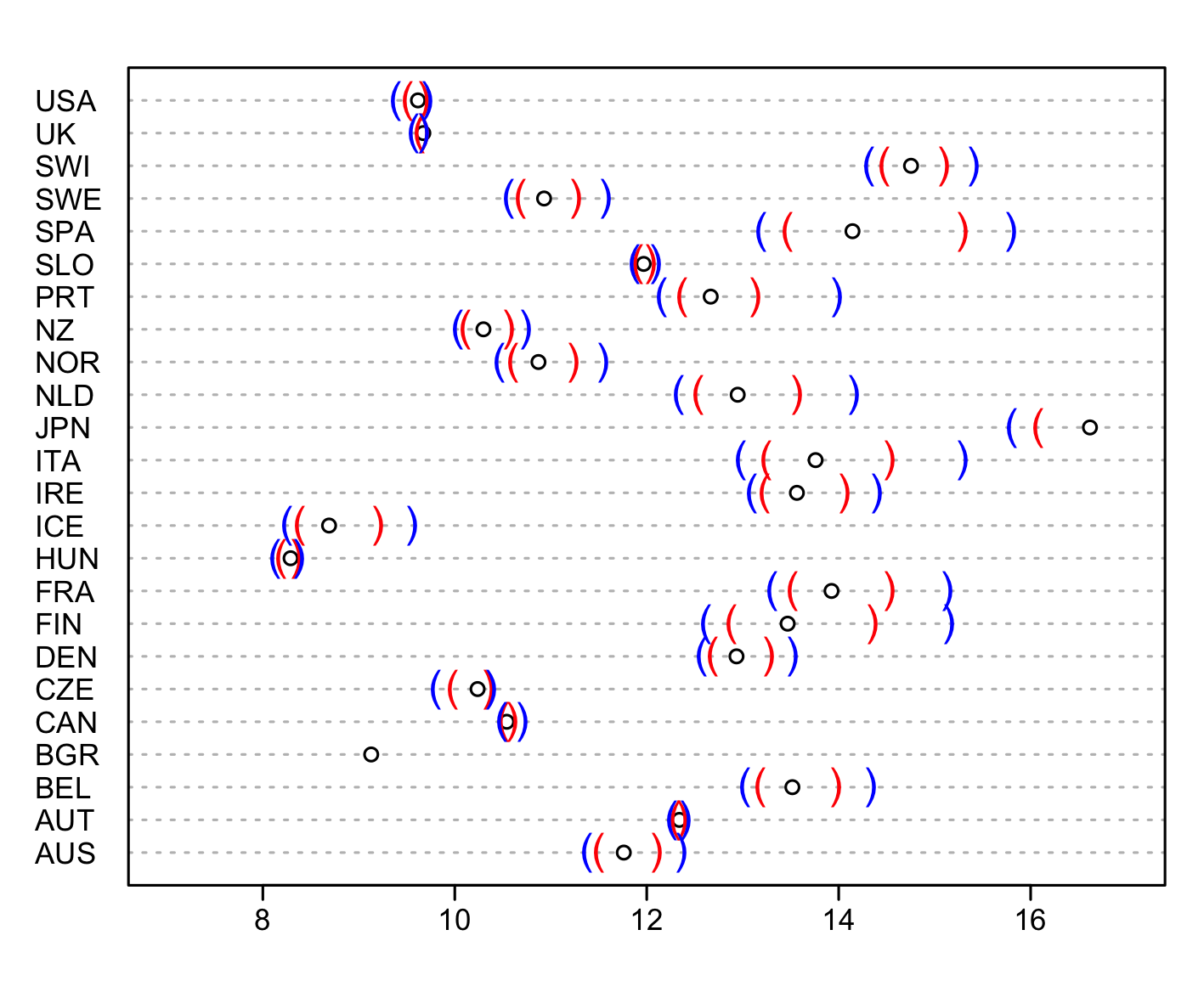}\label{fig:9c}}
\quad
\subfloat[Entry age 70 with 20-year maturity]
{\includegraphics[width=8.55cm]{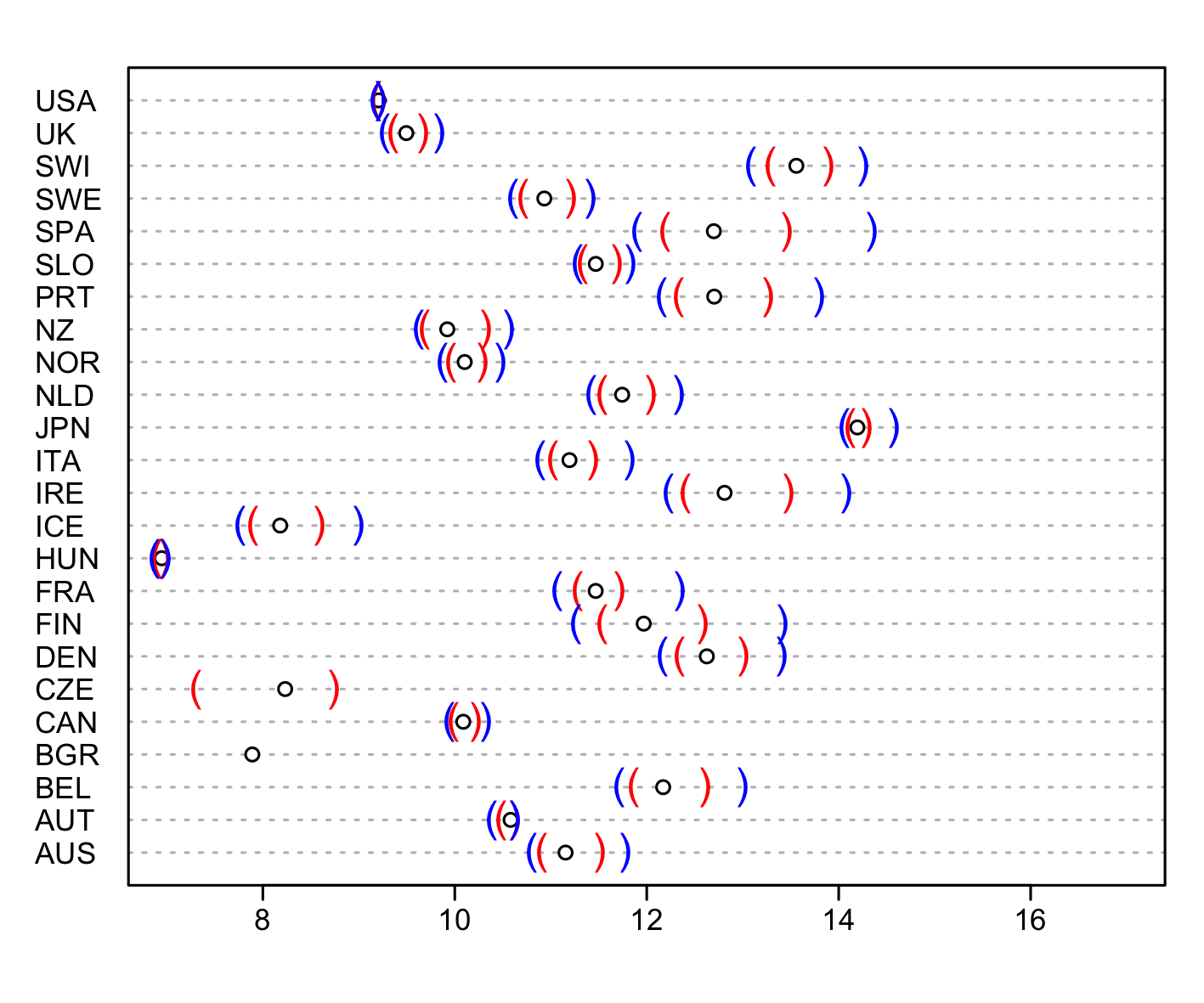}\label{fig:9d}}
\caption{With a return of one monetary amount annually, the 80\% and 95\% pointwise prediction intervals of the annuity prices of the female and male populations in the 24 countries.}\label{fig:9}
\end{figure}

The forecast uncertainty increases the width of the prediction interval as maturity increases from $T=5$ to $T=30$ for a given age. They also increase as the initial ages at contract entry range from 60 to 105 for a given maturity. Figures~\ref{fig:9a} and~\ref{fig:9b} show the 80\% and 95\% prediction intervals for annuity prices at entry age 60 with 5-year maturity in 24 countries, while Figures~\ref{fig:9c} and~\ref{fig:9d} show the results for entry age 70 with 20-year maturity. 

With a return of one monetary unit, estimates are shown as open circles, with open intervals indicating 80\% and 95\% pointwise prediction intervals in red and blue, respectively. The forecast uncertainty is noticeably higher for Iceland and lower for Hungary. As maturity increases, the prediction intervals widen, reflecting greater long-term uncertainty.

\section{Conclusion}\label{sec:6}

We present a comprehensive cross-country validation study of compositional mortality forecasting methods. For modeling and forecasting the age distribution of death counts, we consider the centered log-ratio and cumulative distribution function transformations. Both transformations allow us to map a constrained object into an unconstrained space, where linear techniques can be applied. Such a technique is principal component analysis, which can summarize and model variation in the unconstrained data. Between the two transformations, we demonstrate an additional use of the CDF transformation, providing a scale-free measure to compare the probabilities of dying across genders and countries. 

Using data from 24 countries, we evaluate point forecast accuracy (measured by the Kullback-Leibler and Jensen-Shannon divergences) and interval forecast accuracy (measured by the coverage probability difference between empirical and nominal coverage probabilities and the mean interval score). For producing both point and interval forecasts, the cumulative distribution function transformation systematically provides smaller point and interval forecast errors than the centered log-ratio transformation. Between the two methods for selecting the number of principal components, we advocate $K=6$ for forecasting. 

We apply the cumulative distribution function transformation to generate 50-years-ahead forecasts of age-specific life-table death counts for 24 countries. From the forecasted age distribution of death counts, we evaluate and compare the life expectancy at birth across multiple countries. From the forecasted life-table death counts, we convert them into age-specific survival probabilities and obtain temporary annuity prices. As expected, we observe that the survival probability has a pronounced impact on actuarially fair annuity prices, reflecting gender gap and cross-country heterogeneity. Coupling with the current interest rates in Table~\ref{tab:1}, we quantify the uncertainty of annuity prices via pointwise prediction intervals. These prediction intervals primarily reflect model uncertainty rather than uncertainty around interest rates themselves. For reproducibility, the computer \Rlogo \ code is available at \url{https://github.com/hanshang/CLR_vs_CDF_transformation}.

There are several ways the methodology presented in the paper could be extended, and we briefly discuss six. 
\begin{inparaenum}[1)]
\item We consider a univariate functional time-series method to forecast each series individually. While it provides a common approach for comparing the two transformations, other joint modeling methods could also be used. Also, when forecasting principal component scores, we implement exponential smoothing. Other time-series forecasting methods may also be used.
\item We consider two transformations, namely the centered log-ratio and cumulative distribution function transformations. There exist other transformations that could be considered, namely additive or isometric log-ratio \citep{GGB+23}. While these transformations are variants of the centered log-ratio, we could also consider intrinsic statistical models, such as Wasserstein autoregression \citep{ZKP22} or optimal transport \citep{ZM23}, for modeling density-valued time series.
\item We could also consider testing for time-stochastic dominance between populations, as studied by \cite{LLW23}.
\item For all countries, the sample period encompasses the COVID-19 pandemic, but we do not examine the changes in the age distribution of deaths before and after the pandemic.
\item Although we calculate temporary annuities, other types of annuity prices, such as the whole-life or deferred annuity, are possible. 
\item We could consider cohort life-table death counts for modeling a particular group of individuals, such as baby boomers.
\end{inparaenum}

\section*{Data availability statement}

The primary source for the analysis is the freely available data in the Human Mortality Database (HMD), which is publicly accessible after creating a free HMD account.

\section*{Funding statement}

This research was supported by the Australian Research Council Future Fellowship (grant no. FT240100338). The first author is grateful for the insightful comments and suggestions from the seminar participants at the University of New South Wales.

\section*{Competing interests}

No potential conflict of interest was reported by the authors.

\newpage
\bibliographystyle{agsm}
\bibliography{LTDC_transformation}

\end{document}